\documentclass[11pt,twoside,a4paper,cmspaper,final,collab]{cms-tdr}
\def\svnVersion{b27ba21-D}\def\svnDate{2026/11/02}\def\cmsCernNoTag{CERN-EP-2026-262}\def\cmsCernDate{\today}\def\cmsMessage{Submitted to the Journal of High Energy Physics}
\begin{document}\cmsNoteHeader{B2G-24-023}

\newcommand{\eV}{\ensuremath{\,\text{e\hspace{-.08em}V}}\xspace}
\newcommand{\meV}{\ensuremath{\,\text{me\hspace{-.08em}V}}\xspace}
\newcommand{\mN}{\ensuremath{m_{\PN}}\xspace}
\newcommand{\mW}{\ensuremath{m_{\PW}}\xspace}
\newcommand{\mZ}{\ensuremath{m_{\PZ}}\xspace}
\newcommand{\VmNSq}{\ensuremath{\abs{V_{\Pgm\PN}}^{2}}\xspace}
\newcommand{\VeNSq}{\ensuremath{\abs{V_{\Pe\PN}}^{2}}\xspace}
\newcommand{\VlN}{\ensuremath{\abs{V_{\Pell\PN}}}\xspace}
\newcommand{\BtlN}{\ensuremath{\mathcal{B}(\PQt\to\PQb\Pell\PN)}\xspace}
\newcommand{\Njets}{\ensuremath{N(\text{j})}\xspace}
\newcommand{\Nbjets}{\ensuremath{N(\PQb)}\xspace}
\providecommand{\PAf}{\HepAntiParticle{\Pf}{}{}\xspace}
\newlength\cmsTabSkip\setlength{\cmsTabSkip}{1ex}
\providecommand{\cmsTable}[1]{\resizebox{\textwidth}{!}{#1}}

\cmsNoteHeader{B2G-24-023}

\title{Search for heavy neutral leptons in top quark decays with same-sign, same-flavor dilepton final states in proton-proton collisions at \texorpdfstring{$\sqrt{s}=13\TeV$}{sqrt(s) = 13 TeV}}

\date{\today}

\abstract{A search for a Majorana-type heavy neutral lepton (HNL) produced in top quark decays is reported. The search is based on proton-proton collision data at $\sqrt{s}=13\TeV$, corresponding to an integrated luminosity of 138\fbinv, collected during 2016--2018 with the CMS detector at the CERN LHC. In the type-I seesaw model, a Majorana HNL can mediate lepton-number-violating top quark decays. The search targets top quark pair production events in which one top quark decays into a bottom quark and a $\PW$ boson, which subsequently decays into a charged lepton (electron or muon) and an HNL. The Majorana HNL then decays into a $\PW$ boson and a charged lepton with the same sign and flavor as the one produced with the HNL. The experimental signature features a same-sign, same-flavor lepton pair and jets. The HNL masses between 20 and 100\GeV are probed. The data agree with the standard model background prediction. Observed (expected) upper limits at 95\% confidence level are set on the branching fraction of the top quark decay into an HNL, a charged lepton, and a bottom quark, varying between $[1.1\,(0.71)]\times10^{-5}$ and $[4.8\,(3.0)]\times10^{-5}$ in the dielectron channel, and between $[3.7\,(3.4)]\times10^{-6}$ and $[2.4\,(1.3)]\times10^{-5}$ in the dimuon channel.}

\hypersetup{
pdfauthor={CMS Collaboration},
pdftitle={Search for heavy neutral leptons in top quark decays with same-sign, same-flavor dilepton final states in proton-proton collisions at sqrt(s)=13 TeV},
pdfsubject={CMS},
pdfkeywords={CMS, neutrino, top quark, type-I seesaw, heavy neutral lepton}}

\maketitle 

\section{Introduction}
\label{Intro}

The observation of neutrino oscillations has confirmed the nonzero masses for at least two mass eigenstates of the standard model (SM) neutrinos ($\PGnell$)~\cite{PDG2024}. 
The mass-squared differences measured from neutrino oscillation experiments have set lower limits on the mass values of the two heavier neutrinos at 8.7 and 50\meV~\cite{PDG2024}. 
Cosmological measurements of the cosmic microwave background and baryon acoustic oscillations also place an upper limit on the sum of the neutrino masses at 64\meV~\cite{Elbers:2025vlz}. 
An important question is the origin of the smallness of the neutrino masses. 
Since neutrinos are assumed to be massless in the SM, several beyond-the-SM (BSM) scenarios have been suggested to resolve this issue.
For example, in the type-I seesaw mechanism~\cite{Minkowski_T1Seesaw,GellMann_T1Seesaw,Yanagida_T1Seesaw,Mohapatra_T1Seesaw}, Majorana mass ($m_{\text{M}}$) terms are introduced for right-handed chiral neutrino fields, in addition to Dirac masses ($m_{\text{D}}$) originating from the Yukawa couplings of neutrino fields to the Higgs field. 
 This results in two physical states for each neutrino flavor: a light SM-like neutrino $\PGnell$ and a heavy neutral lepton (HNL) \PN. 
 The mixing between $\PGnell$ and \PN can be quantified by a mixing parameter $V_{\Pell\PN}$, which---along with the physical masses---is determined by $m_{\text{D}}$ and $m_{\text{M}}$. This mixing parameter determines the HNL component of the neutrino flavor eigenstates and hence determines the interaction between the HNL and the \PW boson. 
In the simplest seesaw realization, the light-neutrino masses scale approximately as $m_{\text{D}}^{2}/m_{\text{M}}$, where $m_{\text{M}}$ is effectively the HNL mass (\mN). 
More general realizations can allow larger HNL-SM mixing for a given \mN~\cite{HNLSummary}. 

Searches for HNL span a broad $\mN$ range. At low masses of $\mathcal{O}$(100)\eV, beta-decay spectra have been examined for spectral kinks that would indicate the presence of an HNL~\cite{Galeazzi:2001py,Hiddemann:1995ce}. 
At collider-accessible \mN values up to $\mathcal{O}$(1)\TeV, the CERN LHC experiments searched for the decay of the on- and off-shell \PW bosons, directly produced in proton-proton ($\Pp\Pp$) collisions, into an HNL ($\Pp\Pp\to \PWpm\to \PN \Pellpm$), where the hadronic ($\PN \to \Pellpm\PQq\PAQq'$) and leptonic ($\PN \to \Pellp\Pellm'\PGn$) decay chains of an HNL into quarks (\PQq), charged leptons ($\Pell$) including the tau leptons (\PGt), and $\PGnell$ lead to dilepton and trilepton final states, respectively~\cite{CMS_HN2l2j,CMS_HN3lv,ATLAS_HN3lv, CMS:2024bni}. The LHCb experiment searched for HNLs produced in B meson decays and decaying into a muon (\PGm) and a pion~\cite{LHCb:2025ymr}. 
Throughout the paper, \Pell indicates electrons (\Pe) and muons, excluding \PGt leptons unless mentioned otherwise. 
These direct searches for HNLs have focused on analyzing their decay products or the energy spectrum of the charged lepton produced with the HNL.

 The hypothetical production of HNLs via top quark (\PQt) decays has recently received attention in the theory community and provides a complementary probe of HNLs below the \PW boson mass (\mW)~\cite{BarShalom_TopHN,Si_TopHN,Delepine_TopHN,Liu_TopHN,Lu_TopHNCPV}. In some BSM scenarios, such as supersymmetric models with \emph{R}-parity violation or models with leptoquarks, destructive interference can reduce the sensitivity of direct-search channels, increasing the complementarity of HNL production via top quark decays. In addition, the presence of both $\PWp$ and $\PWm$ bosons in top quark pair (\ttbar) events makes this production mode potentially sensitive to charge-parity violating effects in the HNL sector~\cite{Lu_TopHNCPV}. The ATLAS Collaboration recently searched for Majorana HNLs produced in top quark decays in the same-sign (SS), same-flavor (SF) dilepton final state, using 140\fbinv of data at $\sqrt{s} = 13\TeV$, and set limits for HNL masses between 15 and 75\GeV~\cite{ATLAS:TopHNL}. In this article, we present a search for HNLs in top quark decays using SS, SF dilepton events in the $\Pp\Pp$ collision data at $\sqrt{s} = 13\TeV$, collected using the CMS detector between 2016 and 2018, corresponding to an integrated luminosity of 138\fbinv. 

\begin{figure*}[htbp]
  \centering
  \includegraphics[width=0.35\textwidth]{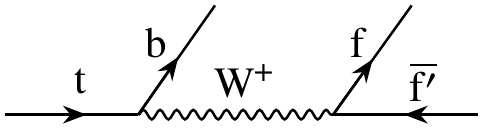}
  \hfill
  \includegraphics[width=0.58\textwidth]{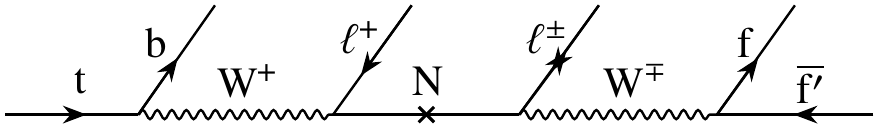}

  \caption{Representative Feynman diagrams of the SM top quark decays (left) and BSM decays involving an HNL (right), where $\Pell$ represents $\Pe$ or $\Pgm$, and $\Pf,\Pf^\prime$ represent SM fermions.}
  \label{fig:SignalFeynman}
\end{figure*}

The search targets lepton number violating (LNV) processes in \ttbar production, where the top quark decays into an HNL, which further decays into SM particles, $\PQt\to\PQb\PWp\to\PQb\Pellp\PN\to\PQb\Pellp\Pellp\PWm\to\PQb\Pellp\Pellp\Pf\PAf'$, while the top antiquark decays via the SM process $\PAQt\to\PAQb\PWm\to\PAQb \Pf\PAf'$, where \Pf denotes an SM fermion. Charge-conjugated processes are implied throughout the text. If $\mN<\mW$, the $\PW$ boson produced in the decay of the HNL is off-shell; if $\mN >\mW$, the $\PW$ boson from which the HNL originates is off-shell. Representative Feynman diagrams of these processes are depicted in Fig.~\ref{fig:SignalFeynman}. In this search, only mixings of an HNL with the electron and muon neutrinos are considered, with strengths \VeNSq and \VmNSq, respectively. Each mixing parameter is considered separately, with one mixing parameter nonzero and the other set to zero. 

Searching for LNV decay modes is of paramount interest, as they provide a characteristic signature of Majorana HNLs. The discovery of a particle with a Majorana mass would provide evidence for the existence of different mass generation mechanisms in nature. In the context of an LNV search, the hadronic decay mode $\PN\to\Pell \PQq\PAQq'$ is the most powerful one, as there is no ambiguity between the lepton number conserving and violating decays. In the case of the leptonic decay mode, $\PN\to\Pell\Pell'\PGn$, it is impossible to conclusively determine whether LNV occurred, since the lepton number of the undetected neutrino through a decay involving a $\PZ$ boson, $\PN\to\PZ\PGn\to\Pell\Pell\PGn$, cannot be known. The possibility of an HNL mixing with more than one generation of SM neutrinos also complicates the interpretation of events in the fully leptonic decay modes, even for the $\PW$ boson mediated cases.

In this search, \mN values between 20 and 100\GeV are considered. For masses below \mW, the HNL decay width scales approximately as $\mN^5$ for a fixed mixing parameter, so a smaller \mN results in a longer lifetime of the HNL~\cite{Si_TopHN}. For \mN values below 20\GeV, the decay length of an HNL cannot be neglected within the allowed mixing parameter region and must be addressed using specialized search techniques optimized for long-lived particles. For masses above \mW, this HNL production channel cannot be mediated by an on-shell $\PW$ boson from a top quark decay. The production rate decreases rapidly with increasing \mN values, and this analysis therefore restricts the search to $\mN<100\GeV$. A machine-learning technique is used to classify events as background or signal, based on their distinguishing characteristics. The backgrounds are composed of nonprompt-lepton, conversion, charge mismeasurement, and prompt-lepton components, as discussed in Section~\ref{BkgEst}.

This paper is structured as follows. The CMS detector and object reconstruction are described in Section~\ref{Detector}, and the data and simulation samples are discussed in Section~\ref{Datasets}. Event reconstruction and selection are presented in Section~\ref{Events}. The background estimation techniques are summarized in Section~\ref{BkgEst}. The systematic uncertainties considered in this analysis are described in Section~\ref{Systematic}. Lastly, the results and interpretations are presented in Section~\ref{Result}, with an overall summary in Section~\ref{Summary}. Tabulated results are available in the HEPData record~\cite{hepdata} for this analysis.

\section{The CMS detector and event reconstruction}
\label{Detector}
The CMS apparatus~\cite{CMS:2008xjf,CMS:2023gfb} is a multipurpose, nearly hermetic detector, designed to trigger on~\cite{CMS:2020cmk,CMS:2016ngn,CMS:2024aqx} and identify \Pe, \PGm, photons (\PGg), and (charged and neutral) hadrons~\cite{CMS:2020uim,CMS:2018rym,CMS:2014pgm}. Its central feature is a superconducting solenoid of 6\unit{m} internal diameter, providing a magnetic field of 3.8\unit{T}. Within the solenoid volume are a silicon pixel and strip tracker, a lead tungstate crystal electromagnetic calorimeter (ECAL), and a brass and scintillator hadron calorimeter (HCAL), each composed of a barrel and two endcap sections. Forward calorimeters extend the pseudorapidity ($\eta$) coverage provided by the barrel and endcap detectors. Muons are reconstructed using gas-ionization detectors interleaved with the layers of the steel flux-return yoke outside the solenoid. More detailed descriptions of the CMS detector, together with a definition of the coordinate system used and the relevant kinematic variables, can be found in Refs.~\cite{CMS:2008xjf,CMS:2023gfb}.

Events of interest are selected using a two-tiered trigger system. The first level (L1), composed of custom hardware processors, uses information from the calorimeters and muon detectors to select events at a rate of around 100\unit{kHz} within a fixed latency of 4\mus~\cite{CMS:2020cmk}. The second level, known as the high-level trigger, consists of a farm of processors running a version of the full event reconstruction software optimized for fast processing, and reduces the event rate to a few kHz before data storage~\cite{CMS:2016ngn,CMS:2024aqx}.

The global event description with the particle-flow (PF) algorithm~\cite{CMS:2017yfk} aims to reconstruct and identify each individual particle in an event, with an optimized combination of all subdetector information. In this process, the identification of the particle type (photon, electron, muon, charged hadron, neutral hadron) plays an important role in the determination of the particle direction and energy. Photons are identified as ECAL energy clusters not linked to the extrapolation of any charged-particle trajectory to the ECAL. Electrons are identified as a primary charged-particle track and potentially many ECAL energy clusters corresponding to this track extrapolation to the ECAL and to possible bremsstrahlung photons emitted along the way through the tracker material. Muons are identified as tracks in the central tracker consistent with either a track or several hits in the muon system, and associated with calorimeter deposits compatible with the muon hypothesis. Charged hadrons are identified as charged-particle tracks neither identified as electrons, nor as muons. Finally, neutral hadrons are identified as HCAL energy clusters not linked to any charged-hadron trajectory, or as a combined ECAL and HCAL energy excess with respect to the expected charged-hadron energy deposit. The primary vertex (PV) of the interaction is taken to be the vertex corresponding to the hardest scattering in the event, evaluated using tracking information alone, as described in Section 9.4.1 of Ref.~\cite{CMS-TDR-15-02}.

 Electrons reconstructed within the tracker coverage ($\abs{\eta}<2.5$), except for the calorimeter gap region ($1.442<\abs{\eta}<1.566$), are used in the analysis. We consider electrons with transverse momentum $\pt> 15\GeV$ and require them to be isolated from other particles in the event to reject nonprompt leptons from decays in jets. Two electron identification (ID) working points are used in the analysis, referred to as loose and tight ID in the following. Electron ID is based on track quality criteria, matching between the electron trajectory and the associated ECAL cluster, and shower shape observables~\cite{CMS:2020uim}. The relative isolation ($I_\text{rel}$) is defined as the scalar \pt sum of all PF particles reconstructed within a cone around the lepton direction divided by the lepton \pt. The cone radius is defined by $\DR = \sqrt{\smash[b]{(\Delta\eta)^2+(\Delta\phi)^2}}$, where $\phi$ is the azimuthal angle in radians. We use a \pt-dependent cone radius of 0.2 for leptons with $\pt<50\GeV$, of $10\GeV/\pt$ for $50<\pt<200\GeV$, and of 0.05 for $\pt>200\GeV$, which improves the efficiency for high-\pt leptons by removing the accidental overlap with other particles~\cite{MiniIso}. The loose and tight electron IDs require $I_\text{rel}< 0.4$ and 0.1, respectively. Furthermore, we require that there be at most one tracker layer that contributes no hit on the track, to reduce contributions from photon conversions~\cite{CMS:2020uim}. To pass the tight ID selection, an electron must also be compatible with originating from the PV, requiring the ratio of the three-dimensional impact parameter of the associated track to its uncertainty ($\text{SIP}_\text{3D}$) to be less than 4. The loose electron ID is used to veto additional leptons and to estimate the nonprompt-lepton background. The tight electron ID is used for the signal region (SR) and background validation region (VR) definitions. The tight electron ID efficiency for the signal event selection varies between 40 and 90\%, depending on the data-taking era, and on the electron \pt and $\eta$. Because the electron charge is crucial and has a higher mismeasurement rate than that of muons, we use a tight charge ID criterion requiring agreement in the charge measurements from three different methods. These methods are based on the Gaussian sum filter (GSF) track curvature~\cite{GSF}, the closest Kalman filter track curvature~\cite{CMS:2017yfk}, and the $\Delta \phi$ between the ECAL cluster and the GSF track extrapolated to the vertex. An event is discarded if it contains an electron for which the three charge measurements do not agree.
 
Muons are reconstructed in the range $\abs{\eta} < 2.4$, and we consider muons with $\pt>15\GeV$. Two muon ID working points are defined in the analysis, referred to as loose and tight ID in the following, serving the same purposes as the electron ID working points. Muon ID is based on the number of reconstructed energy deposits in the tracker and in the muon system, and on the fit quality of the muon track~\cite{CMS:2018rym}. Additionally, the loose (tight) muon ID requires $I_\text{rel}<0.6$ (0.1) and $\text{SIP}_\text{3D}<5$ (3). The muon tight ID efficiency for the signal event selection varies between 70 and 98\% depending on the data-taking era, and on the muon \pt and $\eta$.  

{\tolerance=800 For each event, hadronic jets are clustered from the PF particles using the infrared- and collinear-safe anti-\kt algorithm~\cite{Cacciari:2008gp, Cacciari:2011ma} with a distance parameter of 0.4. Jet momentum is determined as the vectorial sum of all particle momenta in the jet, and is found from simulation to be, on average, within 5--10\% of the true momentum over the entire \pt spectrum and detector acceptance. The jet energy resolution amounts typically to 15--20\% at 30\GeV, 10\% at 100\GeV, and 5\% at 1\TeV~\cite{CMS:2016lmd}. Additional $\Pp\Pp$ interactions within the same or nearby bunch crossings (pileup) can contribute extra tracks and calorimetric energy depositions, increasing the apparent jet momentum. To mitigate this effect, tracks identified to be originating from pileup vertices are discarded, and an offset correction is applied to correct for the remaining pileup contributions~\cite{CMS:2020ebo}. Jet energy corrections are derived from simulation studies so that the average measured energy of jets becomes identical to that of particle-level jets. In situ measurements of the momentum balance in dijet, $\PGg + \text{jet}$, $\PZ + \text{jet}$, and multijet events are used to determine any residual differences between the jet energy scale in data and simulation, and appropriate corrections are made~\cite{CMS:2016lmd}. Jets reconstructed with \pt larger than 25\GeV and $\abs{\eta}$ smaller than 2.4 are considered in the analysis. Additional selection criteria are applied to each jet to remove jets potentially dominated by instrumental effects or reconstruction failures~\cite{CMS:2020ebo}. Lastly, for jets with $\pt<50\GeV$, an additional identification algorithm based on a boosted decision tree technique, referred to as the pileup jet ID, is used to remove jets likely originating from pileup interactions~\cite{CMS-DP-2020-020}.\par}

To identify jets originating from the hadronization of \PQb quarks, the \textsc{DeepJet}~\cite{Bols:2020bkb, CMS-DP-2018-058} \PQb tagging algorithm is applied to each reconstructed jet in an event. The threshold on the discriminant score of the \textsc{DeepJet} algorithm for a jet to be considered a \PQb-tagged jet is chosen such that the tagging efficiency for \PQb  quark initiated jets is approximately 80\% with a misidentification probability of 1\% for jets initiated by light-flavor quarks or gluons. The tagging efficiencies measured in simulation are corrected as functions of the \pt, $\eta$, and flavor of the jet to match those measured in data~\cite{BTV-16-002}.

The missing transverse momentum vector \ptvecmiss is computed as the negative vector \pt sum of all the PF candidates in an event, and its magnitude is denoted as \ptmiss~\cite{CMS:2019ctu}. The \ptvecmiss is modified to account for corrections to the energy scale of the reconstructed jets, muons, and electrons in the event.

\section{Data and simulated samples}
\label{Datasets}

The analysis is performed using a sample of $\Pp\Pp$ collision data at $\sqrt{s}=13\TeV$, collected with the CMS detector in 2016--2018 and corresponding to a total integrated luminosity of 138\fbinv~\cite{CMS-LUM-17-003,CMS-PAS-LUM-17-004, CMS-PAS-LUM-18-002}. Events used in this analysis are selected by SF dilepton triggers. Each trigger has \pt threshold requirements for the leading ($\pt(\Pell_1)$) and subleading ($\pt(\Pell_2)$) leptons. For each lepton in the trigger, a track-based $I_\text{rel}$ threshold requirement is implemented with a fixed cone radius of 0.2 for electrons and 0.3 for muons. Depending on the data-taking era, the trigger also requires the two leptons to originate from the PV (via a requirement on the distance of closest approach, $d_\text{z}$, along the beam axis) and/or to pass a minimum invariant mass ($m(\Pell\Pell)$) threshold. Additionally, two single-lepton triggers for each lepton generation are used for the nonprompt-lepton background estimation. The thresholds are summarized in Table~\ref{tab:trigger_specs}. 

\begin{table}[htbp]
\centering
\topcaption{Summary of trigger \pt thresholds and additional selection requirements. The dilepton triggers are used to define the SR and VR. For the nonprompt-lepton background estimate, two prescaled single-lepton triggers are used per flavor, depending on the lepton \pt.}
\label{tab:trigger_specs}
\begin{tabular}{lccc}

Trigger type & $\pt(\Pell_1)$ ({\GeVns}) & $\pt(\Pell_2)$ ({\GeVns}) & Other \\ \hline
\multicolumn{4}{c}{\textit{Dilepton triggers}} \\ [\cmsTabSkip]
\multirow{2}{3cm}{Dielectron}& \multirow{2}{0.7cm}{${>} 23$} & \multirow{2}{0.7cm}{${>} 12$} & $I_{\text{rel}} < 0.20$ \\
                  &        &        & $d_\text{z} < 0.2\cm$ (2016) \\[\cmsTabSkip]
\multirow{3}{3cm}{Dimuon}                  &   \multirow{3}{0.7cm}{${>} 17$}     &    \multirow{3}{0.7cm}{${>} 8$}      & $I_{\text{rel}} < 0.40$\\
 &  & & $m(\PGm\PGm) > 3.8\GeV$ (2017, 2018) \\
                  &        &        & $d_\text{z} < 0.2\cm$ \\[\cmsTabSkip]
\multicolumn{4}{c}{\textit{Single-lepton triggers}} \\ [\cmsTabSkip]
\multirow{2}{3cm}{Single electron}
  & $>12$,    \:  $\pt(\Pe)<35\GeV$
  & \multirow{2}{*}{\NA}
  & \multirow{2}{*}{$I_{\text{rel}}<0.20$, $\pt(\text{j}_1)>30\GeV$}  \\
  & $>23$,   \:  $\pt(\Pe)>35\GeV$
  & & \\[\cmsTabSkip]
\multirow{2}{3cm}{Single muon}
  & $>8$,     \:\:\,  $\pt(\PGm)<30\GeV$
  & \multirow{2}{*}{\NA}
  & \multirow{2}{*}{$I_{\text{rel}}<0.40$} \\
  & $>17$,  \:  $\pt(\PGm)>30\GeV$
  & & \\
\end{tabular}
\end{table}
The signal events are simulated with \MGvATNLO~\cite{MG5_aMCatNLO} version 2.6.5  using the Majorana HNL model~\cite{UFO1,UFO2,FeynRules} at leading order (LO) in the four-flavor scheme. The $\ttbar$ cross section is calculated with the \TOPpp version 2.0 program to next-to-next-to-leading order (NNLO) in perturbative quantum chromodynamics (QCD), including soft-gluon resummation to next-to-next-to-leading log (NNLL) order (as discussed in Ref.~\cite{tt1} and references therein), and assuming a top quark mass of 172.5\GeV. The signal events from $\ttbar$ production with subsequent top quark decays into an HNL are generated with up to one additional parton in the matrix element calculations. Subsequently, jets from parton shower are merged with those from the matrix element calculations using the MLM algorithm~\cite{MLM}. Decays of the top quarks, \PW bosons, and HNLs are modeled with the same generator, considering the spin correlations among the decay products. The signal samples are simulated for \mN values of 20, 30, 40, 50, 60, 70, 75, 85, 95, and 100\GeV. The decay width of the HNL in the considered mass range is less than 0.1\MeV. The HNL is assumed to decay promptly. The nonzero HNL mixing parameter \VlN is set to 0.01. In addition to the decay mode introduced in Section~\ref{Intro}, the generated signal samples include all HNL decay modes leading to final states with more than two charged leptons. We exclude \PGt leptons from the primary HNL vertices ($\PW\to\PGt\PN$ and $\PN\to\PGt\PW$), while allowing for \PGt lepton production via \PW and \PZ boson decays. The full decay modes considered are: 
\begin{equation*}
  \Pp\Pp\to\ttbar\to (\PQt\to\PQb\Pellp\PN\to\PQb\Pellp\Pellp \Pf\PAf', \PAQt\to\PAQb\PWm\to\PAQb\Pf\PAf'),
\end{equation*}
where \Pf and $\Pf'$ can be any fermion and \PN decay modes include both charged- and neutral-current. The inclusive signal cross section is set to 10\unit{fb}, and the fitted signal strength limit directly scales this cross section. The branching fraction limit of a top quark into an HNL is then obtained by dividing the fitted signal cross section limit by $2\sigma(\ttbar)$. This branching fraction limit is then compared to the branching fraction calculated in simulation with $\VlN=0.01$ and converted into the mixing parameter limit using the relation $\BtlN \propto \VlN^2$.

Background processes involving leptons produced promptly and from photon conversions are estimated using simulation. Additionally, simulation is used to model the background from lepton charge mismeasurement, causing opposite-sign (OS) lepton pairs to be reconstructed as SS lepton pairs. These processes are generated at next-to-leading order (NLO), and their cross sections are calculated at the same order, unless a specific other order is mentioned. The following processes are simulated using \MGvATNLO: Drell--Yan (DY) production, $\PW\PZ$ diboson production with decays into three charged leptons and a neutrino, triboson production with all possible combinations of vector bosons ($\PV\PV\PV$, $\PV=\PW$ or $\PZ$), \PW boson production in association with a photon ($\PW\PGg$), Higgs boson production in association with a vector boson ($\PV\PH$), vector boson production in association with a \ttbar pair ($\ttbar\PV$), single top quark production via the $s$ channel, and single top quark production in association with a \PZ boson ($\PQt\PZ$). The DY processes with $m(\Pell\Pell)$ below 50\GeV and $\PW\PGg$ production are generated at LO. The decays in $\PV\PH$ and $\ttbar\PW$ processes are simulated using \textsc{Madspin}~\cite{Madspin}. The following processes are generated using \POWHEG v2.0~\cite{Powheg1,Powheg2,Powheg3}: production of a pair of \PZ bosons ($\PZ\PZ$) decaying into four leptons~\cite{PowhegVV}, Higgs boson production in association with a \ttbar pair ($\ttbar\PH$)~\cite{PowhegttH}, Higgs boson production via gluon fusion ($\Pg\Pg\PH$)~\cite{PowhegggH} and via vector boson fusion ($\PQq\PQq\PH$)~\cite{PowhegVBFH}, single top quark production in the $t$ channel and in association with a \PW boson~\cite{PowhegtW}, and \ttbar production~\cite{Powhegtt}. For the normalization of the $\PZ\PZ$ process, the production cross section is calculated at NNLO~\cite{ZZNNLOCalc}. The decays of the Higgs boson in $\Pg\Pg\PH$ and $\PQq\PQq\PH$ production are simulated with \textsc{JHUGEN}~\cite{JHUGen}, while \textsc{Madspin} is used for the top quark decay in the single top quark $t$-channel production. Lastly, the diboson processes, except for the \PZ boson pair production with decays into four leptons and $\PW\PZ$ diboson production with decays into three charged leptons and a neutrino, are generated at LO using \PYTHIA~8.240~\cite{Pythia8}.

All simulated samples are interfaced with \PYTHIA to perform parton shower and hadronization. For all simulated samples, the description of the underlying event is provided by the CP5 tune~\cite{CPTunes}. The NNPDF 3.1 NNLO~\cite{NNPDF31} parton distribution function (PDF) set is used for all simulated samples. Pileup effects are included via simulated minimum bias $\Pp\Pp$ collisions superimposed on the primary simulated interaction, with the multiplicity matching that in data. Simulation of the CMS detector response is performed with \GEANTfour~\cite{GEANT4}.

\section{Event selection and analysis strategy}
\label{Events}
The offline selection requires an electron pair of $\pt > (25,15)\GeV$ or a muon pair of $\pt>(20,15)\GeV$, where the \pt thresholds are chosen such that the selected events lie at the trigger efficiency plateau. The baseline selection of the SR requires an event to have exactly two SF SS leptons to target the leptons from the production and decay of the HNL and vetoes the event if there is a third lepton that passes the loose lepton ID. To remove events involving the \PZ boson resonance with a charge-mismeasured electron, the reconstructed dielectron invariant mass ($m(\Pe\Pe)$) must be outside a 10\GeV window around the \PZ boson mass (\mZ) of 91.2\GeV~\cite{PDG2024}. However, this selection is not required for the dimuon channel because of the significantly lower charge mismeasurement rate for muons. The dimuon mass must be greater than 4\GeV, so that it is above the $m(\Pell\Pell)$ threshold of the trigger. The signal events are expected to have four jets from the \PW boson decays and two \PQb-tagged jets from the top quark decays. However, the events are required to have at least three jets, which must include at least one \PQb-tagged jet. Because signal jets tend to be more energetic and likely to pass the acceptance for $\mN>\mW$, we define two baseline selections. When testing a signal hypothesis with $\mN<\mW$, we require at least one non-\PQb-tagged jet; when testing a hypothesis with $\mN>\mW$, we require at least two. The selected data and background event sets for the two mass regimes can overlap. After the baseline selection, the signal acceptance times efficiency ranges from 0.987 (2.71) to 8.58 (7.54)\%, depending on \mN, in the electron (muon) channel. The dominant source of background events passing the SR selection comes from events with nonprompt leptons. This background is estimated by measuring the misidentification rate of a nonprompt lepton in the background measurement region (BMR) and applying it to the data in the application region (AR), as is further explained in Section~\ref{BkgEst}.

Because of the high multiplicity of partons in the signal process, only a small fraction of signal events has all the final-state partons within the \pt and $\eta$ acceptances. For the simulated signal events passing the baseline selection, the efficiency ranges from 0.02 to 4\% for matching all six partons at the generator level to the reconstructed jets within a $\DR$ of 0.4, depending on \mN. Therefore, instead of reconstructing the full signal decay products, we search for the signal by inspecting partial decay products using a multivariate analysis based on the gradient boosted decision tree (BDT) technique. Since the \ttbar background is the main contribution to the nonprompt-lepton background, only signal samples and dilepton and lepton$+$jets \ttbar background simulation samples are used for the training. As discussed in Section~\ref{Result}, a training based only on $\ttbar$ events still separates other backgrounds from the signal. Adding $\ttbar\PV$ and $\ttbar \PH$ backgrounds to the BDT training did not improve the expected sensitivity and led to instabilities in the nuisance parameter fit because of the small number of events in the SR. Despite the imbalance between the number of signal and background events, we used all $\ttbar$ background and signal events, splitting each of them evenly: half for training and the other half for testing. Since the number of signal events passing the baseline selection varies for each \mN and lepton flavor scenario, the total number of events used in the training process differs among the ten classifier models (for $\mN=20$, 50, 75, 85, and 100\GeV, each for dielectron and dimuon final states). Each model is applied to the signal sample with \mN closest to the training signal sample \mN. 

\begin{figure*}[htbp]
\centering
  \includegraphics[width=0.48\textwidth]{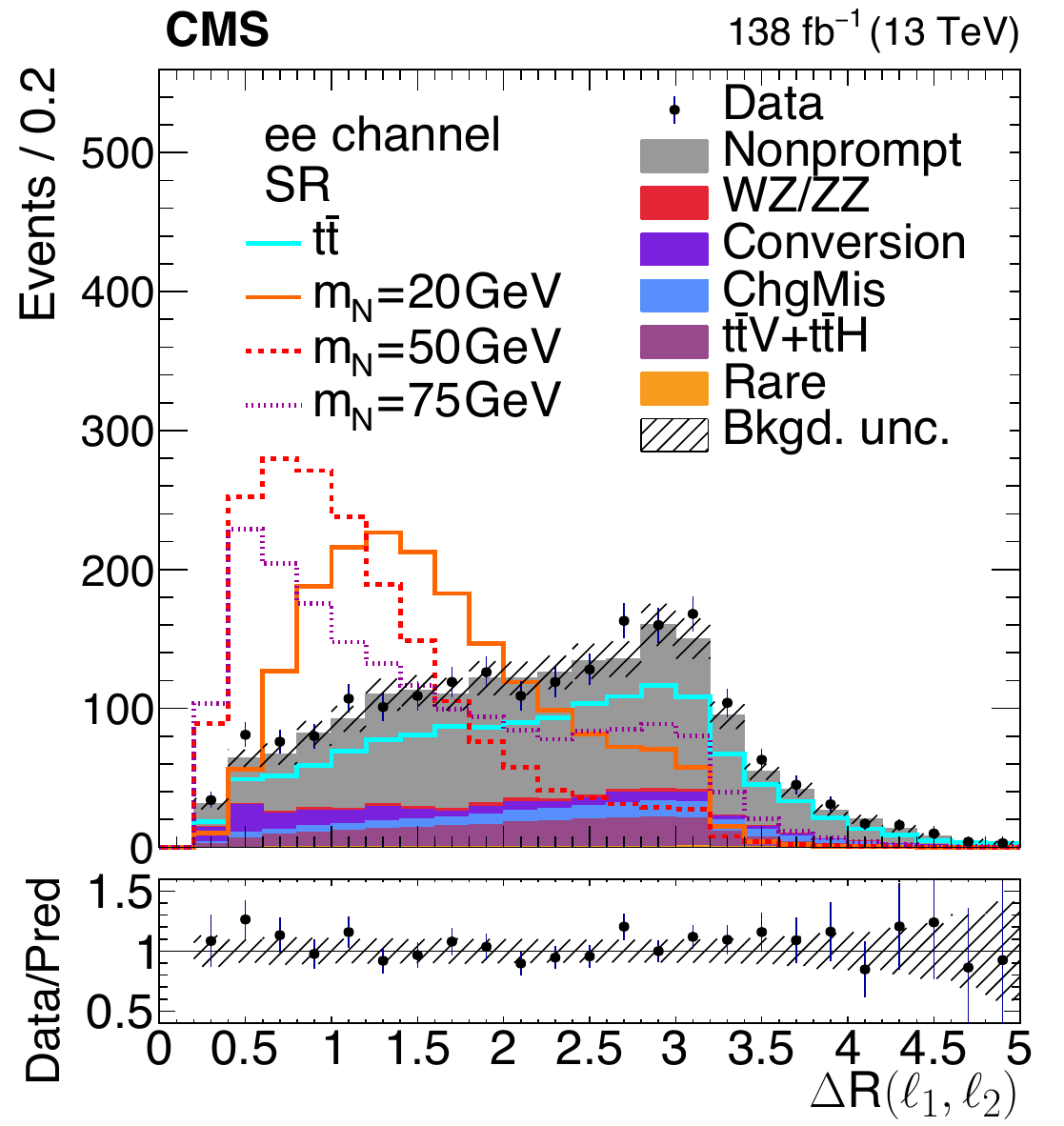}
  \includegraphics[width=0.48\textwidth]{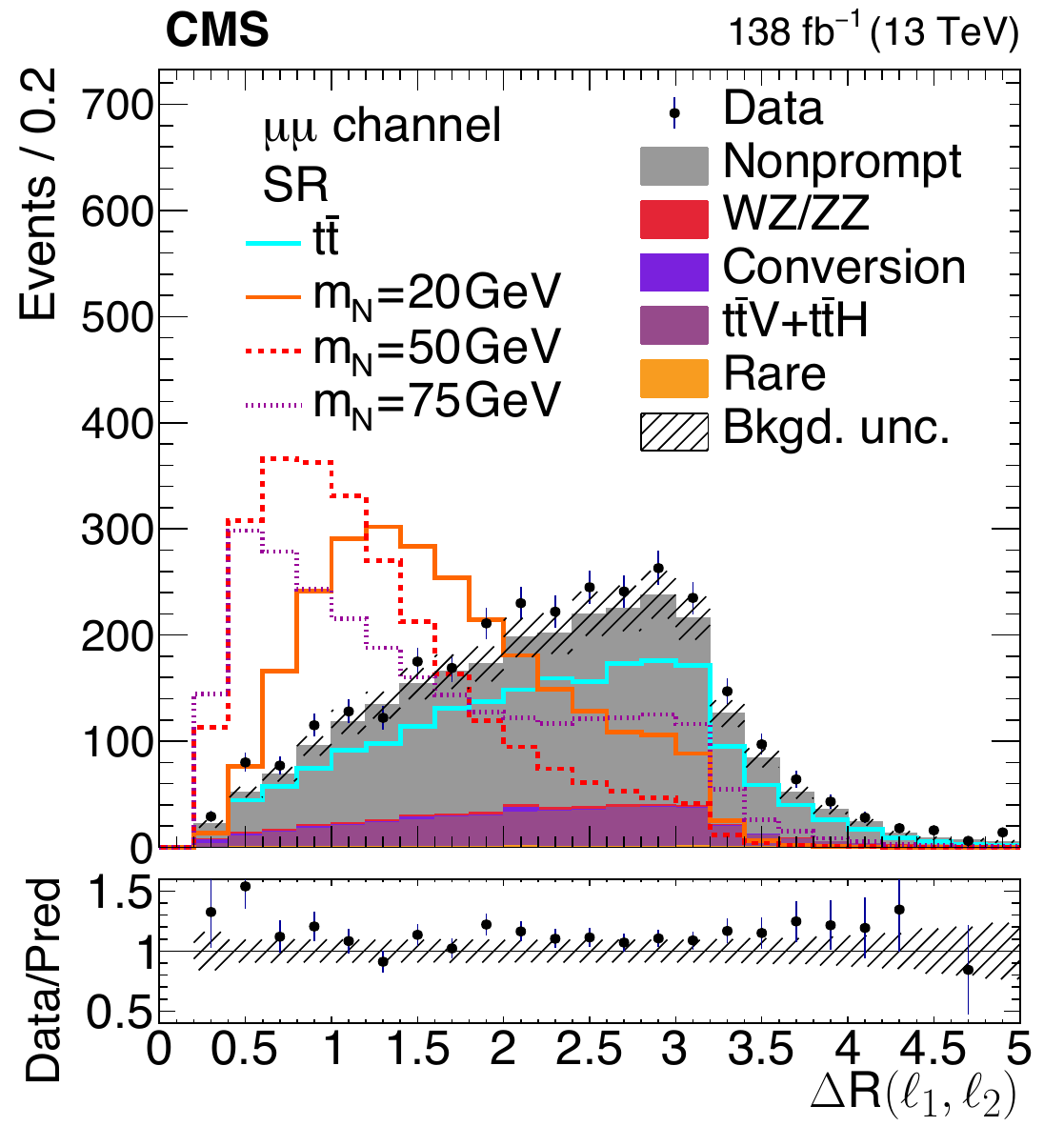}
  \includegraphics[width=0.48\textwidth]{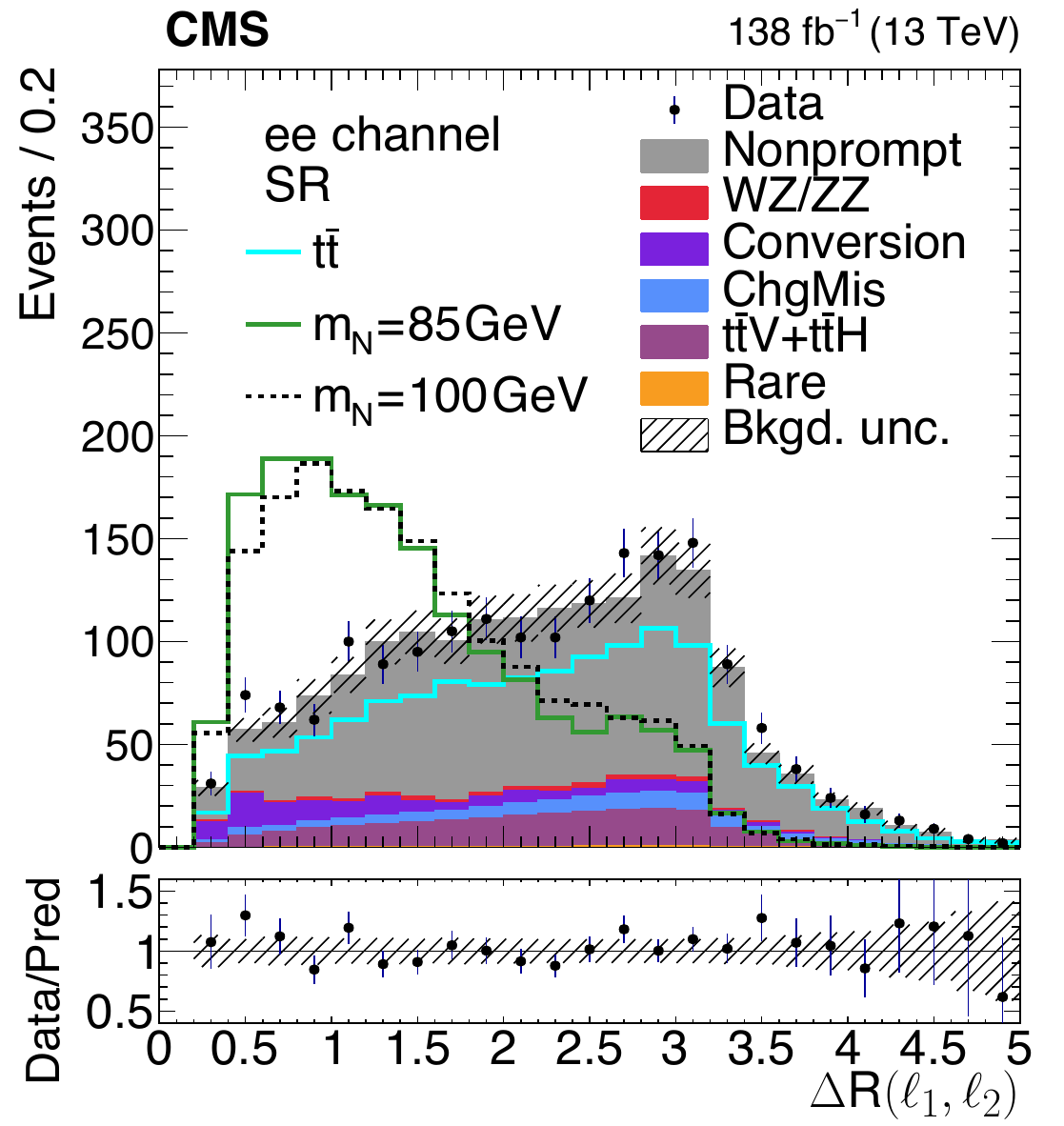}
  \includegraphics[width=0.48\textwidth]{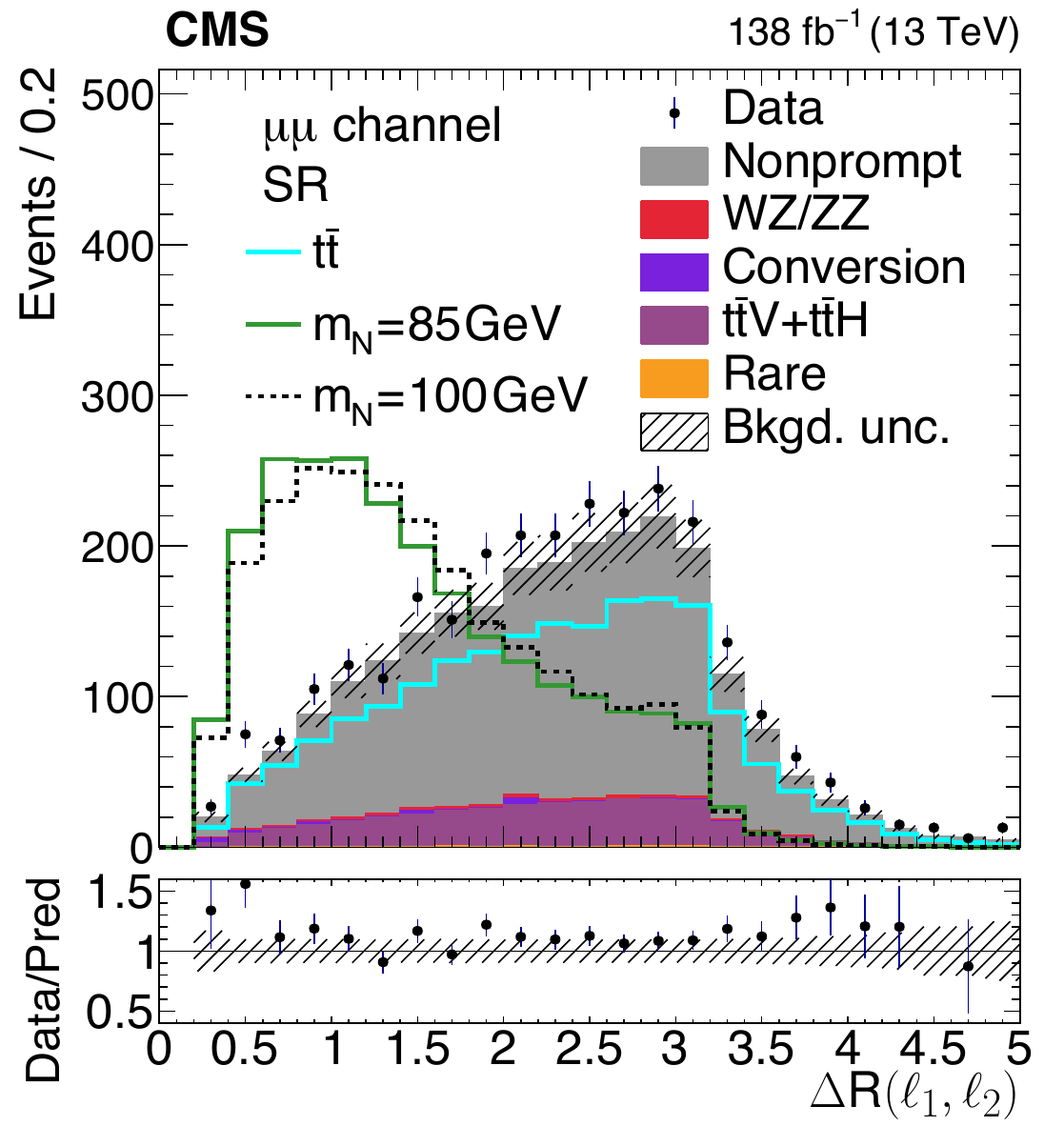}

    \caption{Pre-fit distributions of the $\DR$ between the two leptons are shown for dielectron (left) and dimuon (right) events as examples of the input variables for the training. The upper row and the lower row correspond to the $\mN<\mW$ and $\mN>\mW$ selections, respectively, which differ in the required number of non-\PQb-tagged jets. The cyan line is the \ttbar simulation used for the BDT training. Signal distributions are normalized to the total background. The hatched bands show the total uncertainty, including the systematic component. The last bins include the overflow entries. The lower panels show the ratio of the observed data over the background estimate.}
  \label{fig:VarTrain1}
\end{figure*}

\begin{figure*}[htbp]
\centering
  \includegraphics[width=0.48\textwidth]{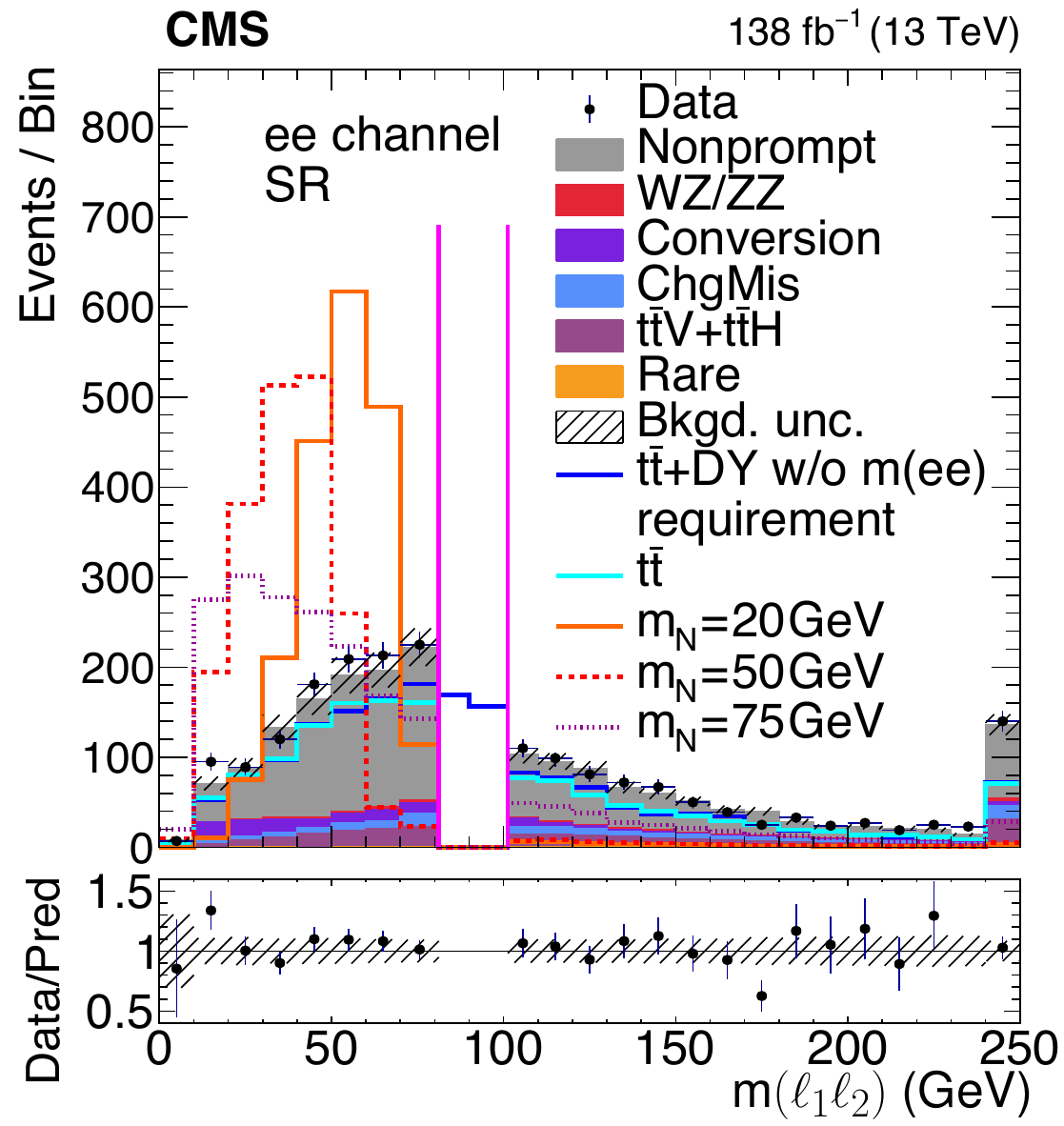}
  \includegraphics[width=0.48\textwidth]{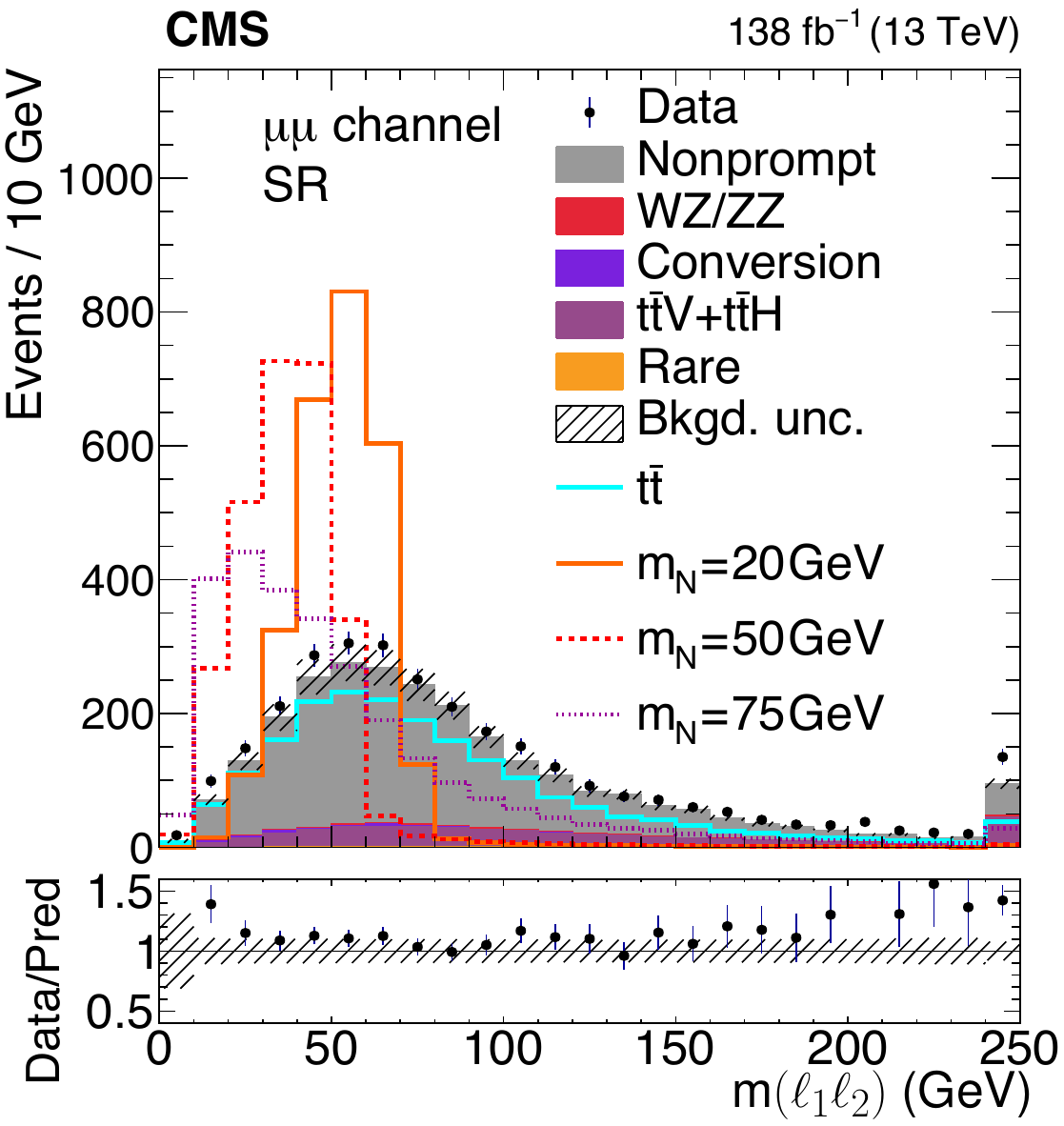}
  \includegraphics[width=0.48\textwidth]{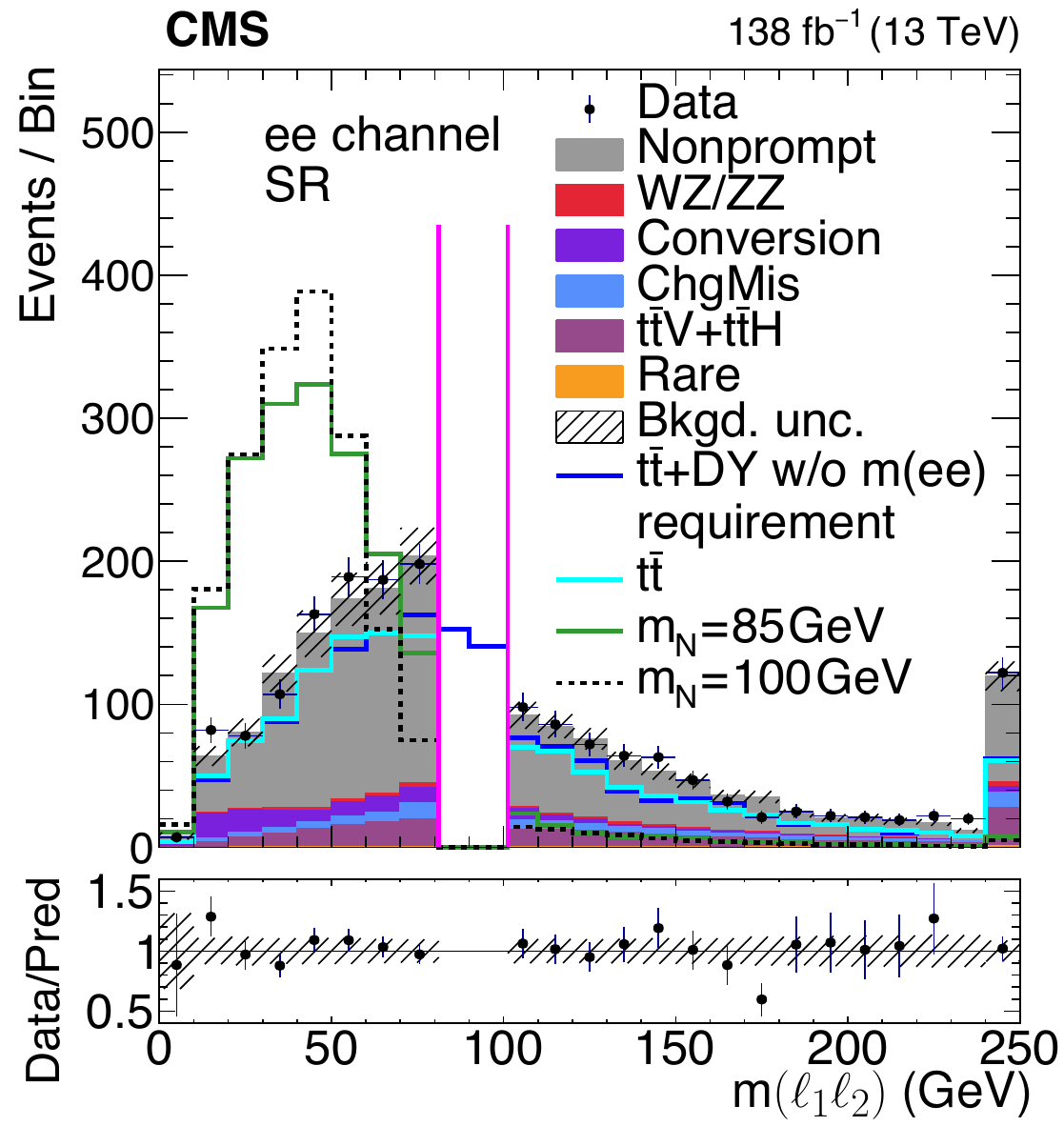}
  \includegraphics[width=0.48\textwidth]{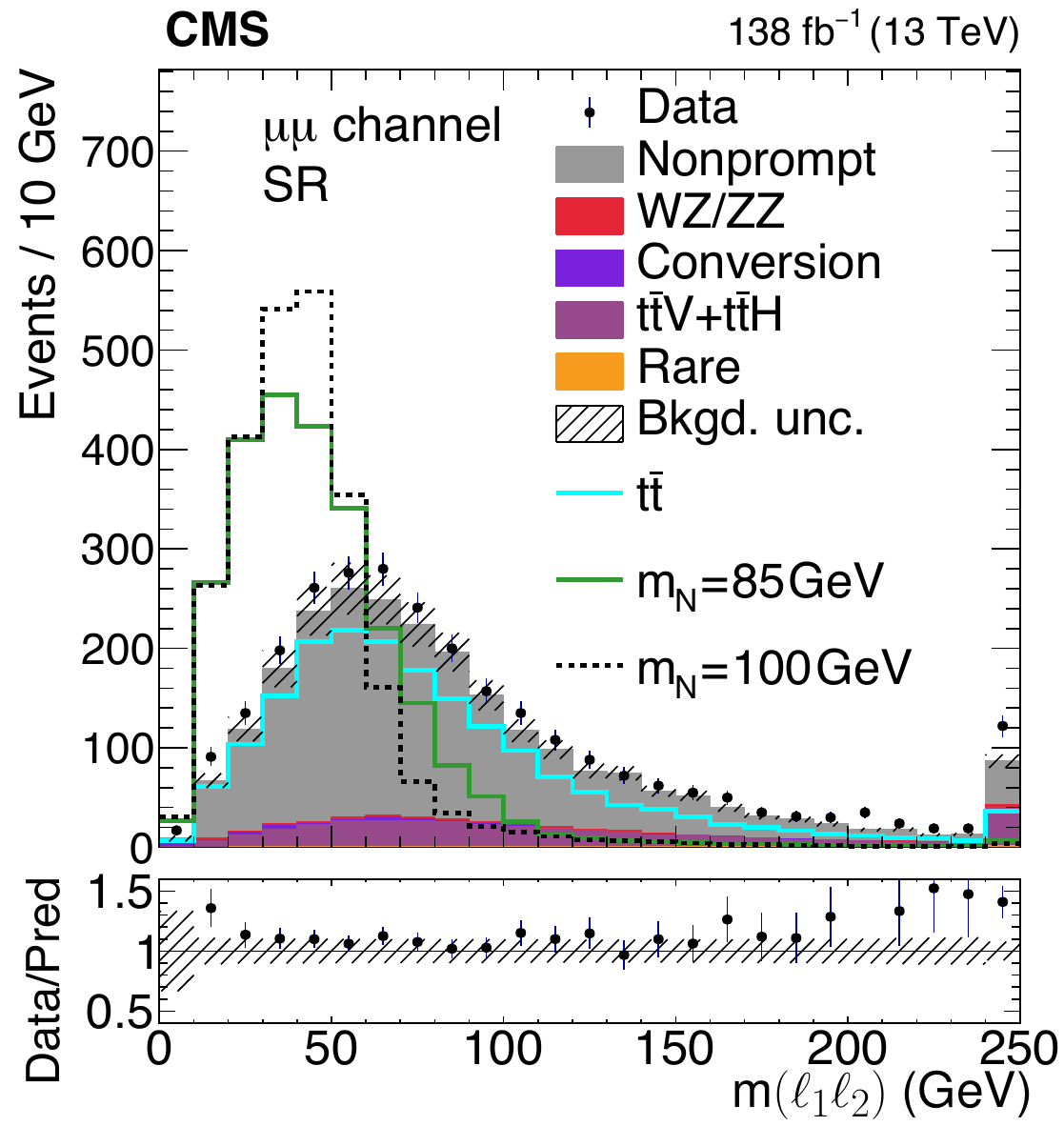}

  \caption{Pre-fit distributions of the invariant mass of the two leptons are shown for dielectron (left) and dimuon (right) events as examples of the input variables for the training. The upper row and the lower row correspond to the $\mN<\mW$ and $\mN>\mW$ selections, respectively, which differ in the required number of non-\PQb-tagged jets. The cyan line is the \ttbar simulation used for the BDT training. Signal distributions are normalized to the total background. The region between two magenta lines depicts what is removed by the $\abs{m(\Pe\Pe)-91.2\GeV}>10\GeV$ requirement in the dielectron channel. The blue line is the \ttbar simulation and DY simulation combined, without the \mZ window veto requirement. In order to match the \mZ window veto requirement to the bin edges, the plots on the left have the bin size of 10\GeV, except for the [70,\,81.2] and [101.2,\,110]\GeV bins. The hatched bands show the total uncertainty, including the systematic component. The last bins include the overflow entries. The lower panels show the ratio of the observed data over the background estimate.}
  \label{fig:VarTrain2}
\end{figure*}

\begin{figure*}[htbp]
\centering
  \includegraphics[width=0.48\textwidth]{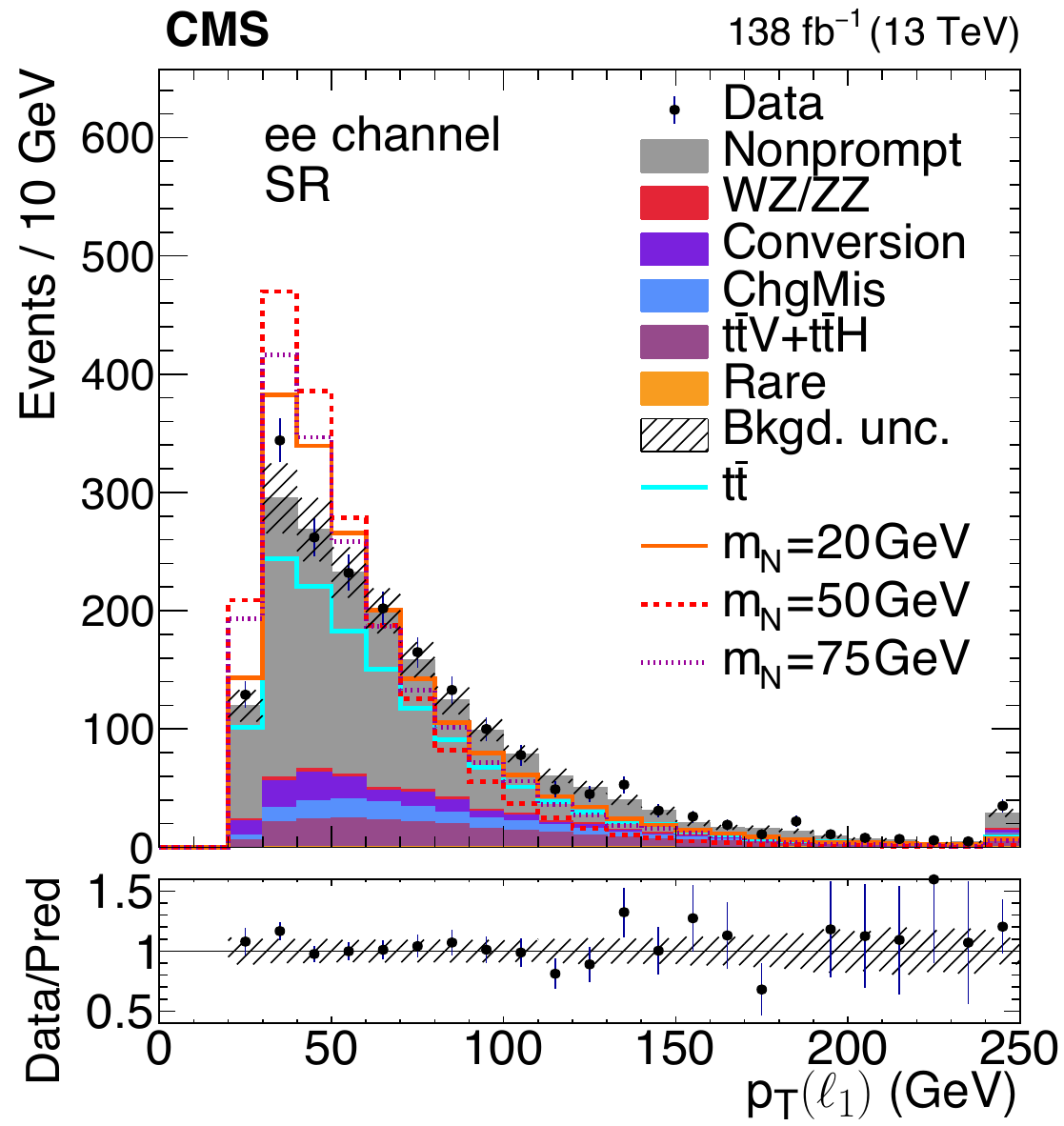}
  \includegraphics[width=0.48\textwidth]{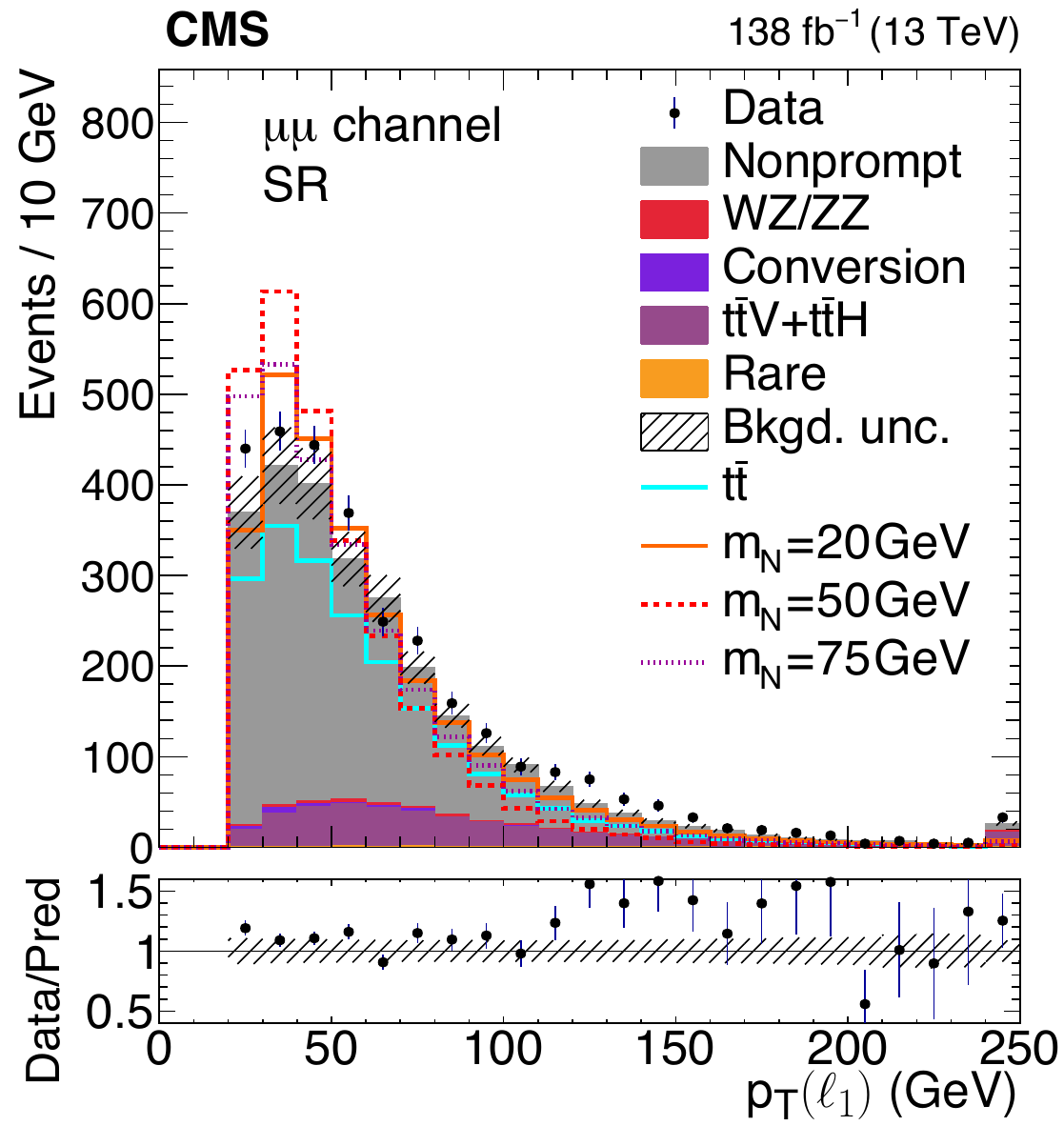}
  \includegraphics[width=0.48\textwidth]{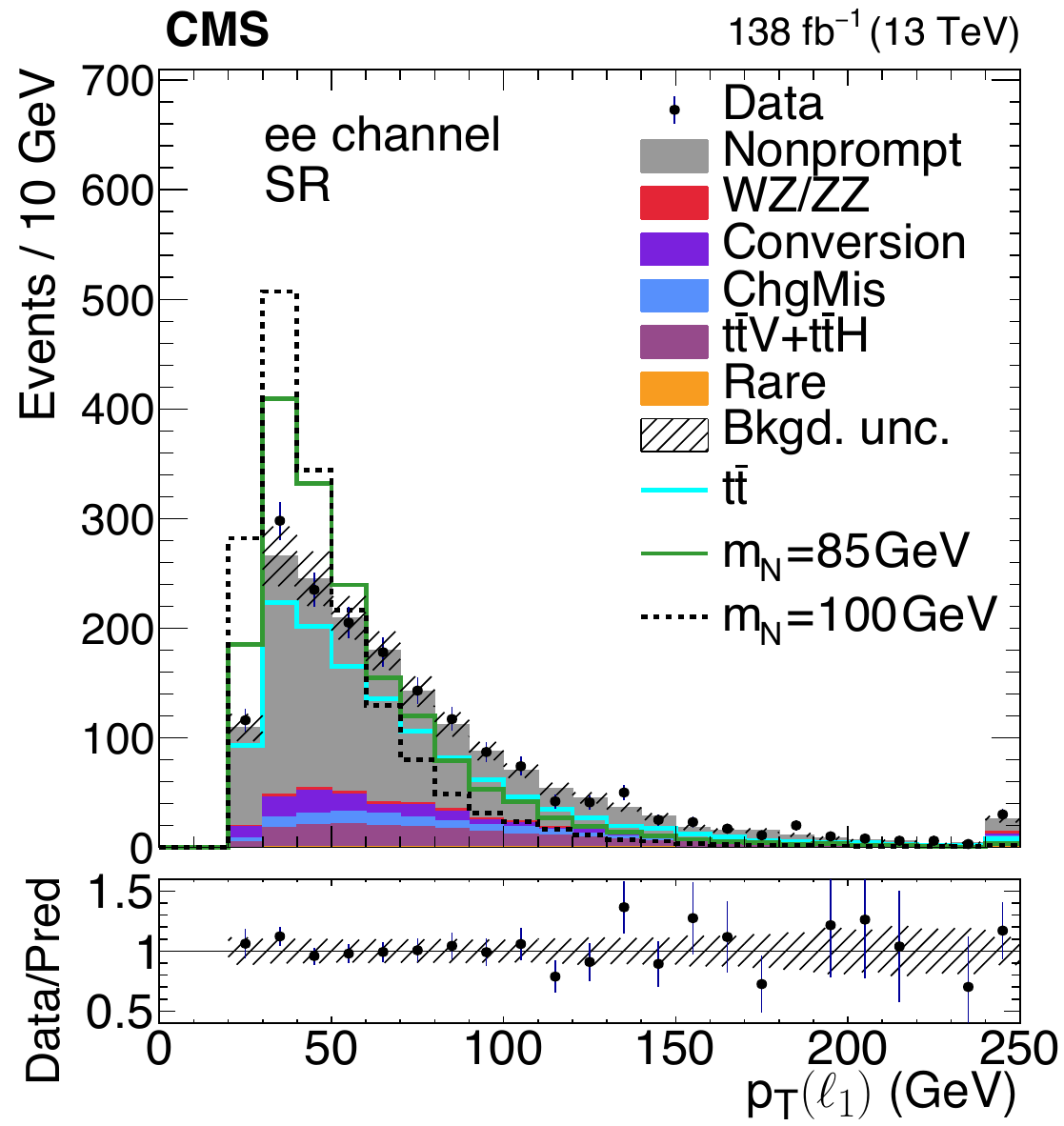}
  \includegraphics[width=0.48\textwidth]{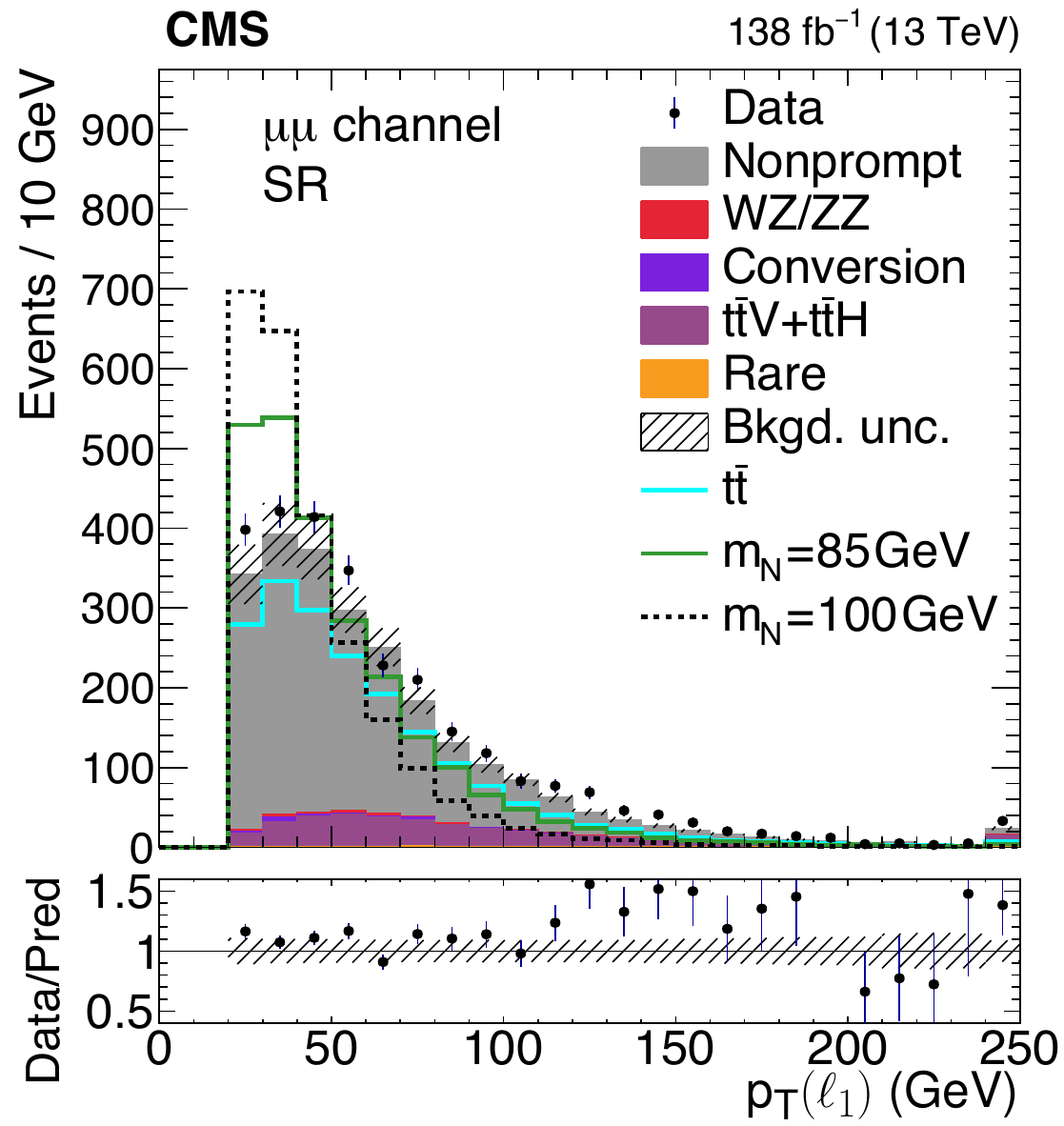}
    \caption{Pre-fit distributions of the leading lepton transverse momentum are shown for dielectron (left) and dimuon (right) events as examples of the input variables for the training. The upper row and the lower row correspond to the $\mN<\mW$ and $\mN>\mW$ selections, respectively, which differ in the required number of non-\PQb-tagged jets. The cyan line is the \ttbar simulation used for the BDT training. Signal distributions are normalized to the total background. The hatched bands show the total uncertainty, including the systematic component. The last bins include the overflow entries. The lower panels show the ratio of the observed data over the background estimate.}
  \label{fig:VarTrain3}
\end{figure*}

\begin{figure*}[htbp]
\centering
  \includegraphics[width=0.48\textwidth]{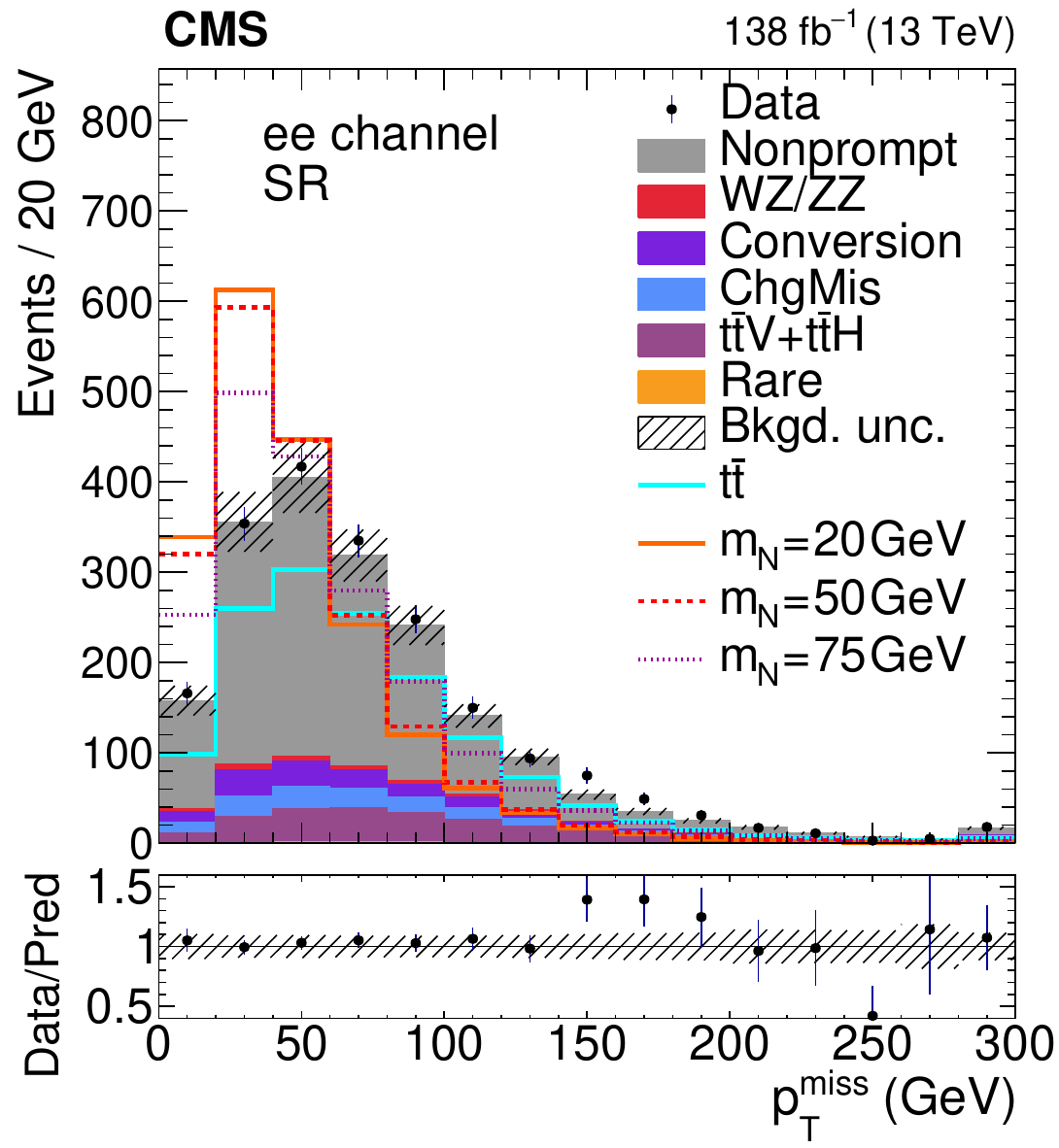}
  \includegraphics[width=0.48\textwidth]{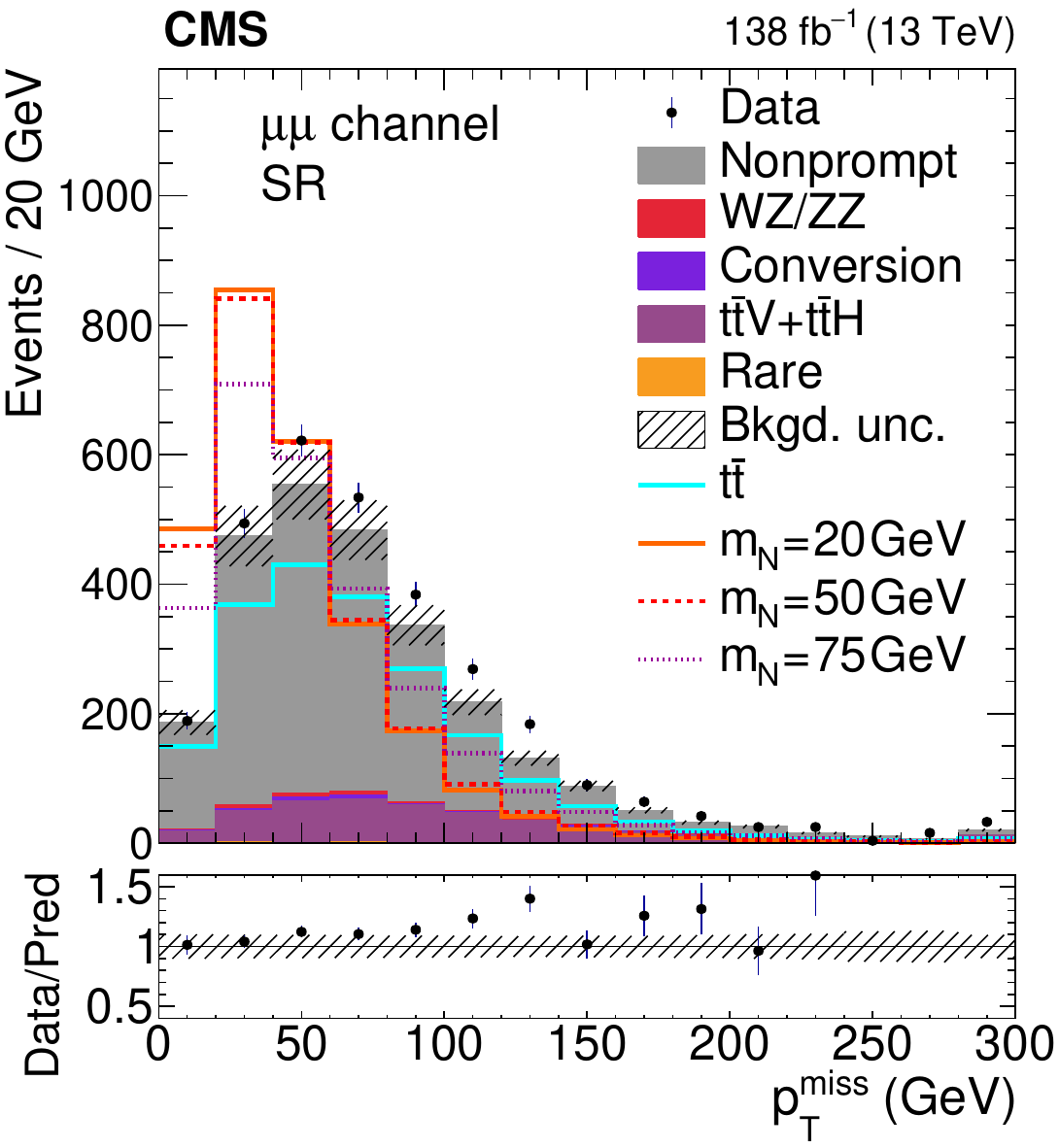}
  \includegraphics[width=0.48\textwidth]{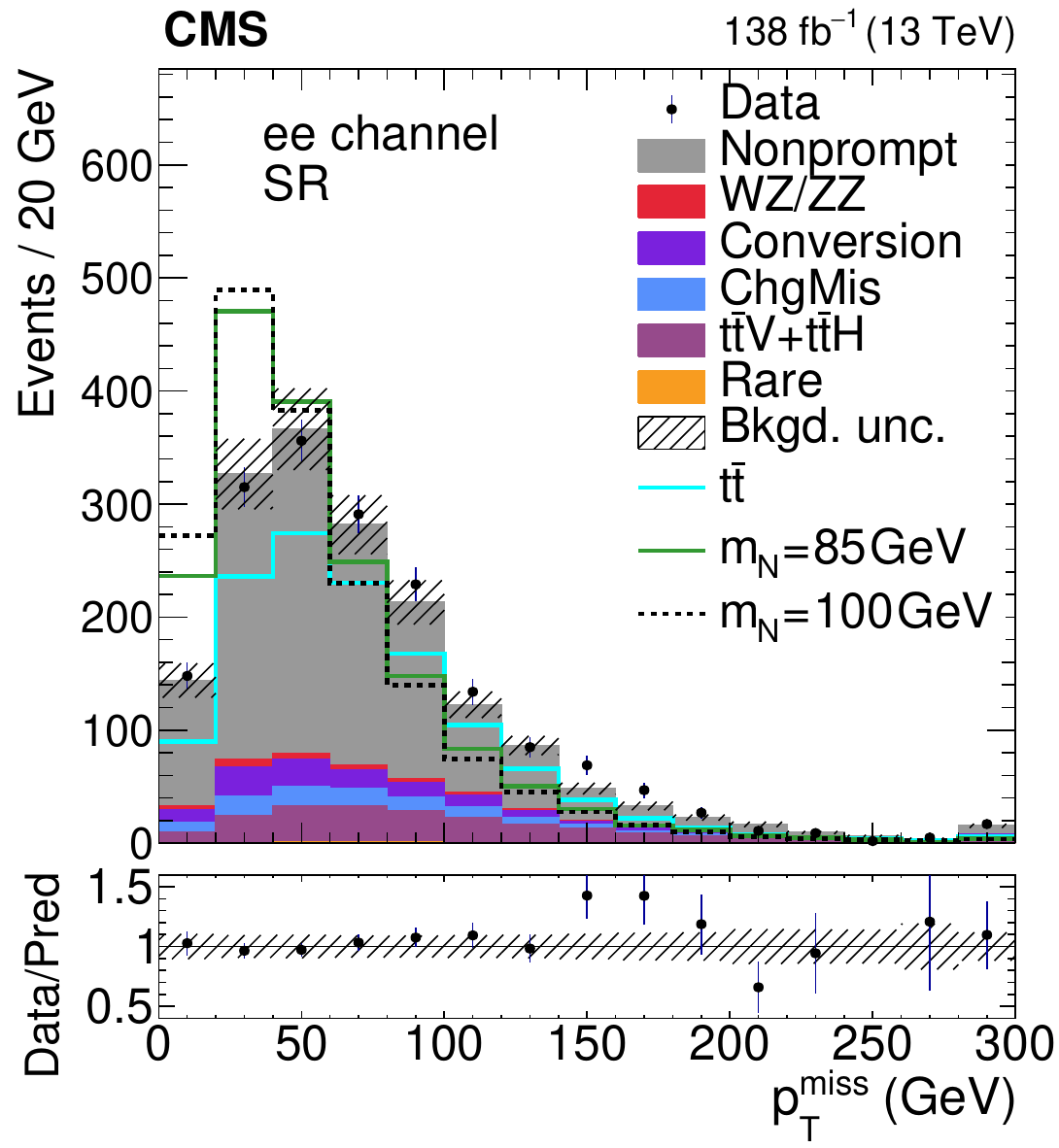}
  \includegraphics[width=0.48\textwidth]{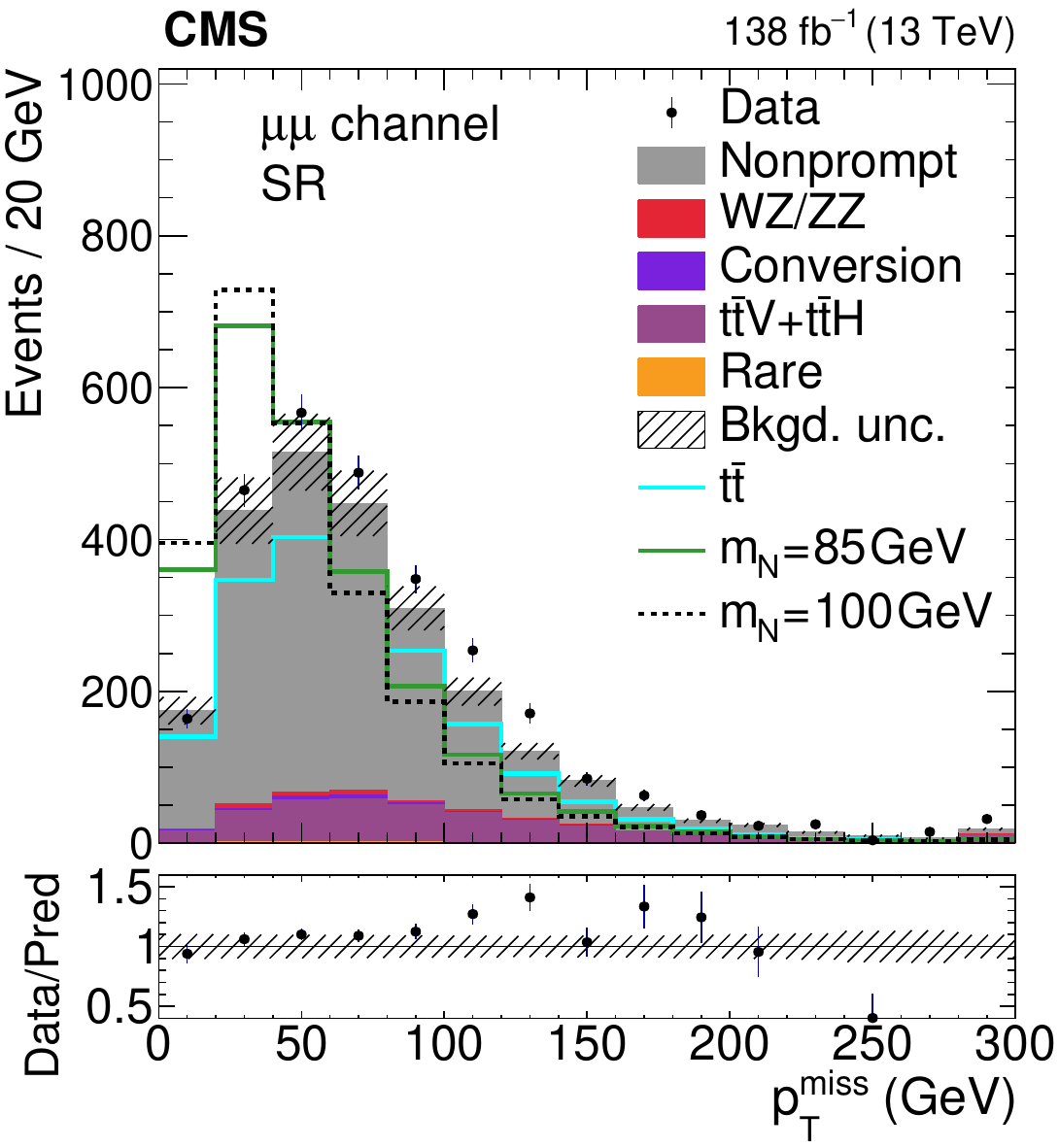}

  \caption{Pre-fit distributions of the missing transverse momentum are shown for dielectron (left) and dimuon (right) events as examples of the input variables for the training. The upper row and the lower row correspond to the $\mN<\mW$ and $\mN>\mW$ selections, respectively, which differ in the required number of non-\PQb-tagged jets. The cyan line is the \ttbar simulation used for the BDT training. Signal distributions are normalized to the total background. The hatched bands show the total uncertainty, including the systematic component. The last bins include the overflow entries. The lower panels show the ratio of the observed data over the background estimate.}
  \label{fig:VarTrain4}
\end{figure*}

The BDT classifier utilizes a set of input variables designed to exploit the kinematic differences between signal and background, specifically focusing on the energy scales, angular separations, and reconstructed masses associated with the final-state leptons and jets, as summarized in Table~\ref{table:BDTInputVar}. For the reconstructed mass variables, jet candidates from the decays of a \PW boson $(\text{j}_\PW)$ and an HNL $(\text{j}_\PN)$ are determined using the combination of jets and leptons that best reconstructs the \PW boson and top quark masses. Applying the classifier model yields a score in the range $[-1,1]$, where signal-like (background-like) events have scores closer to 1 ($-1$). Although the exact order of importance of the input variables differs across different \mN and final states, the $m(\Pell\Pell)$ and angular separation of the two leptons are among the most important variables in all training models. These variables, along with the \pt of the leading lepton and $\ptmiss$ are shown in Figs.~\ref{fig:VarTrain1}--\ref{fig:VarTrain4}. The backgrounds shown in these figures are explained in Section~\ref{BkgEst}. The performance of the training models is slightly better in the dimuon case, because of the larger number of events from higher muon ID efficiency and lower \pt threshold. For both dielectron and dimuon final states, the performance slightly worsens as \mN approaches \mW, resulting in the area under the receiver operating characteristic curve decreasing from 0.90--0.93 to 0.82--0.88. As \mN approaches \mW, one of the two leptons becomes soft and is likely to fail the selection. As a result, the lepton that passes the selection would pair with a lepton from the SM top quark decay or with a nonprompt lepton, leading to background-like BDT outputs. 

\begin{table}[htbp]
\topcaption{Input variables used for the training of classifiers. The usage in the low-mass ($\mN<\mW$) and high-mass ($\mN>\mW$) scenarios is shown in the second and third columns. Subscripts $a$ and $b$ denote leptons and jets, respectively, and $\text{j}_\text{L}$ denotes a non-\PQb-tagged jet. The check marks indicate the corresponding input variable is used for the training. }
\centering
\begin{tabular}{ l c c l }

Variable                                                    & $\mN<\mW$     & $\mN>\mW$    & Definition                                                                       \\\hline
\Njets                                                      & {\checkmark} & {\checkmark} & jet multiplicity                                                                  \\[\cmsTabSkip]
\Nbjets                                                     & {\checkmark} & {\checkmark} & \PQb-tagged jet multiplicity                                                \\[\cmsTabSkip]
$\pt(\Pell_{a})$                                            & {\checkmark} & {\checkmark} & $a^\text{th}$ lepton \pt ($a=1,2$)                                              \\[\cmsTabSkip]
$\ptmiss$                                                   & {\checkmark} & {\checkmark} & missing transverse momentum                                                       \\[\cmsTabSkip]
\HT                                                & {\checkmark} & {\checkmark} & scalar sum of jet \pt                                                           \\[\cmsTabSkip]
$\DR(\Pell_1,\Pell_2)$                                      & {\checkmark} & {\checkmark} & $\DR$ between two leptons                                                         \\[\cmsTabSkip]
\multirow[c]{2}{*}{$\DR(\Pell_{a},\text{j}_{b})$}           & \multirow[c]{2}{*}{\checkmark} & \multirow[c]{2}{*}{\checkmark} & $\DR$ between $a^\text{th}$ lepton and $b^\text{th}$ jet      \\
                                                            &              &              & ($a=1,2$, $b=1,2,3$)                                           \\ [\cmsTabSkip]
$m(\Pell_1\Pell_2)$                                         & {\checkmark} & {\checkmark} & dilepton invariant mass                                                           \\[\cmsTabSkip]
\multirow[c]{2}{*}{$m(\Pell_{a}\text{j}_{\PN,1})$}                              & \multirow[c]{2}{*}{\checkmark} & \multirow[c]{2}{*}{\checkmark} & invariant mass of $a^\text{th}$ lepton and                                           \\
                                                            &              &              & leading HNL jet candidate ($a=1,2$)                                           \\ [\cmsTabSkip]
\multirow[c]{2}{*}{$m(\Pell_1\Pell_2\text{j}_{\PN,1})$}                         & \multirow[c]{2}{*}{\checkmark} &              & invariant mass of two leptons and                                                     \\
                                                            &              &              & leading HNL jet candidate                                                     \\ [\cmsTabSkip]
\multirow[c]{2}{*}{$m(\PQb_1\Pell_1\Pell_2\text{j}_{\PN,1})$}                   & \multirow[c]{2}{*}{\checkmark} & \multirow[c]{2}{*}{\checkmark} & invariant mass of two leptons, leading                               \\
                                                            &              &              & \PQb-tagged jet, and leading HNL jet candidate                                                     \\ [\cmsTabSkip]
\multirow[c]{2}{*}{$m(\Pell_{a}\text{j}_{\PN,1}\text{j}_{\PN,2})$}              &              & \multirow[c]{2}{*}{\checkmark} & invariant mass of $a^\text{th}$ lepton and                                           \\
                                                            &              &              & two HNL jet candidates ($a=1,2$)                                              \\ [\cmsTabSkip]
\multirow[c]{2}{*}{$m(\PQb_1\Pell_1\Pell_2\text{j}_{\PN,1}\text{j}_{\PN,2})$}   &              & \multirow[c]{2}{*}{\checkmark} & invariant mass of two leptons, leading                             \\
                                                            &              &              &\PQb-tagged jet, and two HNL jet candidates                                                        \\[\cmsTabSkip]
${\mT}(\Pell_{1}\ptvecmiss)$                                & {\checkmark} & {\checkmark} & transverse mass of leading lepton and $\ptvecmiss$                                   \\[\cmsTabSkip]
\multirow[c]{2}{*}{${\mT}(\PQb_{1}\Pell_{1}\ptvecmiss)$}                        & \multirow[c]{2}{*}{\checkmark} & \multirow[c]{2}{*}{\checkmark} & transverse mass of leading lepton                                                 \\
                                                            &              &              & leading \PQb-tagged jet, and $\ptvecmiss$                                     \\ [\cmsTabSkip]
\multirow[c]{2}{*}{$m(\Pell_{2}\text{j}_{\text{L},1})$}                         & \multirow[c]{2}{*}{\checkmark} & \multirow[c]{2}{*}{\checkmark} & invariant mass of subleading lepton and                                              \\
                                                            &              &              & leading non-\PQb-tagged jet                                                          \\[\cmsTabSkip]
\multirow[c]{2}{*}{$m(\Pell_{2}\text{j}_{\PW,1}\text{j}_{\PW,2})$}              &              & \multirow[c]{2}{*}{\checkmark} & invariant mass of subleading lepton and                                              \\
                                                            &              &              & two $\PW$ boson jet candidates                                                \\ 

\end{tabular}

\label{table:BDTInputVar}

\end{table}

\clearpage
\section{Background estimation}
\label{BkgEst}
Backgrounds are classified into four categories: nonprompt leptons, lepton charge mismeasurement, conversion, and prompt SS dileptons, as listed below. Each background is discussed in detail after the list. The SR and the VRs used to validate the background estimates are summarized in Table~\ref{tab:SRVR}.
 \begin{enumerate}
  \item \begingroup\emergencystretch=1em
Nonprompt-lepton backgrounds: the contribution of nonprompt leptons originating from jets is estimated using data in the AR and the matrix method~\cite{CMSFRMethod}, which uses the lepton misidentification rate as a function of lepton \pt and $\eta$, as measured in the BMR.\par
\endgroup

  \item Lepton charge mismeasurement backgrounds (ChgMis): a lepton charge can be mismeasured, causing an OS lepton pair to appear as an SS pair. The lepton charge mismeasurement rate (CMR) is measured using simulation and corrected to match that of the data as a function of the lepton \pt and $\eta$.

  \item Conversion backgrounds: leptons from photon conversion can pass the ID requirements and pair with a prompt lepton of the same charge. The contribution from this background source is estimated using simulations.

  \item Prompt-lepton backgrounds: backgrounds that contain only the prompt leptons from \PW or \PZ bosons decays are estimated using simulation.

\end{enumerate}
\subsection{Nonprompt-lepton backgrounds}
 Nonprompt-lepton backgrounds refer to processes with fewer than two prompt leptons that still pass the baseline selection by including at least one jet-induced nonprompt lepton. This background constitutes the dominant contribution, accounting for approximately 74 (87)\% of the total background in the electron (muon) final state. The nonprompt-lepton backgrounds are estimated with the matrix method. This method extrapolates the number of events in the SR from AR, defined by the same baseline selection as the SR but using the loose lepton ID requirements instead of the tight ID ones. The AR thus serves as the sideband region in the lepton ID variables. The extrapolation from the AR to the SR is done through an extrapolation factor, defined as the fraction of leptons that pass the tight ID requirements relative to those that pass the loose ID ones. The extrapolation factor is measured in an independent BMR enriched with dijet events.

 The extrapolation factor for nonprompt leptons is called the ``misidentification rate'', while that for prompt leptons is called the ``prompt rate''. The composition of the observed events ($\vec{N}_{\text{obs}}$) is categorized by how many leptons pass the tight ID requirement (T) or fail the tight ID requirements but pass the loose ID requirements (L). The composition of the underlying events ($\vec{N}_{\text{origin}}$) is categorized by the origin of each lepton as prompt (p) or nonprompt (n). The composition of the observed events and that of the underlying events are connected by a connection matrix ($\mathbb{M}$) constructed from prompt rates ($\varepsilon(\text{p})$) and misidentification rates ($\varepsilon(\text{n})$). Each element of the connection matrix represents the probability that a specific lepton origin configuration is observed in a given ID configuration:
\begin{align*}
  \vec{N}_{\text{obs}}    &= \left(\begin{array}{@{}cccc@{}}N_\text{TT} & N_\text{TL} & N_\text{LT} & N_\text{LL}\end{array}\right)^{\text{T}},\\
  \vec{N}_{\text{origin}} &= \left( \begin{array}{@{}cccc@{}}N_\text{pp}& N_\text{pn}& N_\text{np}& N_\text{nn}\end{array}\right)^{\text{T}},\\
  \vec{N}_{\text{obs}}    &= \mathbb{M}\vec{N}_{\text{origin}}.
\end{align*}
The matrix elements of $\mathbb{M}$, which connect the observed ID composition $N_{\text{ID}_{1}\text{ID}_{2}}$ (with $\text{ID}_{k}=\text{T}$ or L) and the underlying origin composition $N_{\text{o}_{1}\text{o}_{2}}$ (with $\text{o}_{k}=\text{p}$ or n), are given by
\begin{equation}
  \mathbb{M}(\text{ID}_{1},\text{ID}_{2},\text{o}_{1},\text{o}_{2}) =\prod_{k=1}^{2} \left[ \varepsilon(\text{o}_k) \right]^{\delta_{\text{ID}_k, \text{T}}} 
\left[ 1 - \varepsilon(\text{o}_k) \right]^{\delta_{\text{ID}_k, \text{L}}}.
\end{equation}
The underlying composition can be obtained by applying the inverse of the matrix ($\mathbb{M}^{-1}$) to the observed ID composition of events. Because the contribution from the prompt background is small, $\varepsilon(\text{p})$ is assumed to be unity, as its effect on the nonprompt background is negligible.

\begin{figure*}[htbp]
\centering

  \includegraphics[width=0.48\textwidth]{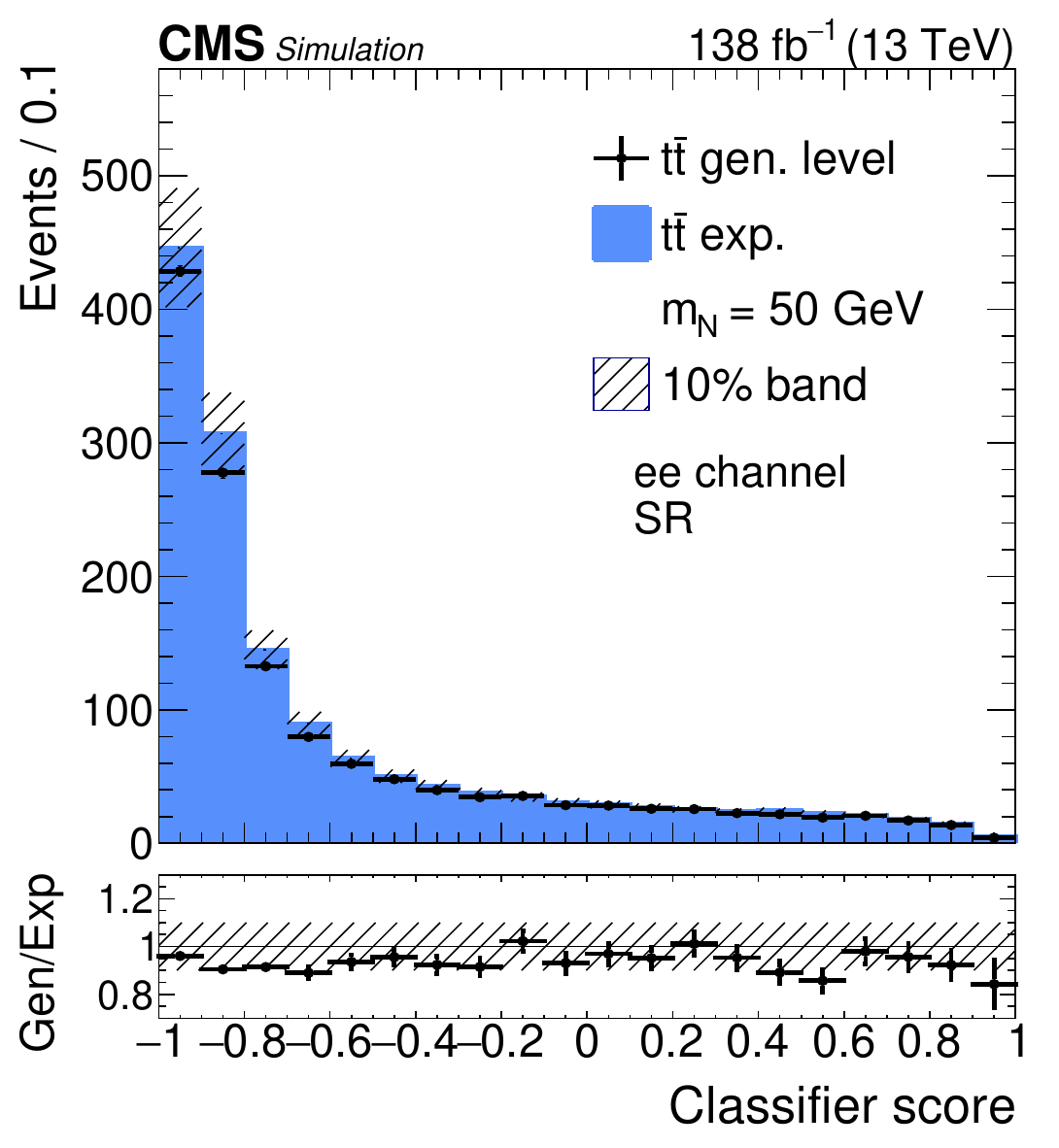}
  \includegraphics[width=0.48\textwidth]{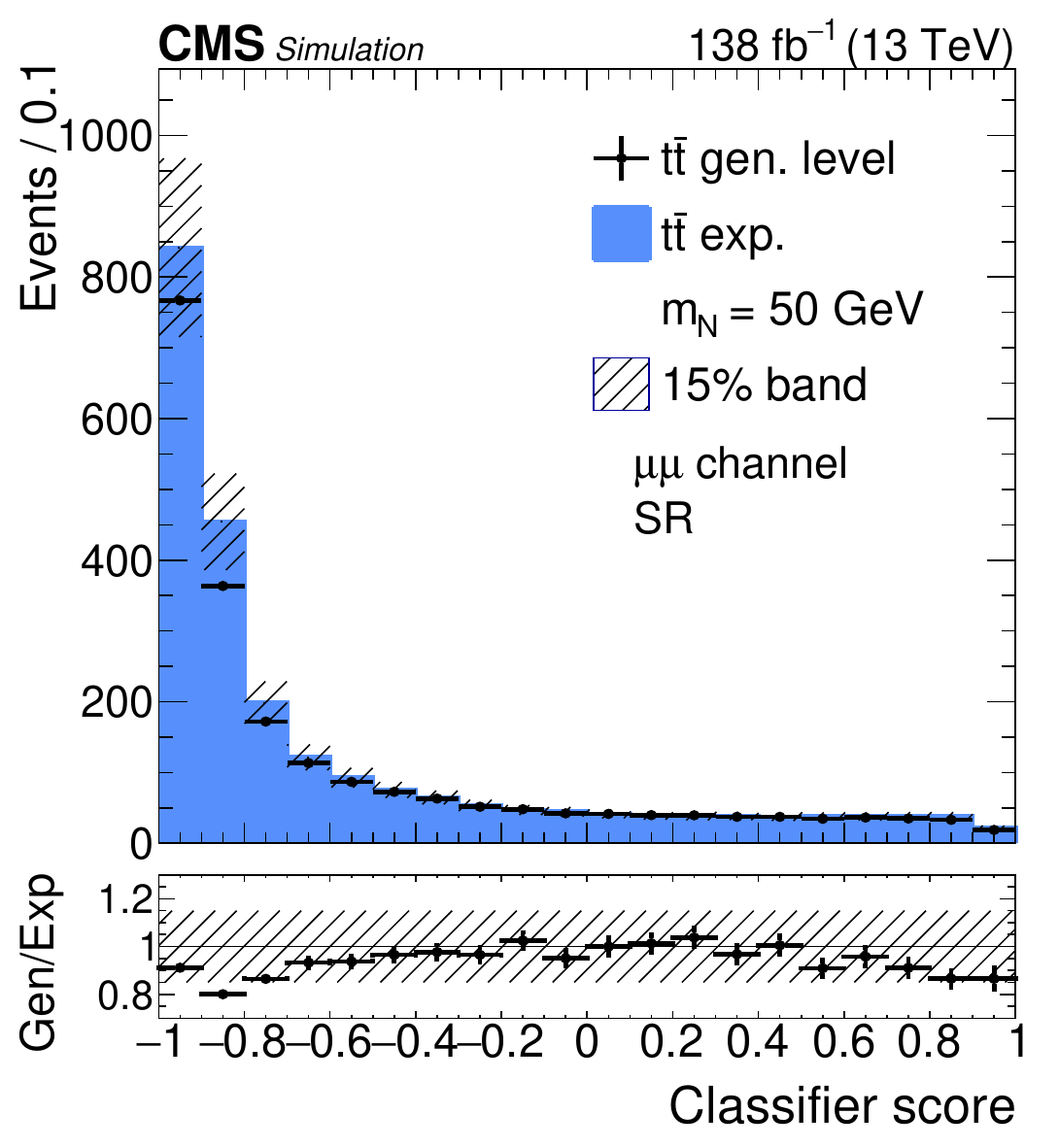}
  \includegraphics[width=0.48\textwidth]{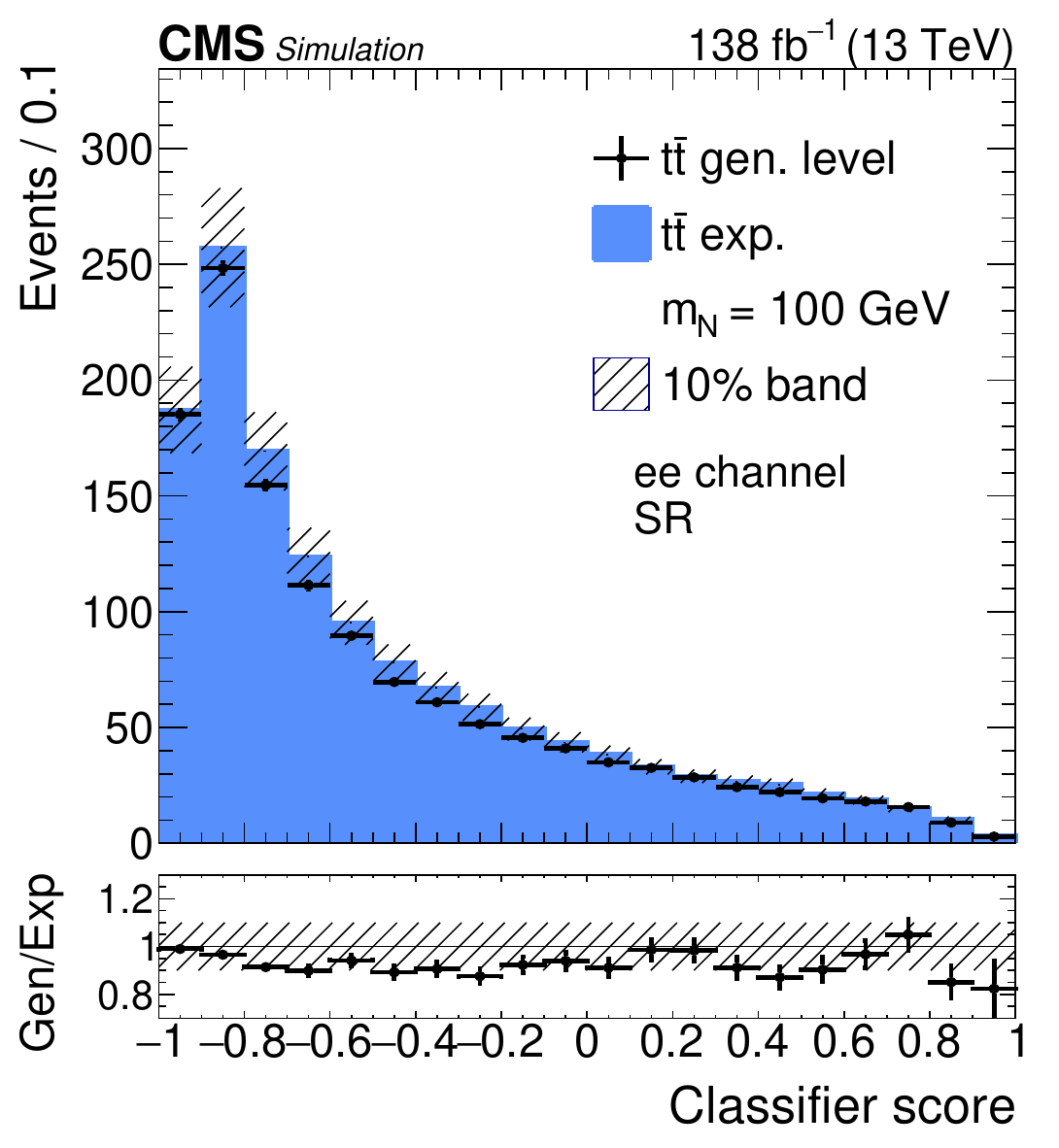}
  \includegraphics[width=0.48\textwidth]{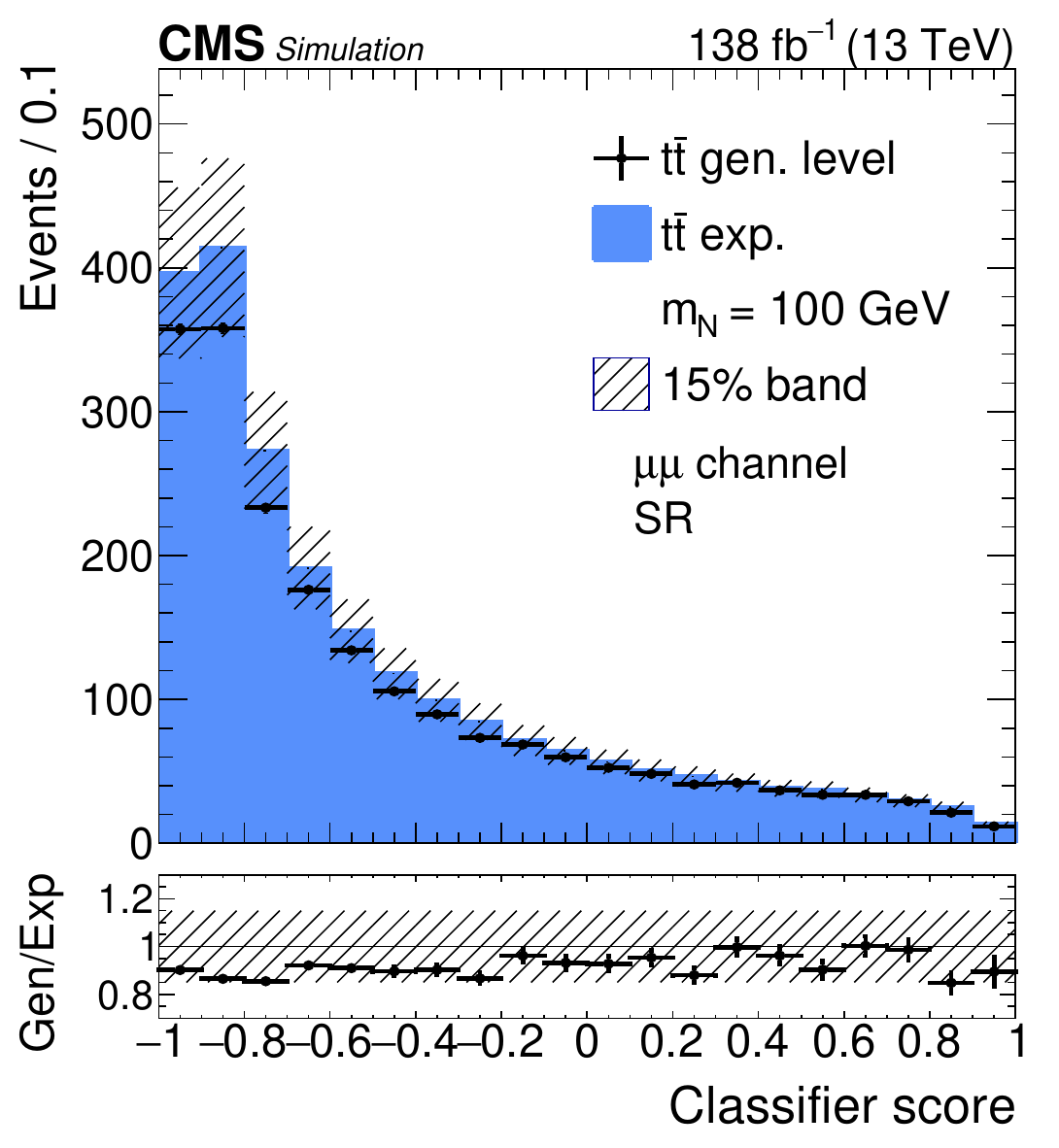}
  \caption{Comparison of the BDT classifier distributions between the $\ttbar$ simulated events with nonprompt leptons from the generator-level information passing the tight ID working point and the expected distribution of the  $\ttbar$ simulation for dielectron (left) and dimuon (right) events in the SR for $\mN=50\GeV$ (upper) and for $\mN=100\GeV$ (lower). The hatched uncertainty bands are $\pm$10 ($\pm$15)\% for the dielectron (dimuon) channel.  The lower panels show the ratio of the $\ttbar$ simulated events with nonprompt leptons from the generator-level information over the expected distribution.}
  \label{fig:BDTGClosureMu}

\end{figure*}

\begin{figure*}[htbp]
\centering
  \includegraphics[width=0.48\textwidth]{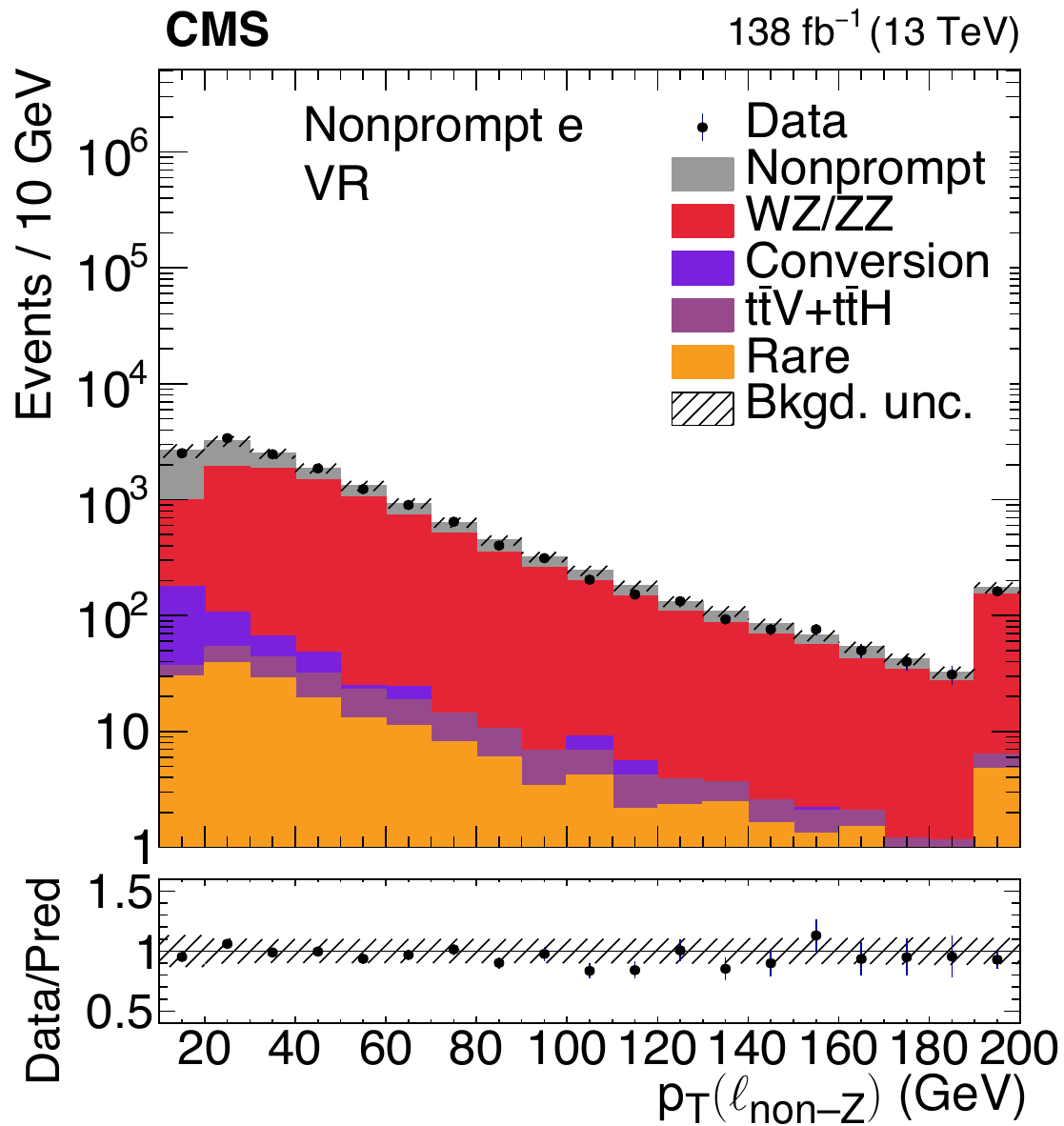}
  \includegraphics[width=0.48\textwidth]{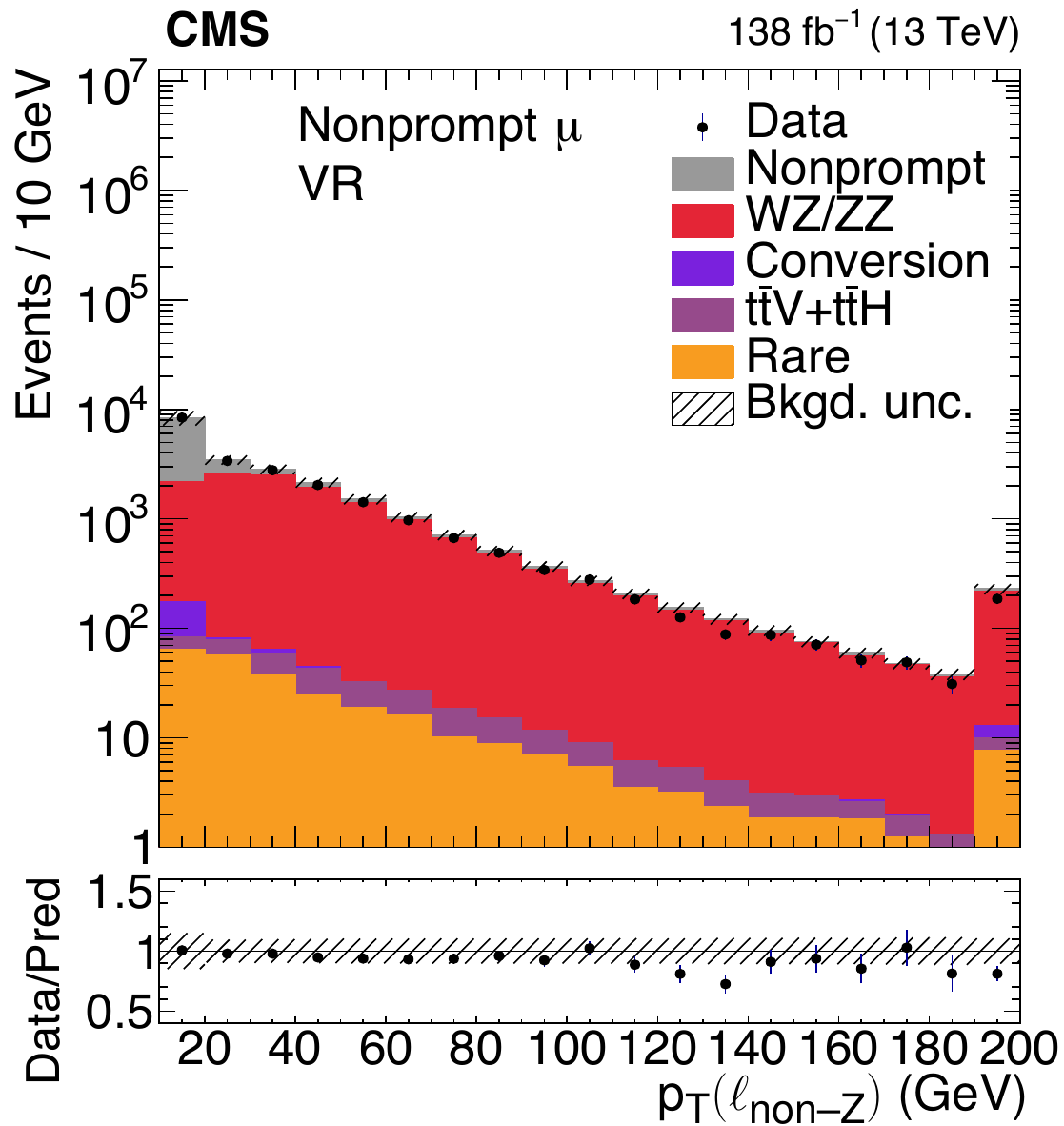}

  \caption{Distributions of \pt of the electron (left) or muon (right) not originating from the $\PZ$ boson decay in the trilepton VR are shown for the nonprompt-lepton background validation. The hatched bands show the total uncertainty, including the systematic component. The last bins include the overflow entries. The lower panels show the ratio of the observed data over the background estimate. }
  \label{fig:FakeValidation}

\end{figure*}

Lepton misidentification rates in data are measured in single-lepton events passing single-lepton triggers. Two different single-lepton triggers described in Table~\ref{tab:trigger_specs} are used depending on the lepton \pt. Events containing only jets produced through the strong interaction, referred to as QCD multijet events, are selected for the measurement, requiring exactly one loose ID lepton and at least one jet with $\pt>40\GeV$ and $\DR(\Pell,\text{j})>1$. To reduce the contamination from processes with prompt leptons, such as \PW boson production in association with jets, $\ptmiss<25\GeV$ and $\mT(\Pell\ptvecmiss)<25\GeV$ are also required. Residual prompt-lepton contributions after the selection are estimated using simulation and are subtracted in the calculation of the misidentification rates. The misidentification rate ranges from 0.18 (0.10) to 0.46 (0.25) for an electron (muon), depending on the lepton \pt and $\eta$. A closure test is performed on the \ttbar simulation. For this closure test, the nonprompt-lepton events estimated using misidentification rates obtained from simulation and the true nonprompt-lepton events identified using generator-level information are compared in the SR. The result of the closure test is presented in Fig.~\ref{fig:BDTGClosureMu}. The closure test shows agreement within 10 (15)\% for dielectron (dimuon) events. The nonprompt-lepton background estimation is validated by comparing it to the data in a separate VR with three leptons, where an OS SF lepton pair from the \PZ boson decay is selected by requiring $\abs{m(\Pell\Pell)-\mZ}<10\GeV$, and the third lepton is assumed to be the nonprompt lepton. In the case of three SF leptons, among two possible combinations of OS lepton pairs, the lepton pair from \PZ boson is defined as the pair with $m(\Pell\Pell)$ closer to \mZ. The \pt distribution of the nonprompt lepton in the VR is shown in Fig.~\ref{fig:FakeValidation}, with good agreement between data and the background prediction.

\subsection{Lepton charge mismeasurement backgrounds}

 Lepton charge mismeasurement backgrounds are processes with two OS leptons in which the reconstructed charge of one of the leptons is mismeasured. From the earlier measurement in CMS, the CMR for muons is known to be less than $10^{-5}$ and does not produce a measurable amount of background, so it is ignored~\cite{CMS:2011bet}. The SM processes that contribute to this background include DY, $\ttbar$, diboson, and single top quark production processes. A direct estimate of the events with two SS leptons from the respective simulated samples suffers from low event counts and results in large statistical uncertainties. Consequently, we estimate this background by measuring the CMR in data and applying it to events in the simulated samples that pass the SR selection, except for the SF OS leptons requirement. The electron CMR is determined from simulated $\PZ\to\Pe\Pe$ events as a function of the electron \pt and $\eta$, and is subsequently corrected to match the observation in data. The electron CMR in simulation is measured by comparing the number of charge-mismeasured electrons and the number of electrons with the correct charge using the generator-level information. The electron CMR in data is measured using the $\PZ\to\Pe\Pe$ resonance. Events must have exactly two electrons, with \pt larger than 25 and 15\GeV, respectively, and pass the tight ID with the $m(\Pe\Pe)$ between 60 and 120\GeV. The number of SS and OS $\PZ\to\Pe\Pe$ events is extracted by fitting the $m(\Pe\Pe)$ distributions to the template shape from simulation. 

\begin{figure}[htbp]
\centering
  \includegraphics[width=0.48\textwidth]{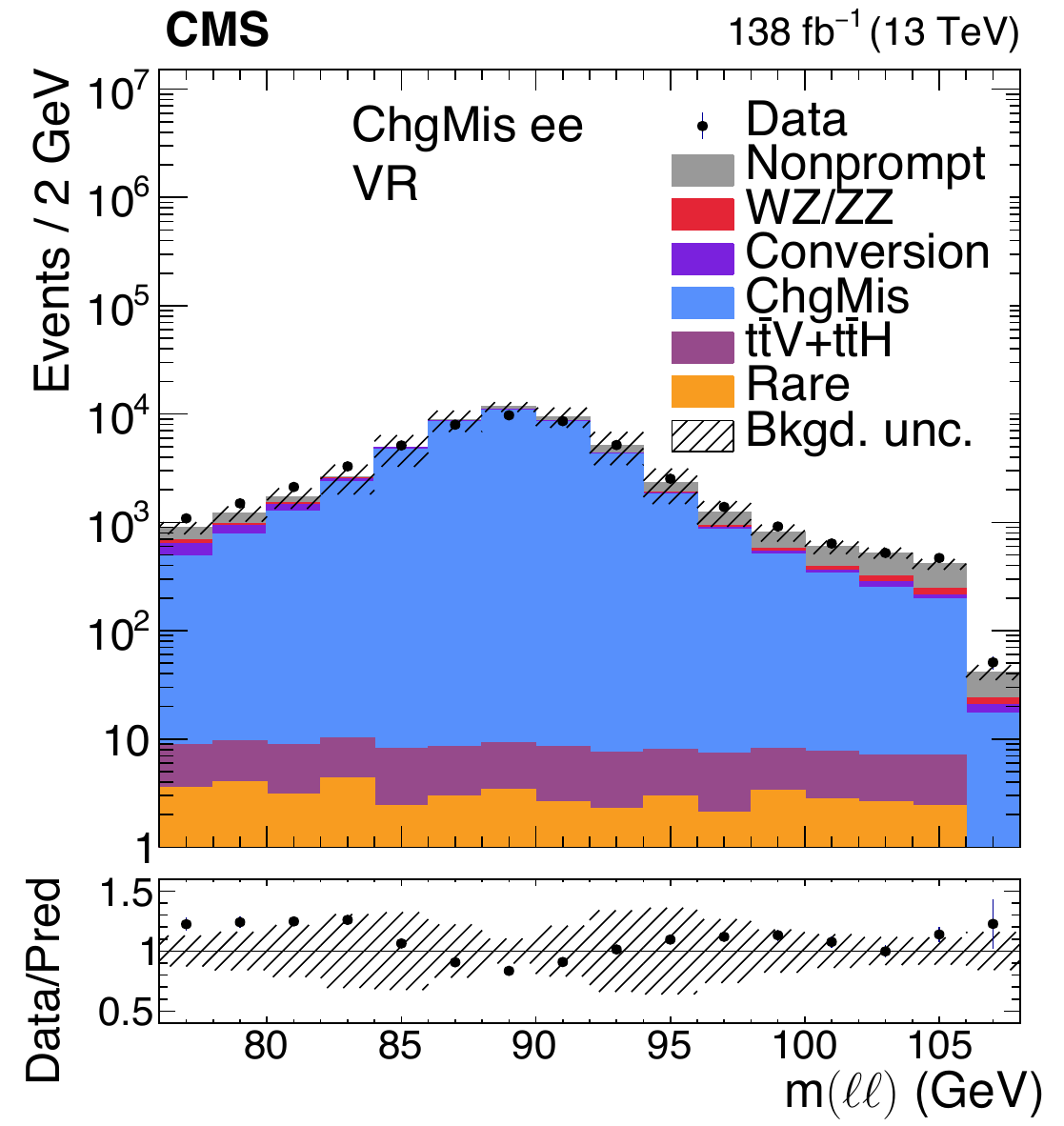}
  \caption{The invariant mass distribution for the SS electron pairs in the lepton charge mismeasurement VR. The hatched bands show the total uncertainty, including the systematic component. The lower panels show the ratio of the observed data over the background estimate.}
  \label{fig:CflipVal}

\end{figure}

 An iterative approach is used to correct the CMR in simulation to match that in data. At each step of the iteration, the CMR ratio between data and simulation is measured and used to scale the number of simulated events with a charge-mismeasured electron. Because both electrons in each event contribute to  their corresponding CMR bin depending on whether their charge is mismeasured or not, the new CMR for the simulation does not perfectly agree with the data CMR. The iteration of scaling the CMR in simulation stops when the magnitude of the corrections for the CMR falls below 1.5\% for all \pt-$\eta$ bins. After the iteration, the product of the correction factors at each step becomes the total correction factor. The total correction factor ranges from 0.74 to 1.61, depending on the electron \pt, $\eta$, and the data-taking era. The electron charge mismeasurement background is validated in data using charge-mismeasured DY events with tighter selection on the $m(\Pe\Pe)$ than the one used in the data CMR measurement region: the two electrons from the \PZ boson decay are measured to have the same charge, and their invariant mass is within 15\GeV of \mZ. The invariant mass distribution shows good agreement between data and the background prediction within the uncertainty, as shown in Fig.~\ref{fig:CflipVal}.

\subsection{Conversion backgrounds}

\begin{figure*}[htbp]
\centering

  \includegraphics[width=0.48\textwidth]{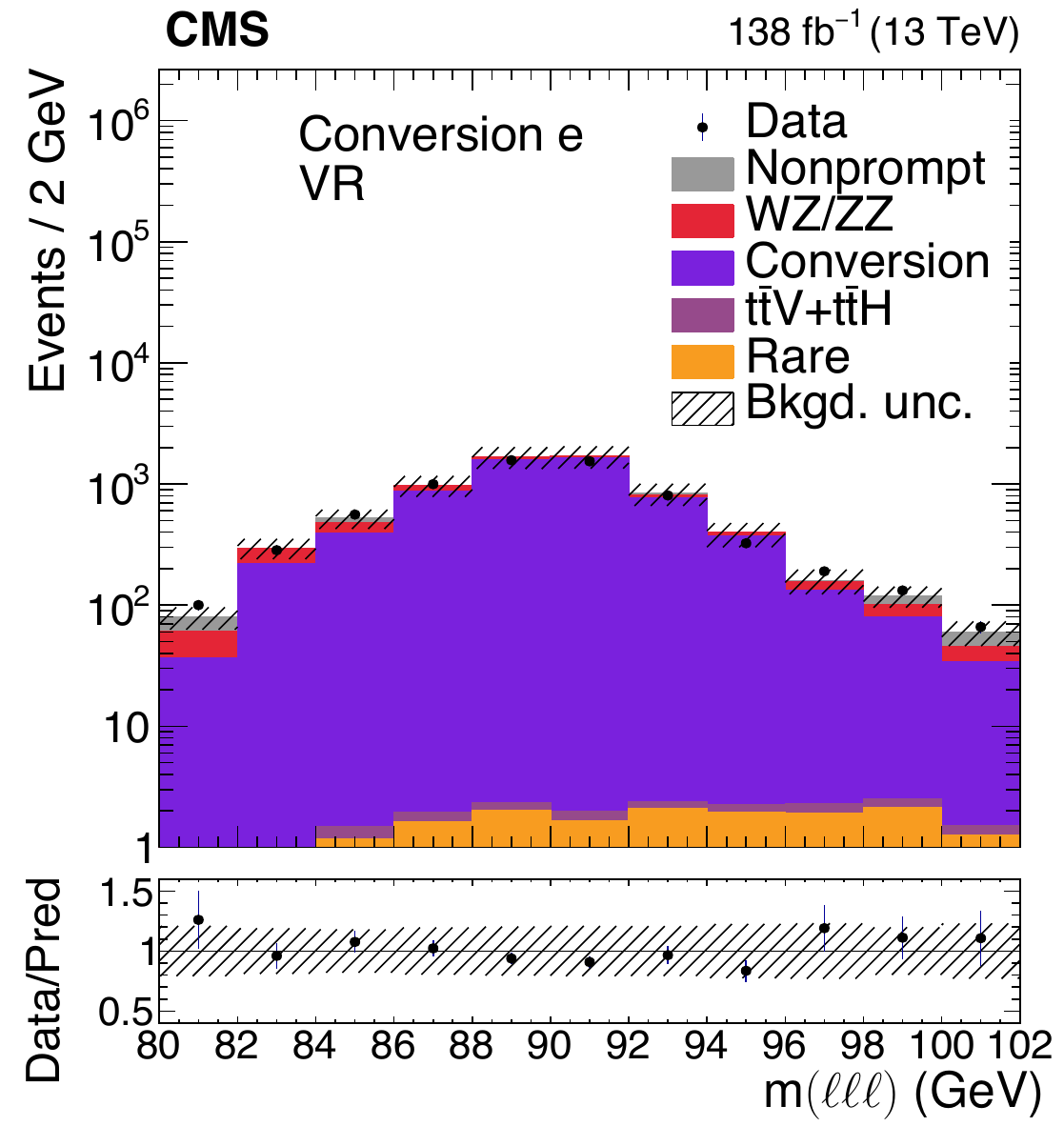}
  \includegraphics[width=0.48\textwidth]{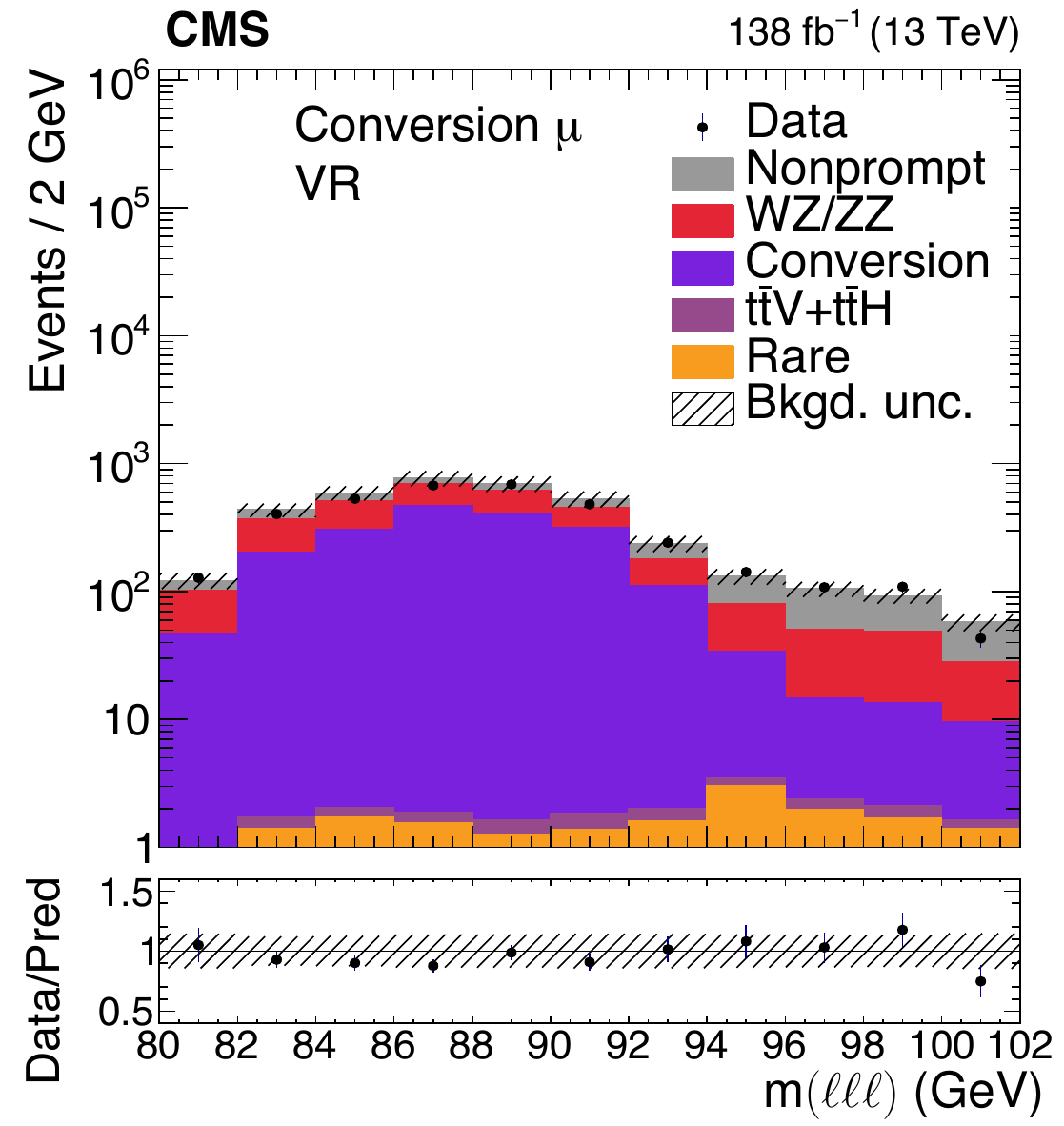}
  \caption{The invariant mass distribution of three leptons is shown for the $\PZ$ boson plus conversion electron (left) or muon (right) events for the validation of the conversion background. The hatched bands show the total uncertainty, including the systematic component. The lower panels show the ratio of the observed data over the background estimate.}
  \label{fig:ConvValidation}

\end{figure*}

The conversion backgrounds include internal and external photon conversions, depending on whether the conversion occurs at the collision point or within the detector, respectively. Since the rate of external conversion is proportional to the inverse of the produced lepton mass squared~\cite{Bethe:1934za}, it is only significant for electrons. For the validation of this background, we select events with three leptons: two from a \PZ boson decay and an additional lepton from a photon conversion. The invariant mass of the three leptons is required to be within a 10\GeV window around \mZ. In this three-lepton VR, we observe reasonable agreement between data and the background prediction within the uncertainty, as shown in Fig.~\ref{fig:ConvValidation}. Therefore, the conversion backgrounds are taken from the simulation.

\subsection{Prompt-lepton backgrounds}

Prompt-lepton backgrounds originate from processes involving at least two SS leptons produced in the decays of \PV bosons, including leptons from decays of \PGt leptons originating from these bosons. These contributions are estimated using simulated samples. Major contributions come from $\ttbar\PV$, $\ttbar\PH$, $\PW\PZ$, and $\PZ\PZ$ production. Processes with small expected contributions are combined as ``rare'' background, dominated by the production of $\PQt\PZ$, $\PV\PV\PV$, $\Pg\Pg\PH$, and $\PQq\PQq\PH$.

\begin{table}[htbp]
\centering
\topcaption{Summary of event selection criteria for the SR and VRs.}
\cmsTable{
\begin{tabular}{lcccc}
{Selection} & {SR} & {Nonprompt VR} & {Conversion VR} & {ChgMis VR} \\ \hline
$N$($\Pell$) & 2  & 3 & 3  & 2 \\ 
Lepton charge & SS& \NA & \NA & SS \\ 
\multirow{2}{3cm}{Lepton \pt ({\GeVns})} & $\Pe_{1,2}>25, 15$  & $\Pe_{1,2}>25, 15$& $\Pe_{1,2}>25, 15$ & \multirow{2}{2.1cm}{$\Pe_{1,2}>25, 15$}\\ 
&$\PGm_{1,2}>20, 15$ & $\PGm_{1,2}>20, 15$ & $\PGm_{1,2}>20, 15$ & \\ [\cmsTabSkip]
\multirow{2}{3cm}{Lepton flavor} & $\Pe\Pe$ & $\Pe\Pe\PGm$ or $\Pe\Pe\Pe$  & $\Pe\Pe\PGm$ or $\Pe\Pe\Pe$   & \multirow{2}{0.4cm}{$\Pe\Pe$} \\ 
& $\PGm\PGm$ &$\PGm\PGm \Pe$ or $\PGm\PGm\PGm$ & $\PGm\PGm \Pe$ or $\PGm\PGm\PGm$ & \\ [\cmsTabSkip]
\Njets & $\geq$3 & \NA & \NA & \NA \\ 
\Nbjets & $\geq$1& 0 & 0 & \NA \\ [\cmsTabSkip]
\multirow{2}{1cm}{$N$($\text{j}_{\text{L}}$)} & $\geq$1 if $\mN < \mW$& \multirow{2}{*}{\NA} & \multirow{2}{*}{\NA} & \multirow{2}{*}{\NA} \\ 
 & $\geq$2 if $\mN > \mW$&  &  &  \\ [\cmsTabSkip]
Mass window ({\GeVns}) & $\abs{m(\Pe\Pe)-\mZ}>10$ & OS SF pair & $m(\Pell\Pell\Pell)\in $   & $m(\Pe\Pe)\in $  \\ 
& $m(\PGm\PGm)  > 4$ & mass $\in [81.2,101.2]$ & $[81.2,101.2]$ & $[76.2, 106.2]$\\ 
\end{tabular}
}
\label{tab:SRVR}
\end{table}

\section{Systematic uncertainties}
\label{Systematic}
  Various sources of systematic uncertainties are considered in the analysis, categorized as background, experimental, and theoretical uncertainties. Experimental and theoretical uncertainties are applied to both signal and simulated backgrounds. The systematic uncertainties are implemented as nuisance parameters in the analysis. Depending on their nature, the uncertainties are treated as correlated or uncorrelated between the data-taking eras. The summary of the uncertainties is shown in Table~\ref{tab:systematics}.

\subsection{Background systematic uncertainties}
 The misidentification rates in nonprompt-lepton backgrounds contribute the largest uncertainty in the total background estimation. We use two methods for assessing the uncertainties in the nonprompt-lepton background. First, from the closure test described in Section~\ref{BkgEst}, an uncertainty of 10 (15)\% is set in the dielectron (dimuon) channel. Secondly, the uncertainty is assessed by analyzing three sources of uncertainty in the misidentification rate itself. The normalization of the prompt-lepton background is varied up and down by the theoretical cross section uncertainty when subtracting it from data. The minimum jet \pt requirement for the BMR selection is varied up and down by 10 or 5\GeV, depending on the single-lepton trigger threshold. The dependence of the measured misidentification rate on the jet flavor is quantified by requiring at least one \PQb-tagged jet among the jets. The relative contributions of these three sources of uncertainty vary with the lepton \pt and $\eta$. In the SR, the combined uncertainty in the nonprompt-lepton background after propagating the effects of the three uncertainty sources is within a 20\% band around the nominal BDT score distribution. To conservatively account for the uncertainties assessed from the closure test and the misidentification rate, a flat 20\% uncertainty is assigned to the nonprompt-lepton background. This uncertainty is treated as uncorrelated between the data-taking eras.

 Three sources of uncertainty are considered in the estimation of the charge-mismeasured background. The first uncertainty arises from the measurement of the correction factors for the CMR. This uncertainty is quantified by using different fitting models for data CMR extraction, and it is found to vary between 2 and 7\% across the \pt-$\eta$ bins. Since this uncertainty is dominated by the statistical component, it is treated as uncorrelated between the data-taking eras. The second uncertainty accounts for an imperfect parameterization of the CMR due to the limited sample size, and it is estimated to be 6\% by performing a closure test on \ttbar simulated events. The test consists of estimating the number of SS dielectron events by applying the measured CMR to the OS dielectron events and comparing the result to the observed number of SS dielectron events. This uncertainty is treated as correlated between the data-taking eras. The last source of uncertainty originates from the electron energy measurement. Highly energetic bremsstrahlung can deflect the electron path, leading to a mismeasurement of the electron charge. Additionally, the emitted radiation may become spatially separated from the electron and not be taken into account in the electron energy reconstruction. To account for this, we emulate the energy underestimation in the simulated events with OS dielectron pairs by shifting the electron \pt in the VR. The nominal shift is optimized by minimizing the $\chi^2$ difference between the OS and SS $m(\Pe\Pe)$ distributions over a $\pm$3\% energy shift scan. We evaluate the systematic uncertainty by repeating this optimization in two kinematic regions, using events with both electrons $\pt > 40\GeV$ or both $\pt<40\GeV$. The maximum absolute deviation from the nominal shift is assigned as a symmetric systematic uncertainty. The uncertainty in the electron \pt from this correction is between 0.7 and 1.5\%, depending on the electron $\eta$ and data-taking era. This uncertainty is treated as correlated between the data-taking eras.

For the conversion background, a 20\% uncertainty is assigned to the rate based on the agreement in the three-lepton VR. The simulation underestimates the conversion event rate in the three-lepton VR by approximately 20\% in all data-taking eras for dimuon events, while the data-to-simulation difference for electron conversions varies between 0 and 20\%, depending on the data-taking eras. Consequently, the muon conversion background uncertainty is treated as correlated, while the electron one is treated as uncorrelated, between the data-taking eras.

\subsection{Experimental systematic uncertainties}

The uncertainty in lepton ID is included, ranging from 1 to 4\% depending on the lepton \pt and $\eta$, and is treated as correlated between the data-taking eras. Lepton ID correction factors are measured using the ``tag-and-probe'' method~\cite{tagandprobe}, where the efficiency is measured as the fraction of the reconstructed leptons from DY events that pass the ID requirement. The uncertainty is then determined by changing the minimum \pt requirement on the leptons, the fit function for the $m(\Pell\Pell)$ distribution, and the Monte Carlo generator. 

For electrons, there are uncertainties associated with the efficiency of electron reconstruction and energy correction~\cite{CMS:2020uim}. These uncertainties are treated as correlated between different data-taking eras. In the case of muons, the reconstruction efficiency is close to unity, and no statistically significant evidence for the difference between data and simulation is observed. The muon energy scale uncertainty is considered, and it is treated as correlated between the data-taking eras~\cite{CMS:2018rym}.

 Uncertainties in the jet energy scale and resolution are assessed by reevaluating observables in simulated events with the jet energy shifted by its uncertainty. The systematic uncertainties associated with the jet energy scale and resolution corrections are considered uncorrelated between the data-taking eras. These uncertainties are propagated to the $\ptmiss$ uncertainty. Additionally, the uncertainty in the energy scale of PF particles not associated with jets, leptons, and photons, known as unclustered energy, is propagated to the $\ptmiss$ uncertainty~\cite{CMS:2019ctu}. The uncertainties in the \PQb tagging efficiency correction are evaluated by varying the correction factors for light-quark and gluon, \PQc quark, and \PQb quark jets in simulation~\cite{CMS-DP-2018-058}. Each \PQb tagging uncertainty has both correlated and uncorrelated sources between the data-taking eras. The uncertainty in the efficiency correction factor corresponding to the pileup jet identification is also applied and treated as uncorrelated between the data-taking eras~\cite{CMS-DP-2020-020}.

Other experimental uncertainties have small impacts. The uncertainty in the integrated luminosity~\cite{CMS-LUM-17-003,CMS-PAS-LUM-17-004,CMS-PAS-LUM-18-002} has both correlated and uncorrelated components between the data-taking eras. 
The overall integrated luminosity uncertainty for the 2016--2018 data set is 1.6\%. The uncertainty in the number of pileup interactions is estimated by applying a $\pm$4.6\% variation to the total inelastic $\Pp\Pp$ cross section used to reweight simulated events to match the pileup distribution observed in data~\cite{CMS:2018mlc} and is treated as correlated between the data-taking eras. The uncertainties in the trigger efficiencies are measured using the tag-and-probe method described earlier and are treated as correlated between the data-taking eras. During the 2016 and 2017 data-taking periods, a gradual shift in the timing of the inputs of the ECAL L1 trigger caused a trigger inefficiency, referred to as the ``L1-prefiring''. An uncertainty related to the inefficiency from the L1-prefiring in 2016 and 2017 is applied and treated as uncorrelated between the data-taking eras~\cite{L1Prefire}.

The statistical uncertainty of the data used for the nonprompt-lepton background estimations, as well as that of the simulated samples used in all background estimations, is considered. 
\subsection{Theoretical systematic uncertainties}

An uncertainty of ${}^{+3.5}_{-4.4}$\% is assigned to the signal production cross section~\cite{tt1}. Theoretical uncertainties in the signal acceptance from the uncertainties in the PDFs and in the renormalization ($\mu_\text{R}$) and factorization ($\mu_\text{F}$) scales are considered. For the uncertainty arising from PDFs, the changes in the acceptance for 100 replicas of the PDF set are added in quadrature. The uncertainties in the signal acceptance arising from $\mu_\text{R}$ and $\mu_\text{F}$ are studied by varying the scales by factors of 0.5 and 2 individually or simultaneously, excluding the two variations in the opposite directions, and taking the envelope of all variations. The uncertainty in the parton shower modeling is estimated by varying initial- and final-state radiation scales separately by factors of 0.5 and 2 and is defined by the envelope obtained from these variations. Among the signal theoretical uncertainties, the ones from initial- and final-state radiation, and from the \ttbar cross section are dominant. 

Three theoretical uncertainties in the cross sections of the background processes are taken into account: $\mu_\text{R}$ and $\mu_\text{F}$, PDFs, and strong coupling constant (\alpS), added in quadrature. Because the uncertainties resulting from higher and lower scale choices are comparable in magnitude, the larger deviation is assigned as a symmetric uncertainty. For $\PW\PZ$ production, the radiation zero effect is considered, whose magnitude decreases in the diagrams with additional partons in higher-order QCD calculations~\cite{WZNNLOCalc}. We assign an 11\% uncertainty due to the radiation zero effect, resulting in a total uncertainty of 12\%. For $\PZ\PZ$ production, we apply a cross section uncertainty of 6.4\%~\cite{ZZNNLOCalc}. For $\ttbar+\PV/\PH$ we apply the uncertainty from Ref.~\cite{YR4}. For $\ttbar\PH$ processes, the uncertainty in the Higgs boson branching fraction is also included. The total uncertainty values obtained are 11.9\% for $\ttbar\PZ$, 13.3\% for $\ttbar\PW$, and 10.0\% for $\ttbar\PH$. For the $\PQt\PZ$ process, the total uncertainty is 5.2\%~\cite{tXNLOCalc}. For $\PV\PV\PV$, $\Pg\Pg\PH$, $\PQq\PQq\PH$, and \PV\PH production, a conservative 50\% uncertainty is assigned. All uncertainty sources with theoretical origins are considered as fully correlated between the data-taking eras.

\begin{table}[htbp]
\centering
\topcaption{Summary of systematic uncertainties. Each uncertainty is
implemented as either a rate or shape uncertainty.  Uncertainties affecting only the normalization are marked as ``Rate'' in the table. Those affecting
BDT template shapes are indicated as ``Shape.'' For shape uncertainties, ``\NA'' indicates that its effect is propagated to the BDT score distribution through the corresponding up and down templates and may depend on the process, channel, and HNL mass hypothesis.}
\label{tab:systematics}

\begin{tabular}{lcc}
Source & Assigned variation & Type \\ \hline
\multicolumn{3}{c}{\textit{Background uncertainties}}\\[\cmsTabSkip]
  Nonprompt leptons                    & 20\%           & Rate  \\
  CMR fitting model                    & \NA        & Shape \\
  Charge-mismeasurement closure test   & 6\%            & Rate  \\
  Electron energy shifts               & \NA     & Shape \\
  Conversions                          & 20\%           & Rate  \\[\cmsTabSkip]
\multicolumn{3}{c}{\textit{Experimental uncertainties}}\\[\cmsTabSkip]
  Lepton ID efficiency             & \NA       & Shape \\
  Electron reconstruction & \NA & Shape \\
  Electron energy correction & \NA  & Shape \\
  Muon momentum scale                          & \NA  & Shape \\
  Jet energy scale                             & \NA  & Shape \\
  Jet energy resolution                        & \NA  & Shape \\
  Pileup jet identification                    & \NA  & Shape \\
  \PQb tagging efficiency corrections          & \NA  & Shape \\
  Integrated luminosity                        & 1.6\% & Rate \\
  Inelastic $\Pp\Pp$ cross section               & 4.6\% & Rate  \\
  Trigger efficiencies                         & 1\%  & Rate  \\
  L1-prefiring inefficiency                    & \NA  & Shape \\[\cmsTabSkip]
\multicolumn{3}{c}{\textit{Theoretical uncertainties}}\\[\cmsTabSkip]
  Signal PDF and \alpS variations                    & \NA       & Shape \\
  Signal $\mu_\text{R}$ and $\mu_\text{F}$ variations
                                     & \NA  & Shape \\
  Signal parton shower modeling            & \NA  & Shape \\
  \ttbar cross section              & ${}^{+3.5}_{-4.4}\%$              & Rate  \\
  $\ttbar\PZ$ cross section           & 11.9\%             & Rate  \\
  $\ttbar\PW$ cross section           & 13.3\%             & Rate  \\
  $\ttbar\PH$ cross section           & 10.0\%             & Rate  \\
  $\PW\PZ$ cross section              & 12\%               & Rate \\
  $\PZ\PZ$ cross section              & 6.4\%              & Rate \\
  $\PQt\PZ$ cross section             & 5.2\%              & Rate \\
  $\PV\PV\PV$ cross section           & 50\%               & Rate \\
  $\Pg\Pg\PH$ cross section       & 50\%               & Rate \\
  $\PQq\PQq\PH$ cross section       & 50\%               & Rate \\
  $\PV\PH$ cross section            & 50\%               & Rate \\
\end{tabular}
\end{table}

\section{Results}
\label{Result}

\begin{figure*}[htbptbp]
\centering
  \includegraphics[width=0.48\textwidth]{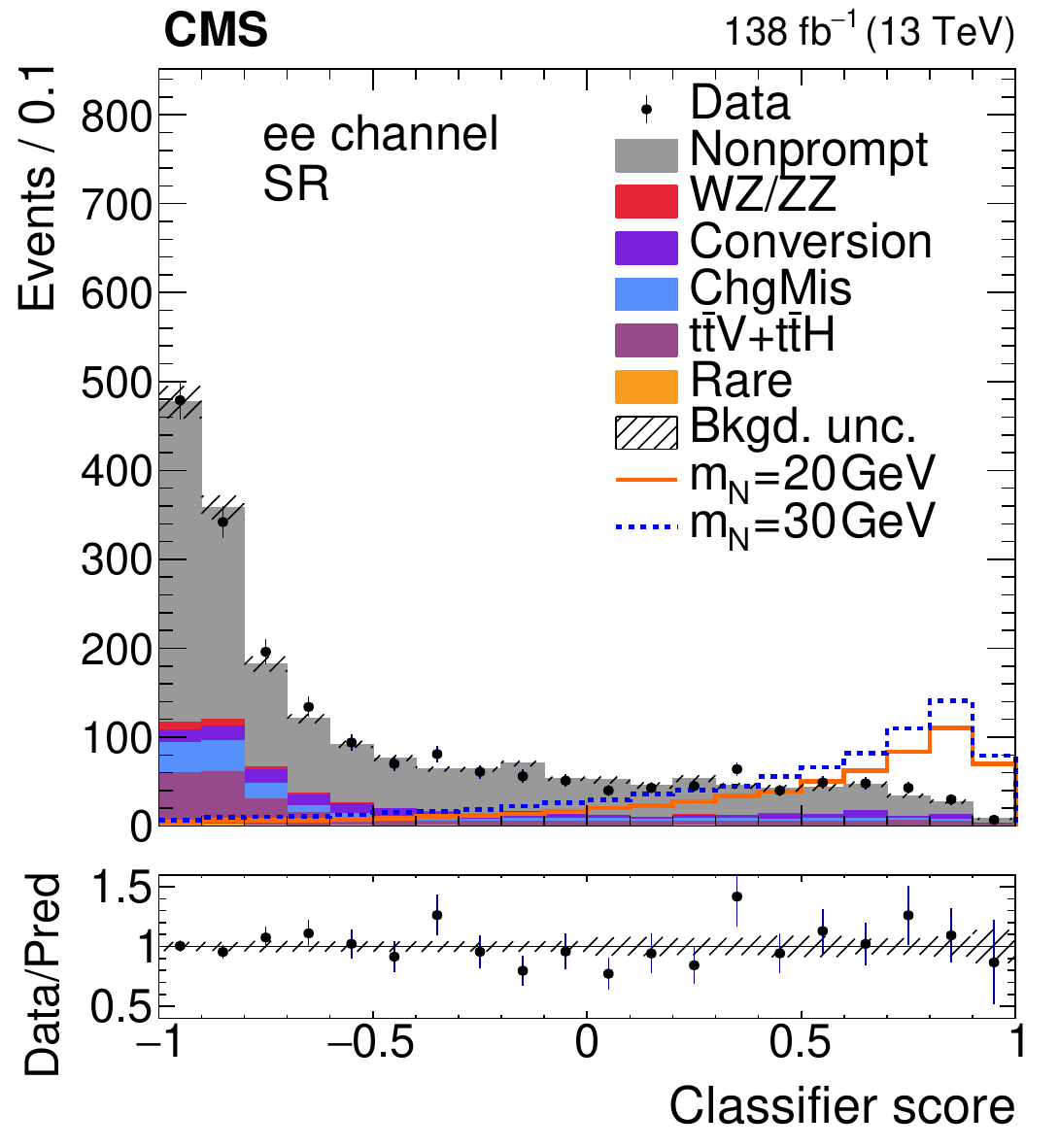}
  \includegraphics[width=0.48\textwidth]{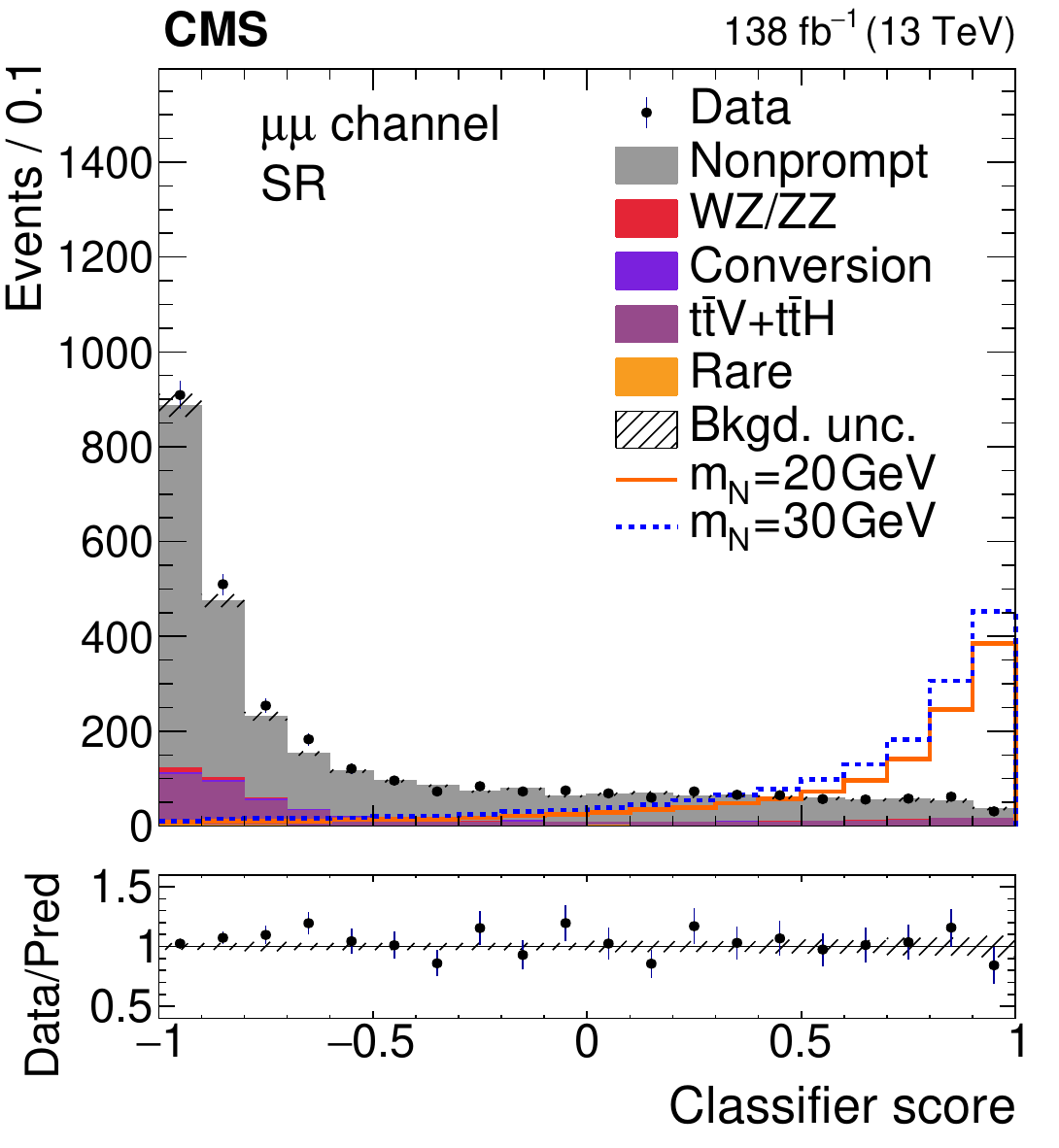}
  \includegraphics[width=0.48\textwidth]{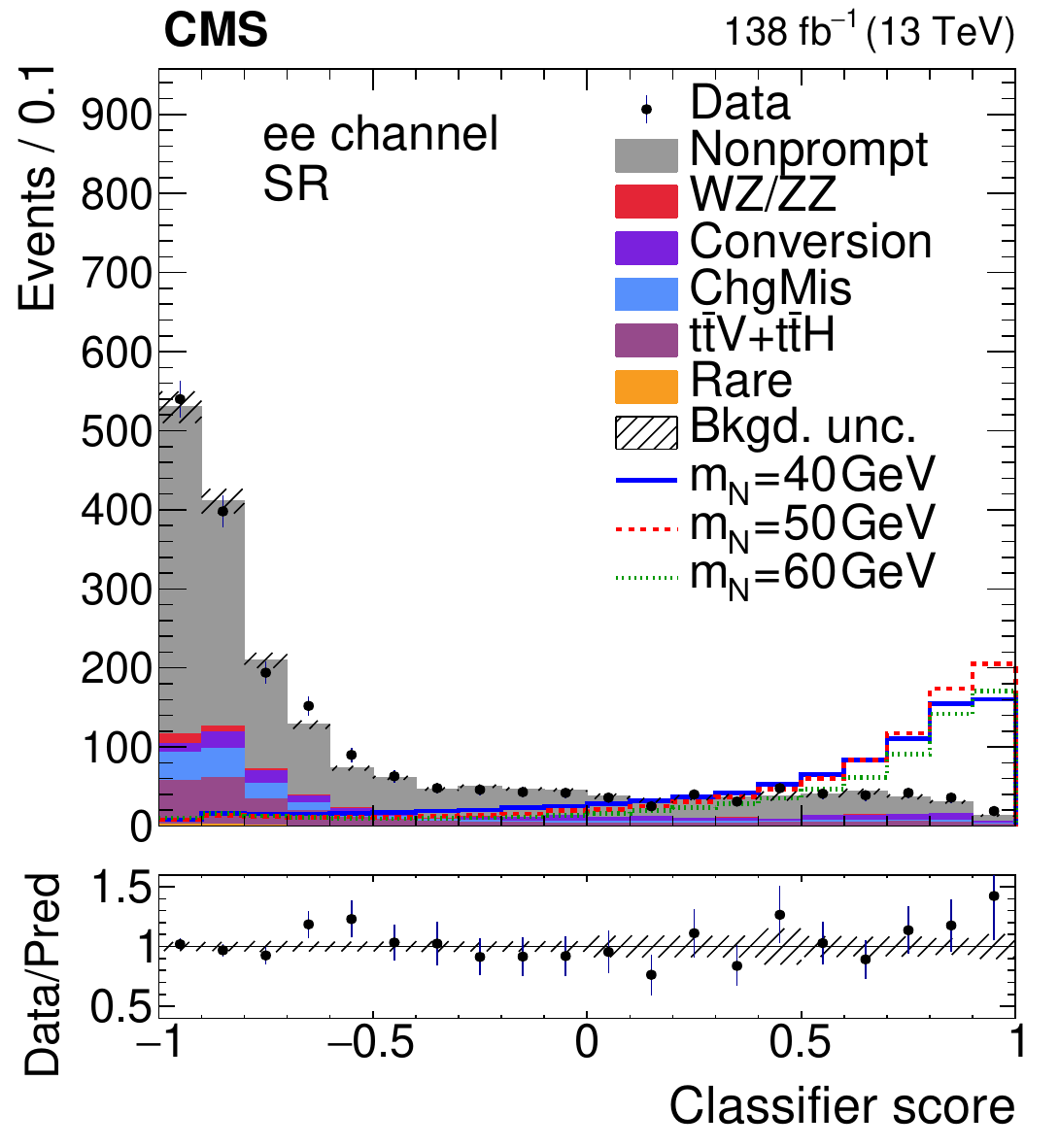}
  \includegraphics[width=0.48\textwidth]{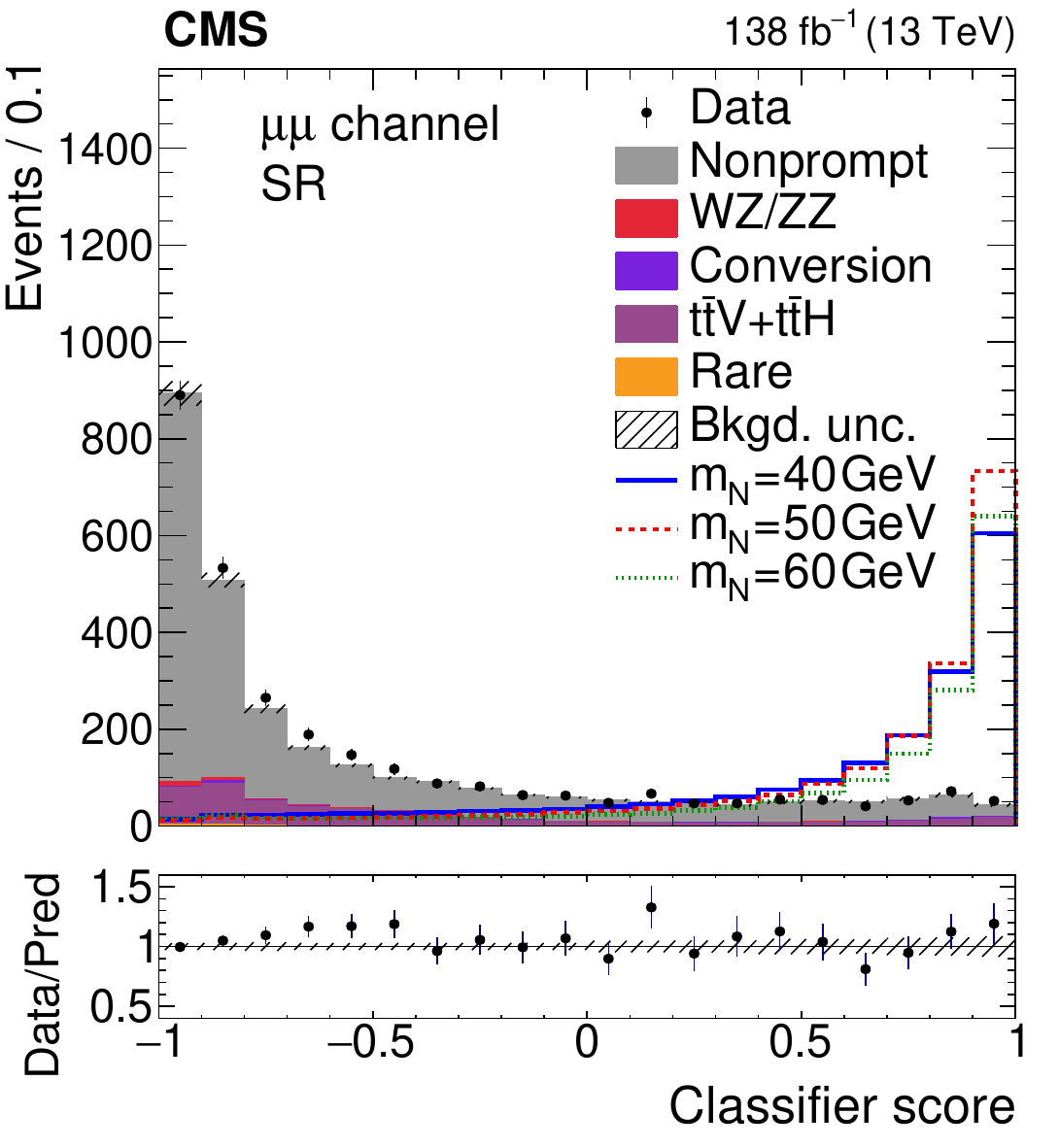}
  \caption{Comparison of the observed (points) and predicted (filled histograms) post-fit BDT score distributions in the dielectron (left) and dimuon (right) channels. The BDT score distributions correspond to classifiers trained with signal mass hypotheses of 20 (upper) and 50\GeV (lower). The expected signal distributions extracted with the same classifier are overlaid on the figures. They are normalized assuming $\BtlN=2\times10^{-4}$, a value chosen to ensure that the signal distributions are visible across all channels and \mN hypotheses. The hatched bands show the total uncertainty, including the systematic component. The lower panels show the ratio of the observed data over the background estimate.}
  \label{fig:ResultBDTLow}
\end{figure*}

\begin{figure*}[htbp]
\centering
  \includegraphics[width=0.48\textwidth]{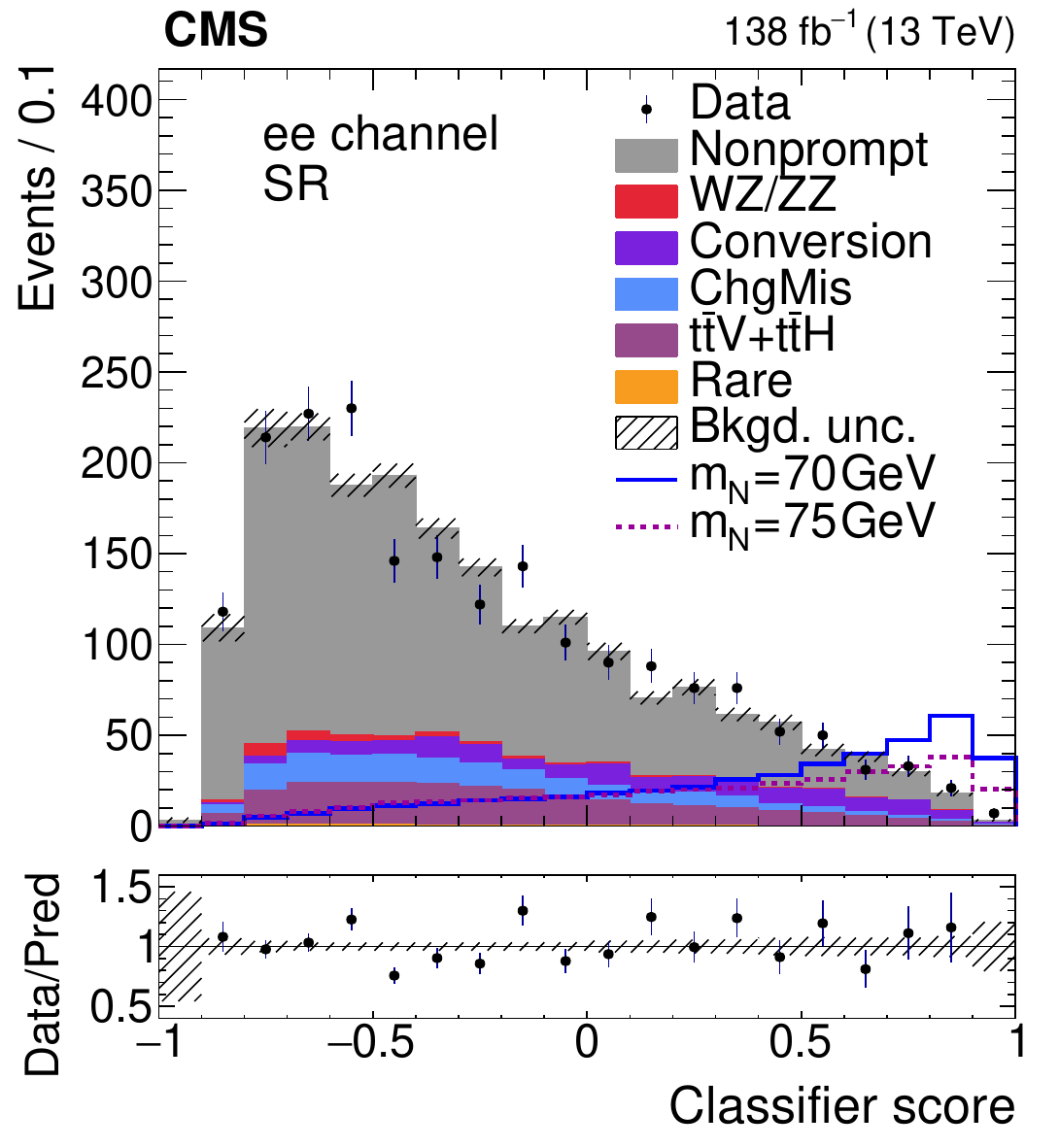}
  \includegraphics[width=0.48\textwidth]{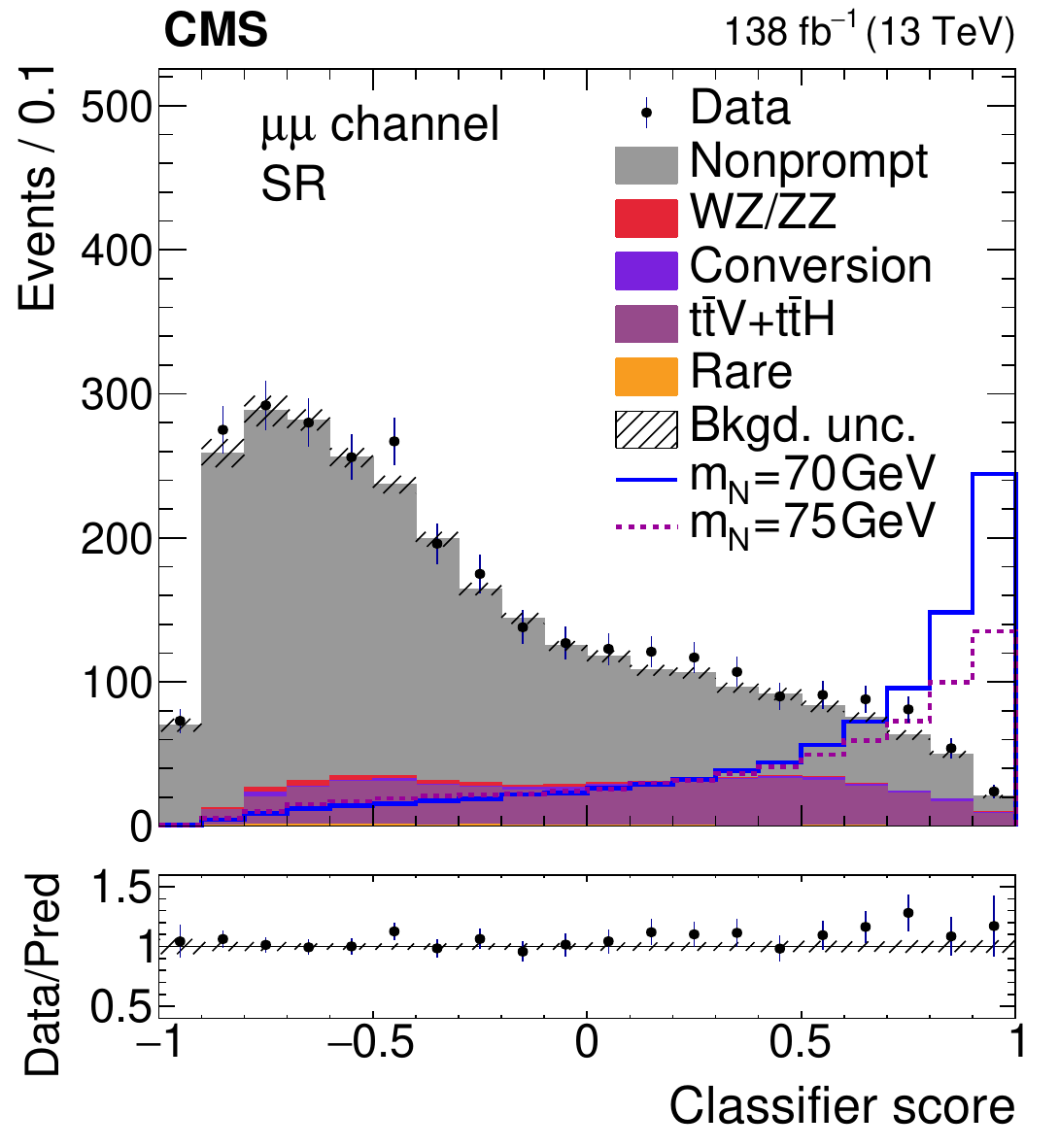}
  \includegraphics[width=0.48\textwidth]{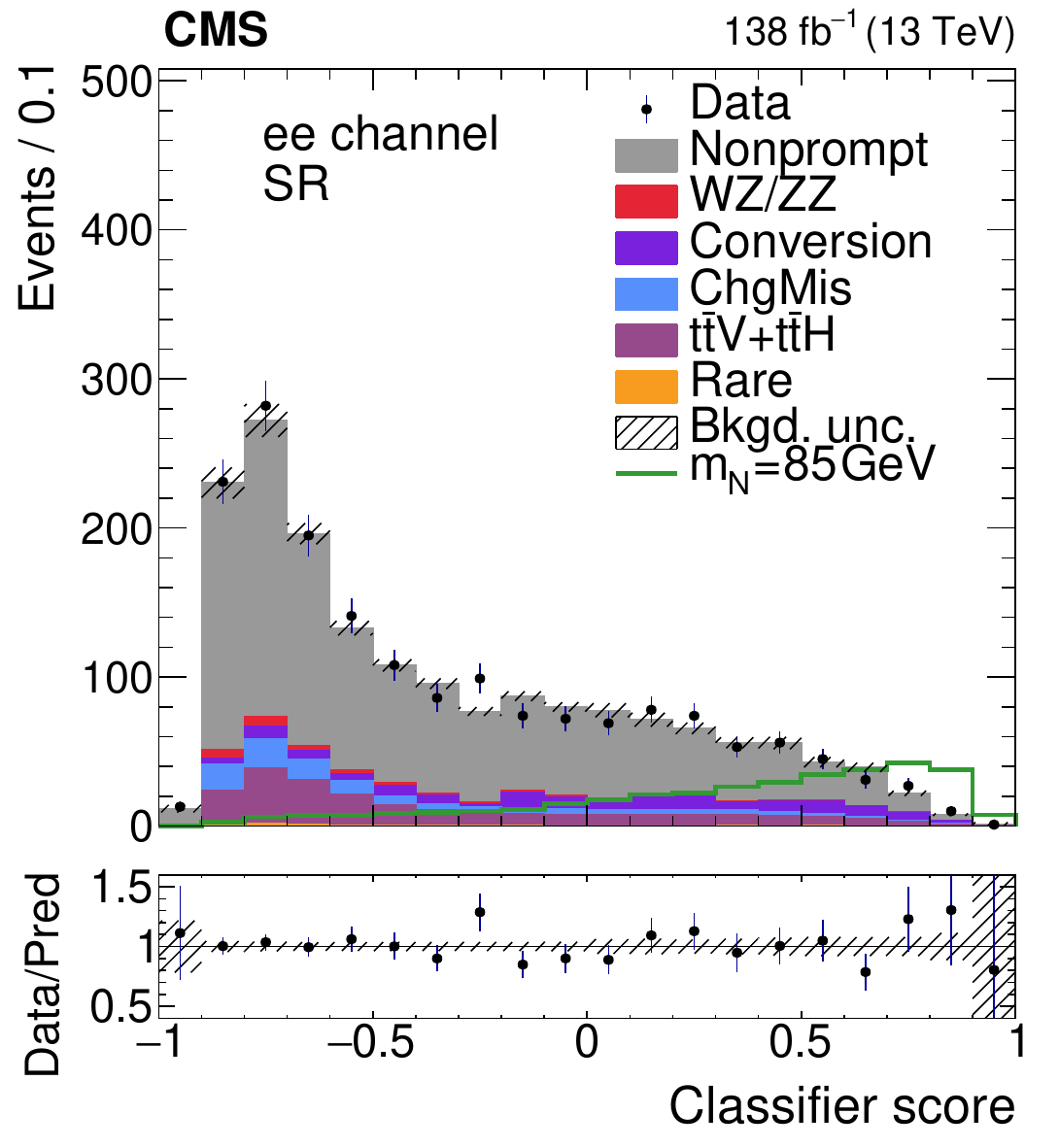}
  \includegraphics[width=0.48\textwidth]{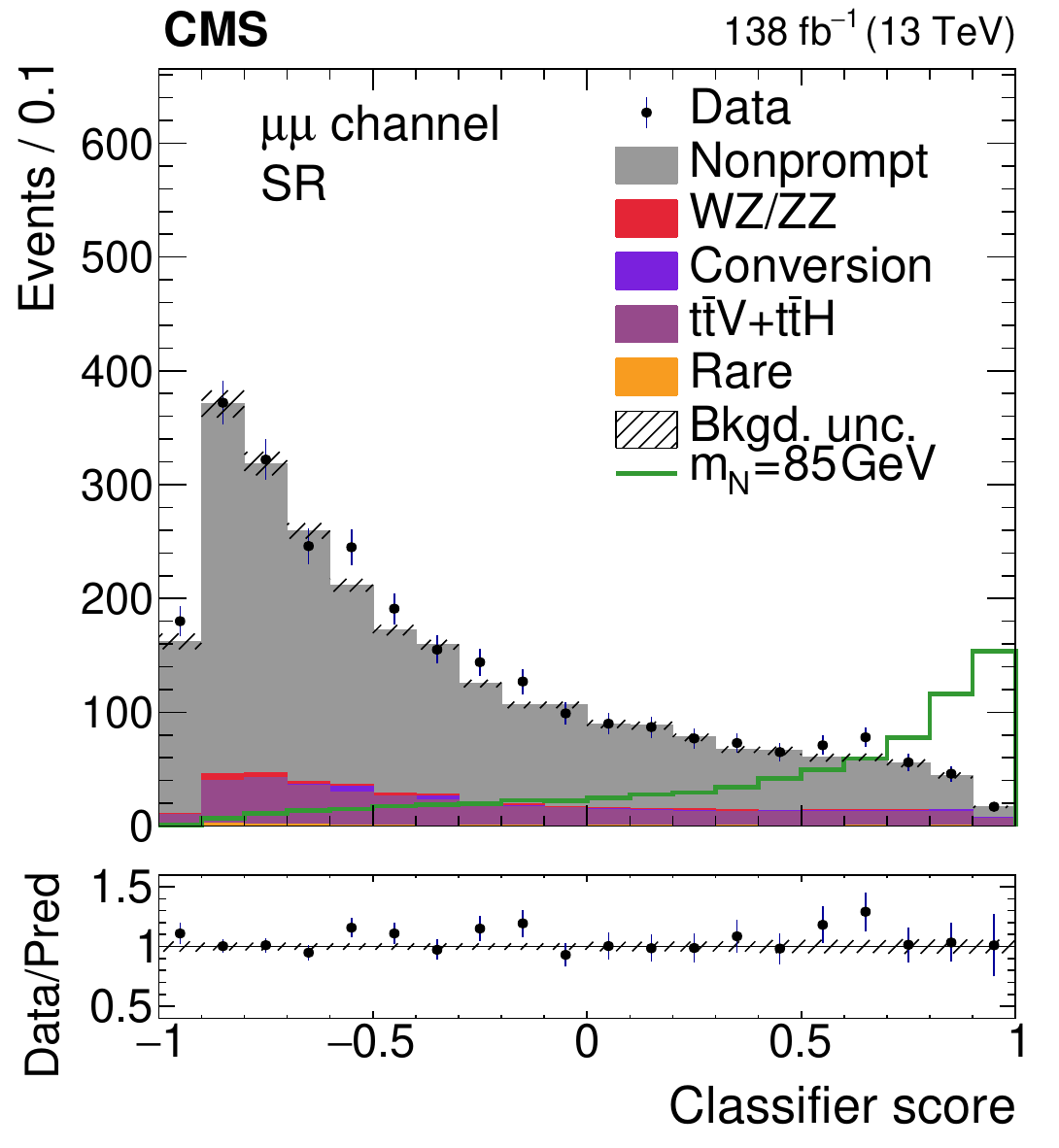}
  \caption{Comparison of the observed (points) and predicted (filled histograms) post-fit BDT score distributions in the dielectron (left) and dimuon (right) channels. The BDT score distributions correspond to classifiers trained with signal mass hypotheses of 75 (upper) and 85\GeV (lower). The expected signal distributions extracted with the same classifier are overlaid on the figures. They are normalized assuming $\BtlN=2\times10^{-4}$, a value chosen to ensure that the signal distributions are visible across all channels and \mN hypotheses. The hatched bands show the total uncertainty, including the systematic component. The lower panels show the ratio of the observed data over the background estimate.}
  \label{fig:ResultBDTMid}
\end{figure*}

\begin{figure*}[htbp]
\centering
  \includegraphics[width=0.48\textwidth]{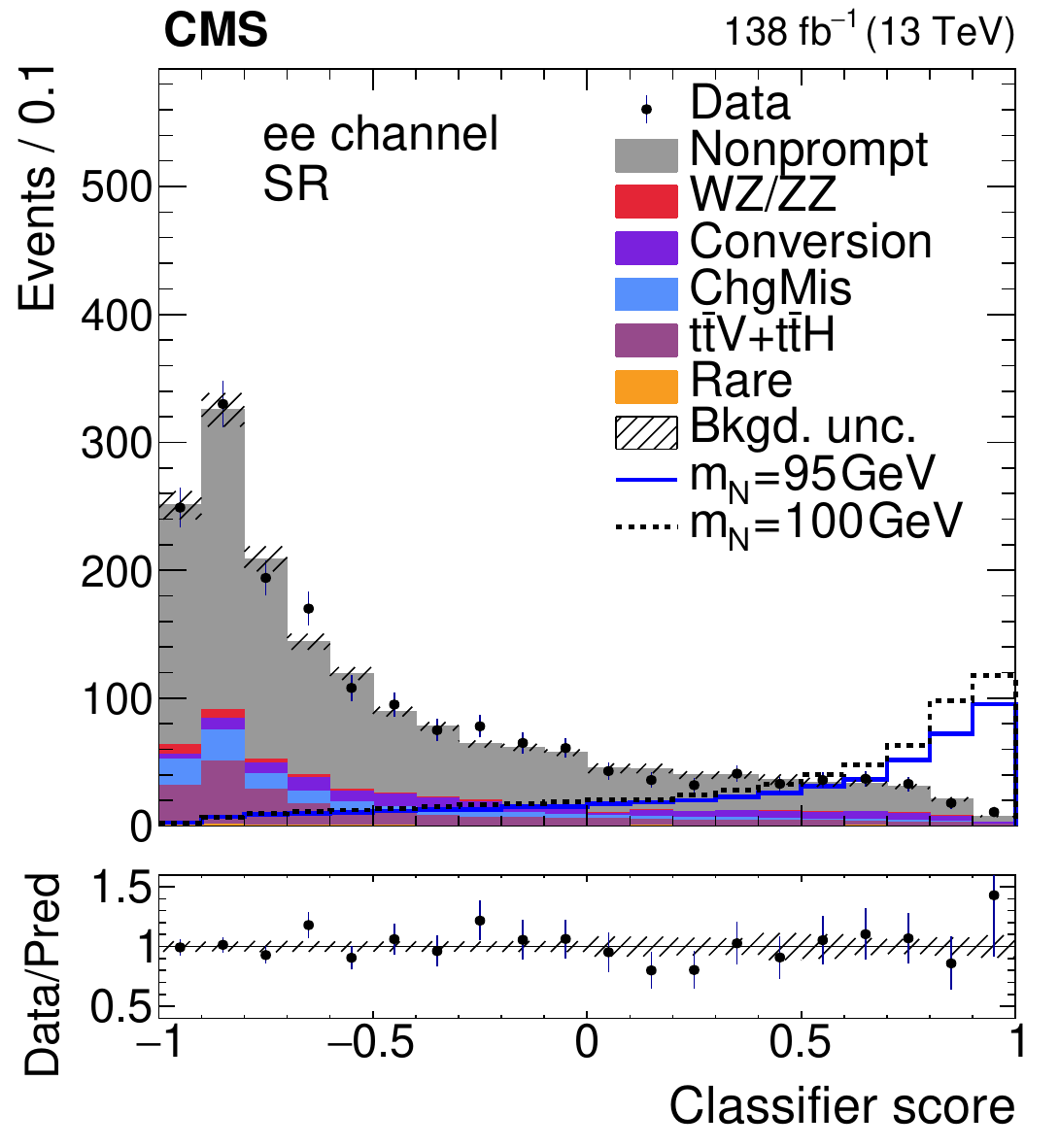}
  \includegraphics[width=0.48\textwidth]{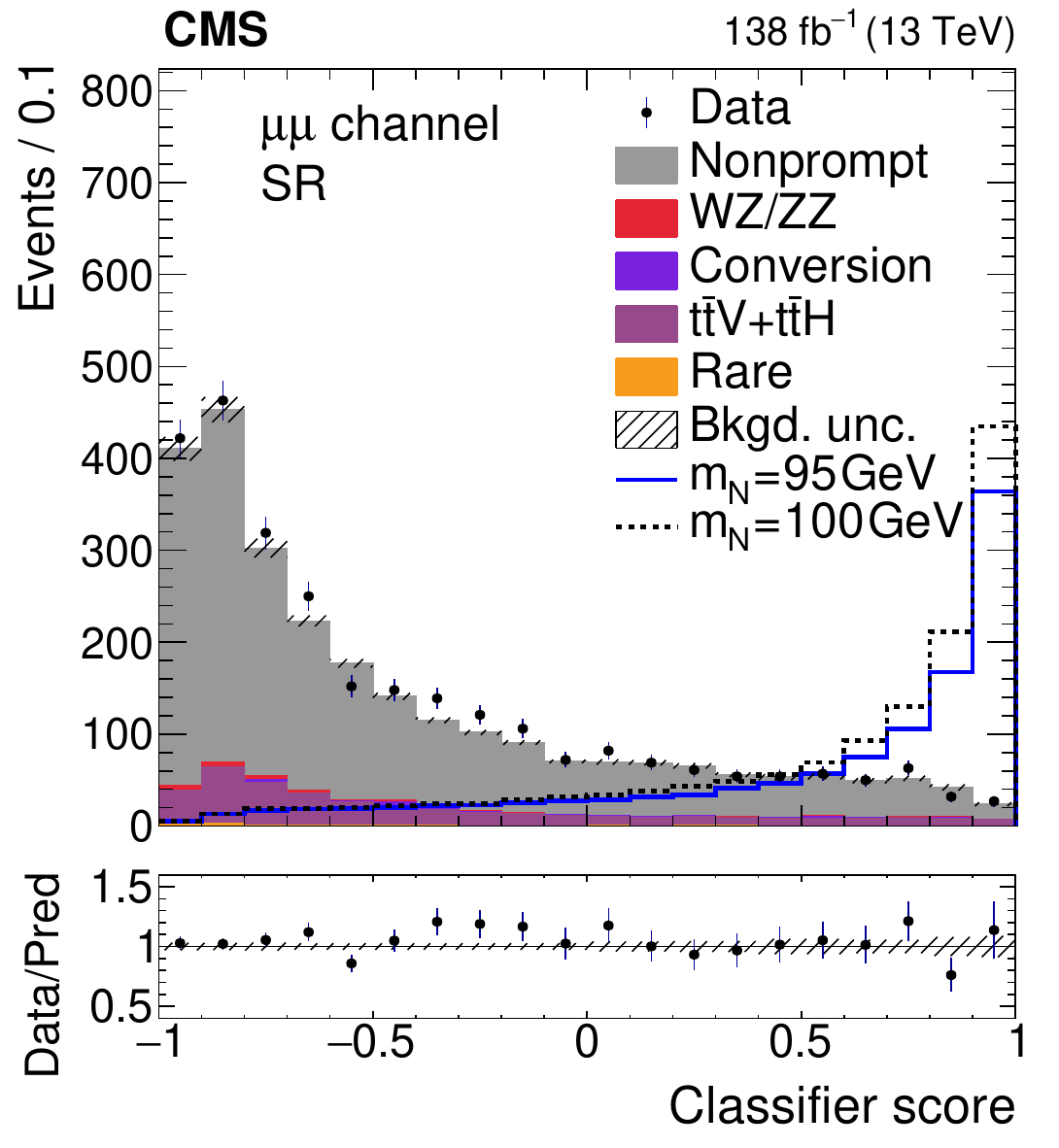}
  \caption{Comparison of the observed (points) and predicted (filled histograms) post-fit BDT score distributions in the dielectron (left) and dimuon (right) channels. The BDT score distributions correspond to classifiers trained with signal mass hypothesis of 100\GeV. The expected signal distributions extracted with the same classifier are overlaid on the figures. They are normalized assuming $\BtlN=2\times10^{-4}$, a value chosen to ensure that the signal distributions are visible across all channels and \mN hypotheses. The hatched bands show the total uncertainty, including the systematic component. The lower panels show the ratio of the observed data over the background estimate.}
  \label{fig:ResultBDTHigh}
\end{figure*}

To extract the final results, a simultaneous binned maximum likelihood fit with profiling of nuisance parameters is performed over the 2016--2018 data set, separately for each \mN hypothesis in the SR, using the BDT classifier score distribution. The following results have been determined using the CMS statistical analysis tool \textsc{Combine}~\cite{CMS:2024onh}, which is based on the \textsc{RooFit}~\cite{Verkerke:2003ir} and \textsc{RooStats}~\cite{Moneta:2010pm} frameworks. Nuisance parameters are categorized into two groups~\cite{CMS-NOTE-2011-005}: rate uncertainties are implemented with log-normal nuisance parameters, while shape uncertainties are incorporated through variations of the BDT templates with Gaussian-constrained nuisance parameters. A simplified Barlow--Beeston approach~\cite{BARLOW1993219}, with a single nuisance parameter per bin of the distribution, is used to treat the statistical uncertainties due to the limited size of the simulated samples. The uncertainties with the greatest impact for all mass hypotheses are the statistical uncertainties of the background templates in the individual bins. With its conservative approach, the nonprompt-lepton background uncertainty gets reduced by a factor of 2--4, depending on the data-taking era, lepton flavor, and \mN hypothesis. To assess the compatibility between the observed data and the background-only hypothesis, a goodness-of-fit test is performed using the $\chi^2$ test statistic. The expected distribution of the test statistic is derived from 2000 background-only pseudo-experiments generated from the post-fit nuisance parameter configurations. The observed data yield a $p$-value~\cite{pvalue} of 0.3 or above for all signal models, indicating that the data are compatible with the background model. The fit is performed for each mass hypothesis to assess its dependence on the HNL mass. The observed and expected BDT score distributions after the fit in the SR are shown in Figs.~\ref{fig:ResultBDTLow}--\ref{fig:ResultBDTHigh} for various \mN hypotheses. The score distributions in data are consistent with those in the background. The figures show that the BDT classifier trained on the $\ttbar$ process also yields low-score values for the majority of other minor backgrounds. Therefore, a classifier trained on dilepton and lepton$+$jets $\ttbar$ events as the only background sample is sufficient to distinguish other backgrounds from the signal.

\begin{figure*}[htbp]
\centering
  \includegraphics[width=0.48\textwidth]{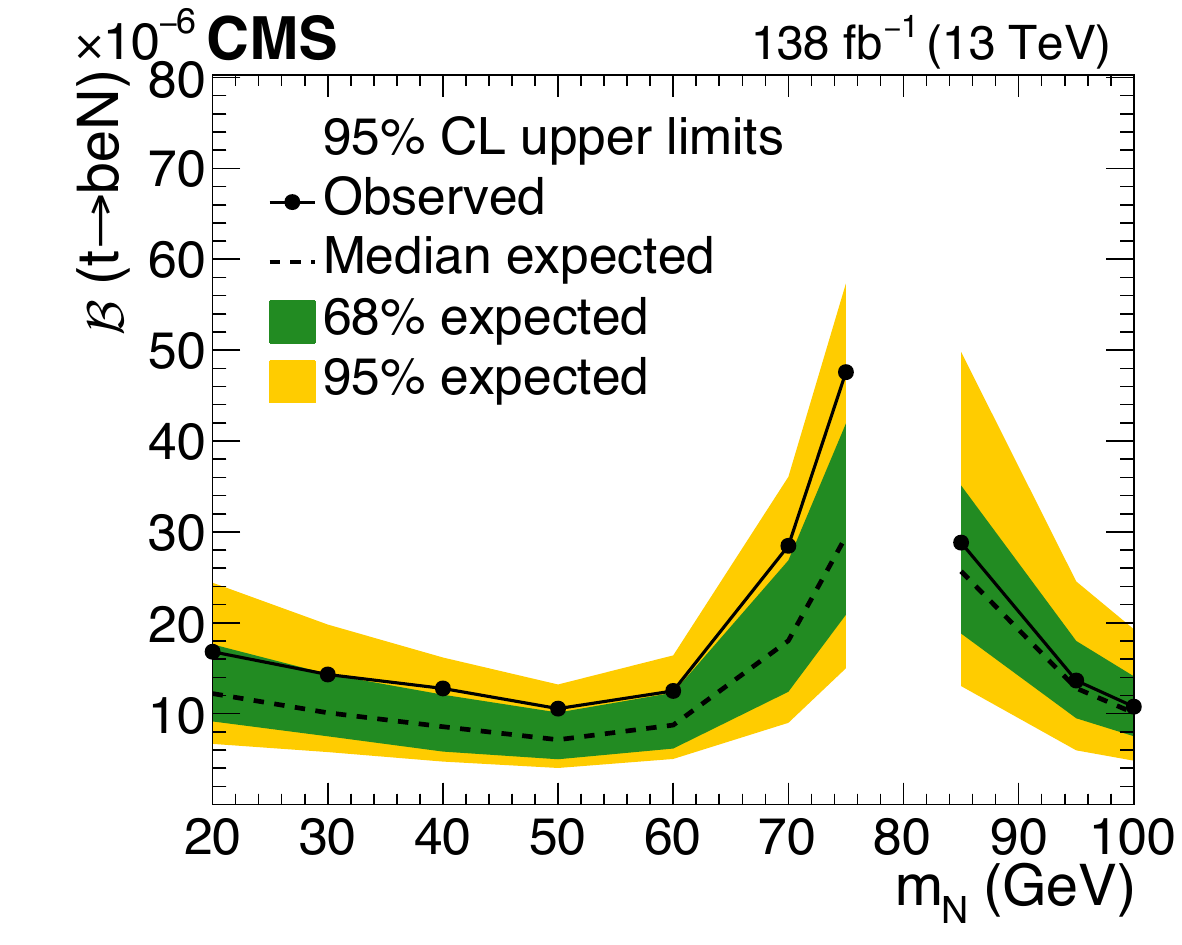}
  \includegraphics[width=0.48\textwidth]{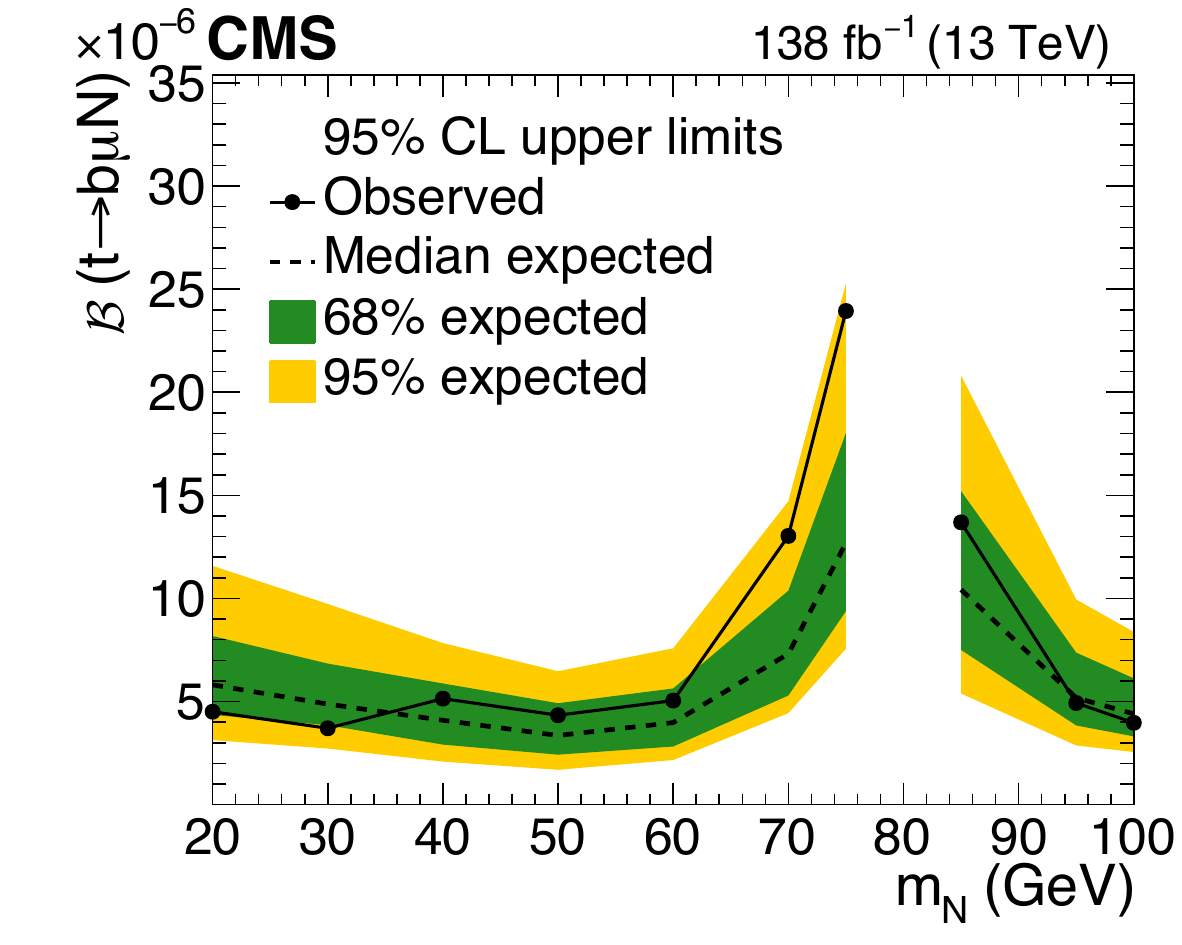}

  \caption{The expected and observed upper limits at 95\% \CL on the branching fraction, \BtlN, as functions of \mN. The left and right figures correspond to the electron and muon neutrino mixing scenarios, respectively.}
  \label{fig:LimitBr}
\end{figure*}

 Based on the classifier score distribution, an upper limit at 95\% confidence level (\CL) on the signal strength is determined using the $\CLs$ criterion~\cite{Junk_1999,Read_2002} in a modified frequentist approach. The test statistic used is the ratio of profiled likelihoods defined according to the LHC-style prescription from Ref.~\cite{CMS:2024onh}. In the conversion to the branching fraction $\BtlN$, the $\ttbar$ production cross section at NNLO+NNLL QCD of 833.9\unit{pb} is used, and the signal decay rates are calculated at LO. The observed and expected limits on $\BtlN$ for the \mN values considered in the search are shown in Fig.~\ref{fig:LimitBr}. The expected upper limits on $\BtlN$ vary from $[7.1\,(3.4)]\times10^{-6}$ to $[3.0\,(1.3)]\times10^{-5}$ in the dielectron (dimuon) channel. The observed upper limits vary from $[11\,(3.7)]\times10^{-6}$ to $[4.8\,(2.4)]\times10^{-5}$ in the dielectron (dimuon) channel, and the results are consistent with the background-only hypothesis. The upper limits increase rapidly when the \mN value approaches the \mW value because of the vanishing lepton signal acceptance and the challenging signal classification. For this reason, we do not present an interpolation of the measured limits in the \mN range $[75,85]\GeV$. The decreasing lepton acceptance is also responsible for the increase in the expected limits shown for $\mN<50\GeV$. The upper limits on the branching fraction for the electron neutrino mixing scenario are larger than those for the muon neutrino mixing scenario by approximately a factor of two. This originates from the lower trigger acceptance and ID efficiency of electrons compared to those of muons.

\begin{figure*}[htbp]
\centering
  \includegraphics[width=0.48\textwidth]{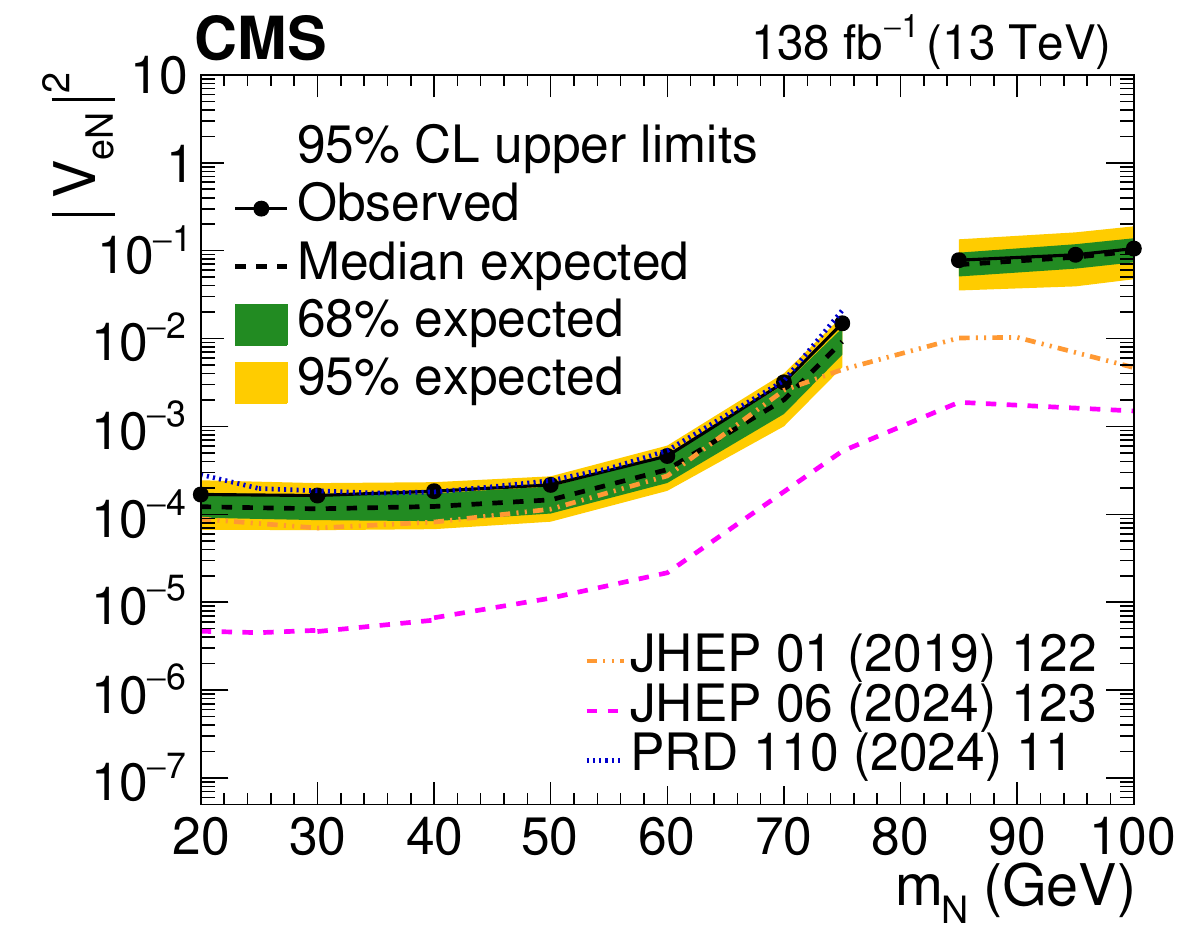}
  \includegraphics[width=0.48\textwidth]{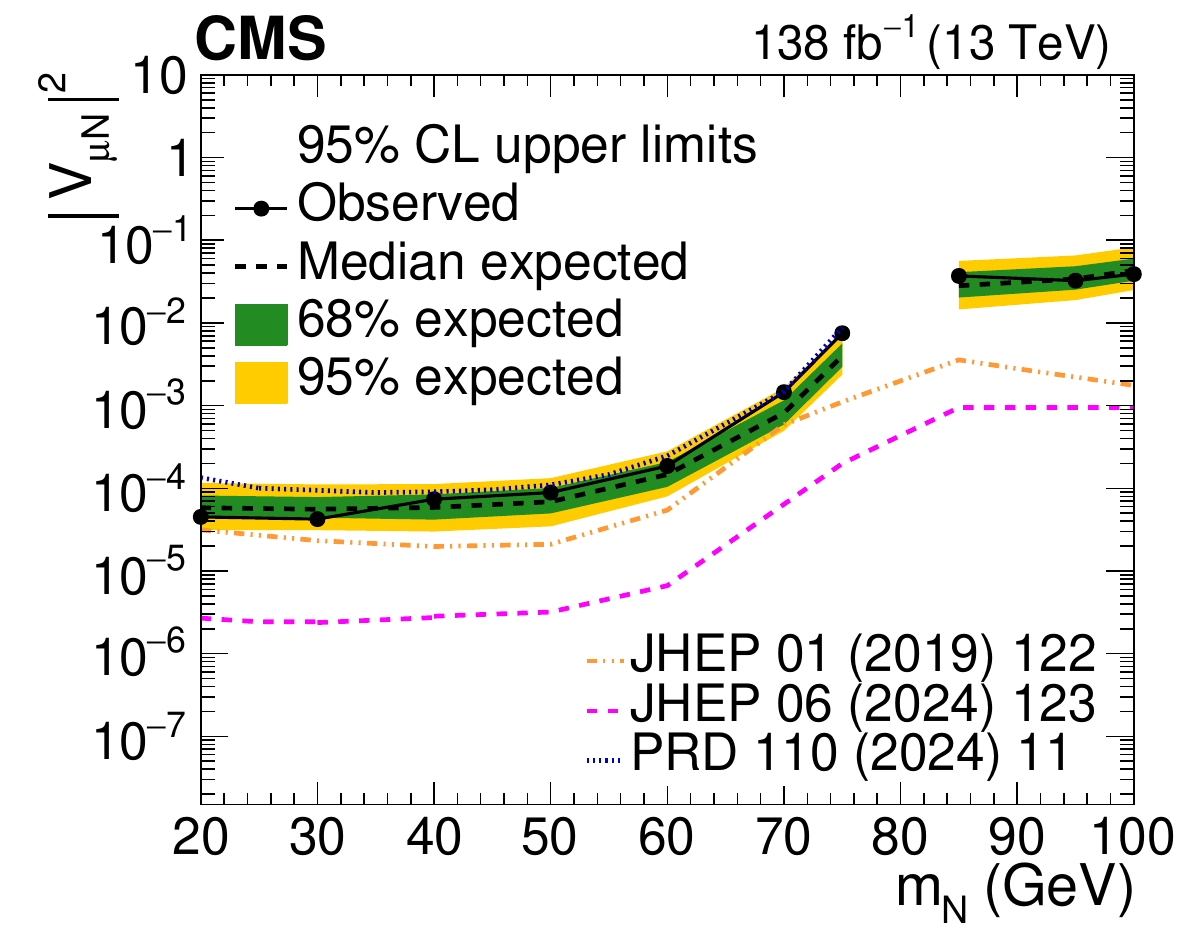}
  \caption{The expected and observed upper limits at 95\% \CL on the model parameters, \VeNSq and \VmNSq, as functions of \mN. The left and right figures correspond to the electron and muon neutrino mixing scenarios, respectively. For comparison, expected limits from the previous CMS searches in charged-current Drell--Yan production~\cite{CMS_HN2l2j,CMS_HN3lv} and the ATLAS \ttbar production search~\cite{ATLAS:TopHNL} are overlaid. The orange dash-dotted line is the SS dilepton channel search in CMS, the magenta dashed line is the trilepton channel search in CMS, and the blue dotted line is the ATLAS search.}
  \label{fig:LimitVlN}
\end{figure*}

For a straightforward comparison with earlier type-I seesaw model search results, the upper limits on the $\BtlN$ are converted into the upper limits on the model parameters, \VeNSq and \VmNSq, as shown in Fig.~\ref{fig:LimitVlN}. The expected upper limits on \VeNSq (\VmNSq) vary between $[12\,(5.6)]\times10^{-5}$ and $[9.9\,(4.3)]\times10^{-2}$. The observed upper limits on \VeNSq (\VmNSq) vary between $[16\,(4.2)]\times10^{-5}$ and $[11\,(3.9)]\times10^{-2}$. The existing results from the direct search channels in CMS and from the ATLAS search for HNLs produced in top quark decays are compared with those from this search. The expected results in this search are comparable to the existing results from the dilepton channel and weaker by approximately a factor of ten than those from the trilepton channel for $\mN < \mW$. At higher mass values, this search becomes less sensitive compared to results from the direct search channels. This is because the signal decay process, $\PQt\to\PQb\Pell\PN$, is no longer mediated by an on-shell $\PW$ boson, \ie, $\PQt\to\PQb\PW$ and $\PW\to\Pell\PN$, meaning the corresponding partial decay width, $\Gamma(\PQt\to\PQb\Pell\PN)$, is significantly reduced. For $20 \leq \mN \leq 75\GeV$, the expected limits on \VeNSq and \VmNSq from this search are stronger by approximately a factor of two than those from the ATLAS search for HNLs produced in top quark decays~\cite{ATLAS:TopHNL}. The size of the difference varies with \mN.

\section{Summary}
\label{Summary}

A search for heavy neutral leptons (HNLs) produced in top quark decays in proton-proton collisions at $\sqrt{s}=13\TeV$, corresponding to an integrated luminosity of 138\fbinv collected with the CMS detector at the LHC, has been presented. The analysis targets HNL masses between 20 and 100\GeV in the type-I seesaw framework. Events are selected by requiring jets and a pair of same-sign, same-flavor leptons. A multivariate classifier is used to enhance the sensitivity of the analysis. No significant excess over the background prediction is observed. Observed (expected) upper limits at 95\% confidence level are set on the branching fraction of the top quark decay into an HNL, a charged lepton, and a bottom quark, varying between $[1.1\,(0.71)]\times10^{-5}$ and $[4.8\,(3.0)]\times10^{-5}$ in the dielectron channel, and between $[3.7\,(3.4)]\times10^{-6}$ and $[2.4\,(1.3)]\times10^{-5}$ in the dimuon channel. The results are also interpreted as upper limits on the squared mixing amplitudes between the HNL and the standard model electron and muon neutrinos, denoted \VeNSq and \VmNSq, respectively. The observed upper limits range from $16\times10^{-5}$ to $11\times10^{-2}$ for \VeNSq and from $4.2\times10^{-5}$ to $3.9\times10^{-2}$ for \VmNSq.

\begin{acknowledgments}
We congratulate our colleagues in the CERN accelerator departments for the excellent performance of the LHC and thank the technical and administrative staffs at CERN and at other CMS institutes for their contributions to the success of the CMS effort. In addition, we gratefully acknowledge the computing centres and personnel of the Worldwide LHC Computing Grid and other centres for delivering so effectively the computing infrastructure essential to our analyses. Finally, we acknowledge the enduring support for the construction and operation of the LHC, the CMS detector, and the supporting computing infrastructure provided by the following funding agencies: SC (Armenia), BMFWF and FWF (Austria); FNRS and FWO (Belgium); CNPq, CAPES, FAPERJ, FAPERGS, and FAPESP (Brazil); MES and BNSF (Bulgaria); CERN; CAS, MoST, and NSFC (China); MINCIENCIAS (Colombia); MSES and CSF (Croatia); RIF (Cyprus); SENESCYT (Ecuador); ERC PRG and PSG, TARISTU24-TK10 and MoER TK202 (Estonia); Academy of Finland, MEC, and HIP (Finland); CEA and CNRS/IN2P3 (France); SRNSF (Georgia); BMFTR, DFG, and HGF (Germany); GSRI (Greece); MATE and NKFIH (Hungary); DAE and DST (India); IPM (Iran); SFI (Ireland); INFN (Italy); MSIT and NRF (Republic of Korea); MES (Latvia); LMTLT (Lithuania); MOE and UM (Malaysia); BUAP, CINVESTAV, CONACYT, LNS, SEP, and UASLP-FAI (Mexico); MOS (Montenegro); MBIE (New Zealand); PAEC (Pakistan); MSHE, NSC, and NAWA (Poland); FCT (Portugal);  MESTD (Serbia); MICIU/AEI and PCTI (Spain); MOSTR (Sri Lanka); Swiss Funding Agencies (Switzerland); NSTC (Taiwan); MHESI (Thailand); TUBITAK and TENMAK (T\"{u}rkiye); NASU (Ukraine); STFC (United Kingdom); DOE and NSF (USA).

\begin{sloppypar}
\setlength\emergencystretch{\hsize}
\hyphenation{Rachada-pisek} Individuals have received support from the Marie-Curie programme and the European Research Council and Horizon 2020 Grant, contract Nos.\ 675440, 724704, 752730, 758316, 765710, 824093, 101115353, 101002207, 101001205, and COST Action CA16108 (European Union); the Leventis Foundation; the Alfred P.\ Sloan Foundation; the Alexander von Humboldt Foundation; the Science Committee, project no. 22rl-037 (Armenia); the Fonds pour la Formation \`a la Recherche dans l'Industrie et dans l'Agriculture (FRIA) and Fonds voor Wetenschappelijk Onderzoek contract No. 1228724N (Belgium); the Beijing Municipal Science \& Technology Commission, No. Z191100007219010, the Fundamental Research Funds for the Central Universities, the Ministry of Science and Technology of China under Grant No. 2023YFA1605804, the Natural Science Foundation of China under Grant No. 12535004, and USTC Research Funds of the Double First-Class Initiative No.\ YD2030002017 (China); the Ministry of Education, Youth and Sports (MEYS) of the Czech Republic; the Shota Rustaveli National Science Foundation (Georgia); the Deutsche Forschungsgemeinschaft (DFG), among others, under Germany's Excellence Strategy -- EXC 2121 ``Quantum Universe" -- 390833306, and under project number 400140256 - GRK2497; the Hellenic Foundation for Research and Innovation (HFRI), Project Number 2288 (Greece); the Hungarian Academy of Sciences, the New National Excellence Program - \'UNKP, the NKFIH research grants K 131991, K 138136, K 143460, K 143477, K 147557, K 146913, K 146914, K 147048, TKP2021-NKTA-64, 2026-4.1.1-MISSZI\'{O}K-2026-00005 and 2025-1.1.5-NEMZ\_KI-2025-00004, and MATE KKP and KKPCs Research Excellence and Flagship Research Groups grants (Hungary); the Council of Science and Industrial Research, India; ICSC -- National Research Centre for High Performance Computing, Big Data and Quantum Computing, FAIR -- Future Artificial Intelligence Research, and CUP I53D23001070006 (Mission 4 Component 1), funded by the NextGenerationEU program, the Italian Ministry of University and Research (MUR) under Bando PRIN 2022 -- CUP I53C24002390006, PRIN PRIMULA 2022RBYK7T (Italy); the Latvian Council of Science; the Ministry of Science and Higher Education, project no. 2022/WK/14, and the National Science Centre, contracts Opus 2021/41/B/ST2/01369, 2021/43/B/ST2/01552, 2023/49/B/ST2/03273, and the NAWA contract BPN/PPO/2021/1/00011 (Poland); the Funda\c{c}\~ao para a Ci\^encia e a Tecnologia (Portugal); the National Priorities Research Program by Qatar National Research Fund; MICIU/AEI/10.13039/501100011033, ERDF/EU, ``European Union NextGenerationEU/PRTR", projects PID2022-142604OB-C21, PID2022-139519OB-C21, PID2023-147706NB-I00, PID2023-148896NB-I00, PID2023-146983NB-I00, PID2023-147115NB-I00, PID2023-148418NB-C41, PID2023-148418NB-C42, PID2023-148418NB-C43, PID2023-148418NB-C44, PID2024-158190NB-C22, RYC2021-033305-I, RYC2024-048719-I, CNS2023-144781, CNS2024-154769 and Plan de Ciencia, Tecnolog{\'i}a e Innovaci{\'o}n de Asturias, Spain; the Chulalongkorn Academic into Its 2nd Century Project Advancement Project, the National Science, Research and Innovation Fund program IND\_FF\_68\_369\_2300\_097, and the Program Management Unit for Human Resources \& Institutional Development, Research and Innovation, grant B39G680009 (Thailand); the Eric \& Wendy Schmidt Fund for Strategic Innovation through the CERN Next Generation Triggers project under grant agreement number SIF-2023-004; the Kavli Foundation; the Nvidia Corporation; the SuperMicro Corporation; the Welch Foundation, contract C-1845; and the Weston Havens Foundation (USA).
\end{sloppypar}
\end{acknowledgments}\section*{Data availability} Release and preservation of data used by the CMS Collaboration as the basis for publications is guided by the  \href{https://doi.org/10.7483/OPENDATA.CMS.1BNU.8V1W}{CMS data preservation, re-use and open access policy}.

\bibliography{auto_generated} 

@ARTICLE{PDG2024,
	AUTHOR=	"{Particle Data Group}",
	TITLE=	"Review of particle physics",
	DOI=	"10.1142/s0217751x26300115",
	JOURNAL=	"Int. J. Mod. Phys. A",
	VOLUME=	"41",
	PAGES=	"2630011",
	YEAR=	"2026",
}

@ARTICLE{Elbers:2025vlz,
	AUTHOR=	"Elbers, W. and others",
	TITLE=	"Constraints on neutrino physics from {DESI DR2 BAO} and {DR1} full shape",
	EPRINT=	"2503.14744",
	ARCHIVEPREFIX=	"arXiv",
	PRIMARYCLASS=	"astro-ph.CO",
	REPORTNUMBER=	"FERMILAB-PUB-25-0168-PPD",
	DOI=	"10.1103/w9pk-xsk7",
	JOURNAL=	"Phys. Rev. D",
	VOLUME=	"112",
	PAGES=	"083513",
	YEAR=	"2025",
}

@ARTICLE{Minkowski_T1Seesaw,
	AUTHOR=	"Minkowski, Peter",
	TITLE=	"$\mu \to e\gamma$ at a rate of one out of $10^{9}$ muon decays?",
	REPORTNUMBER=	"Print-77-0182 (BERN)",
	DOI=	"10.1016/0370-2693(77)90435-X",
	JOURNAL=	"Phys. Lett. B",
	VOLUME=	"67",
	YEAR=	"1977",
	PAGES=	"421",
}

@INPROCEEDINGS{GellMann_T1Seesaw,
	AUTHOR=	"Gell-Mann, Murray and Ramond, Pierre and Slansky, Richard",
	TITLE=	"Complex spinors and unified theories",
	BOOKTITLE=	"Supergravity: Proceedings of the Supergravity Workshop at Stony Brook",
	PUBLISHER=	"North-Holland",
	EPRINT=	"1306.4669",
	ARCHIVEPREFIX=	"arXiv",
	PRIMARYCLASS=	"hep-th",
	REPORTNUMBER=	"PRINT-80-0576",
	PAGES=	"315",
	YEAR=	"1979",
}

@ARTICLE{Yanagida_T1Seesaw,
	AUTHOR=	"Yanagida, Tsutomu",
	TITLE=	"Horizontal symmetry and masses of neutrinos",
	REPORTNUMBER=	"TU-80-208",
	DOI=	"10.1143/PTP.64.1103",
	JOURNAL=	"Prog. Theor. Phys.",
	VOLUME=	"64",
	PAGES=	"1103",
	YEAR=	"1980",
}

@ARTICLE{Mohapatra_T1Seesaw,
	AUTHOR=	"Mohapatra, Rabindra N. and Senjanovi{\'c}, Goran",
	TITLE=	"Neutrino mass and spontaneous parity nonconservation",
	REPORTNUMBER=	"MDDP-TR-80-060, MDDP-PP-80-105, CCNY-HEP-79-10",
	DOI=	"10.1103/PhysRevLett.44.912",
	JOURNAL=	"Phys. Rev. Lett.",
	VOLUME=	"44",
	PAGES=	"912",
	YEAR=	"1980",
}

@ARTICLE{HNLSummary,
	AUTHOR=	"Abdullahi, Asli M. and others",
	TITLE=	"The present and future status of heavy neutral leptons",
	EPRINT=	"2203.08039",
	ARCHIVEPREFIX=	"arXiv",
	PRIMARYCLASS=	"hep-ph",
	REPORTNUMBER=	"FERMILAB-CONF-22-184-T-V",
	DOI=	"10.1088/1361-6471/ac98f9",
	JOURNAL=	"J. Phys. G",
	VOLUME=	"50",
	PAGES=	"020501",
	YEAR=	"2023",
}

@ARTICLE{Galeazzi:2001py,
	TITLE=	"Limits on the existence of heavy neutrinos in the range 50--1000 {eV} from the study of the ${}^{187}\mathrm{Re}$ beta decay",
	AUTHOR=	"Galeazzi, M. and Fontanelli, F. and Gatti, F. and Vitale, S.",
	JOURNAL=	"Phys. Rev. Lett.",
	VOLUME=	"86",
	PAGES=	"1978",
	NUMPAGES=	"0",
	YEAR=	"2001",
	PUBLISHER=	"American Physical Society",
	DOI=	"10.1103/PhysRevLett.86.1978",
}

@ARTICLE{Hiddemann:1995ce,
	AUTHOR=	"Hiddemann, K. H. and Daniel, H. and Schwentker, O.",
	TITLE=	"Limits on neutrino masses from the tritium beta spectrum",
	DOI=	"10.1088/0954-3899/21/5/008",
	JOURNAL=	"J. Phys. G",
	VOLUME=	"21",
	PAGES=	"639",
	YEAR=	"1995",
}

@ARTICLE{CMS_HN2l2j,
	AUTHOR=	"{CMS Collaboration}",
	TITLE=	"Search for heavy {Majorana} neutrinos in same-sign dilepton channels in proton-proton collisions at $\sqrt{s}=13$ {TeV}",
	EPRINT=	"1806.10905",
	ARCHIVEPREFIX=	"arXiv",
	PRIMARYCLASS=	"hep-ex",
	REPORTNUMBER=	"CMS-EXO-17-028, CERN-EP-2018-159",
	DOI=	"10.1007/JHEP01(2019)122",
	JOURNAL=	"JHEP",
	VOLUME=	"01",
	PAGES=	"122",
	YEAR=	"2019",
}

@ARTICLE{CMS_HN3lv,
	AUTHOR=	"{CMS Collaboration}",
	TITLE=	"Search for heavy neutral leptons in final states with electrons, muons, and hadronically decaying tau leptons in proton-proton collisions at $ \sqrt{s} $ = 13 {TeV}",
	EPRINT=	"2403.00100",
	ARCHIVEPREFIX=	"arXiv",
	PRIMARYCLASS=	"hep-ex",
	REPORTNUMBER=	"CMS-EXO-22-011, CERN-EP-2024-032",
	DOI=	"10.1007/JHEP06(2024)123",
	JOURNAL=	"JHEP",
	VOLUME=	"06",
	PAGES=	"123",
	YEAR=	"2024",
}

@ARTICLE{ATLAS_HN3lv,
	AUTHOR=	"{ATLAS Collaboration}",
	TITLE=	"Search for heavy neutral leptons in decays of {W} bosons produced in 13 {TeV} pp collisions using prompt and displaced signatures with the {ATLAS} detector",
	EPRINT=	"1905.09787",
	ARCHIVEPREFIX=	"arXiv",
	PRIMARYCLASS=	"hep-ex",
	REPORTNUMBER=	"CERN-EP-2019-071",
	DOI=	"10.1007/JHEP10(2019)265",
	JOURNAL=	"JHEP",
	VOLUME=	"10",
	PAGES=	"265",
	YEAR=	"2019",
}

@ARTICLE{CMS:2024bni,
	AUTHOR=	"{CMS Collaboration}",
	TITLE=	"Review of searches for vector-like quarks, vector-like leptons, and heavy neutral leptons in proton-proton collisions at $\sqrt{s}=13$ {TeV} at the {CMS} experiment",
	EPRINT=	"2405.17605",
	ARCHIVEPREFIX=	"arXiv",
	PRIMARYCLASS=	"hep-ex",
	REPORTNUMBER=	"CMS-EXO-23-006, CERN-EP-2024-095",
	DOI=	"10.1016/j.physrep.2024.09.012",
	JOURNAL=	"Phys. Rept.",
	VOLUME=	"1115",
	PAGES=	"570",
	YEAR=	"2025",
}

@ARTICLE{LHCb:2025ymr,
	AUTHOR=	"{LHCb Collaboration}",
	TITLE=	"Search for heavy neutral leptons in {B}-meson decays",
	EPRINT=	"2512.14551",
	ARCHIVEPREFIX=	"arXiv",
	PRIMARYCLASS=	"hep-ex",
	REPORTNUMBER=	"LHCb-PAPER-2025-042, CERN-EP-2025-264",
	DOI=	"10.1007/JHEP03(2026)178",
	JOURNAL=	"JHEP",
	VOLUME=	"03",
	PAGES=	"178",
	YEAR=	"2026",
}

@ARTICLE{BarShalom_TopHN,
	AUTHOR=	"Bar-Shalom, Shaouly and Deshpande, Nilendra G. and Eilam, Gad and Jiang, Jing and Soni, Amarjit",
	TITLE=	"{Majorana} neutrinos and lepton-number-violating signals in top-quark and {W}-boson rare decays",
	EPRINT=	"hep-ph/0608309",
	ARCHIVEPREFIX=	"arXiv",
	REPORTNUMBER=	"BNL-HET-06-9, OITS-784",
	DOI=	"10.1016/j.physletb.2006.10.060",
	JOURNAL=	"Phys. Lett. B",
	VOLUME=	"643",
	PAGES=	"342",
	YEAR=	"2006",
}

@ARTICLE{Si_TopHN,
	AUTHOR=	"Si, Zongguo and Wang, Kai",
	TITLE=	"{GeV} {Majorana} neutrinos in top-quark decay at the {LHC}",
	EPRINT=	"0810.5266",
	ARCHIVEPREFIX=	"arXiv",
	PRIMARYCLASS=	"hep-ph",
	REPORTNUMBER=	"MADPH-08-1524, IPMU-08-0068, CAS-KITPC-ITP-072",
	DOI=	"10.1103/PhysRevD.79.014034",
	JOURNAL=	"Phys. Rev. D",
	VOLUME=	"79",
	PAGES=	"014034",
	YEAR=	"2009",
}

@ARTICLE{Delepine_TopHN,
	AUTHOR=	"Quintero, N. and Lopez Castro, G. and Delepine, D.",
	TITLE=	"Lepton number violation in top quark and neutral {B} meson decays",
	EPRINT=	"1108.6009",
	ARCHIVEPREFIX=	"arXiv",
	PRIMARYCLASS=	"hep-ph",
	DOI=	"10.1103/PhysRevD.84.096011",
	JOURNAL=	"Phys. Rev. D",
	VOLUME=	"84",
	PAGES=	"096011",
	YEAR=	"2011",
	NOTE=	"[Erratum: \doi{10.1103/PhysRevD.86.079905}]",
}

@ARTICLE{Liu_TopHN,
	AUTHOR=	"Liu, Ning and Si, Zong-Guo and Wu, Lei and Zhou, Hang and Zhu, Bin",
	TITLE=	"Top quark as a probe of heavy {Majorana} neutrino at the {LHC} and future colliders",
	EPRINT=	"1910.00749",
	ARCHIVEPREFIX=	"arXiv",
	PRIMARYCLASS=	"hep-ph",
	DOI=	"10.1103/PhysRevD.101.071701",
	JOURNAL=	"Phys. Rev. D",
	VOLUME=	"101",
	PAGES=	"071701",
	YEAR=	"2020",
}

@ARTICLE{Lu_TopHNCPV,
	AUTHOR=	"Lu, Peng-Cheng and Si, Zong-Guo and Wang, Zhe and Yang, Xing-Hua",
	TITLE=	"{CP} violation in top quark decay via heavy {Majorana} neutrinos at the {LHC}",
	EPRINT=	"2110.10463",
	ARCHIVEPREFIX=	"arXiv",
	PRIMARYCLASS=	"hep-ph",
	DOI=	"10.1103/PhysRevD.104.115003",
	JOURNAL=	"Phys. Rev. D",
	VOLUME=	"104",
	PAGES=	"115003",
	YEAR=	"2021",
}

@ARTICLE{ATLAS:TopHNL,
	AUTHOR=	"{ATLAS Collaboration}",
	TITLE=	"Search for heavy right-handed {Majorana} neutrinos in the decay of top quarks produced in proton-proton collisions at $\sqrt{s} = 13$ {TeV} with the {ATLAS} detector",
	EPRINT=	"2408.05000",
	ARCHIVEPREFIX=	"arXiv",
	PRIMARYCLASS=	"hep-ex",
	REPORTNUMBER=	"CERN-EP-2024-154",
	DOI=	"10.1103/PhysRevD.110.112004",
	JOURNAL=	"Phys. Rev. D",
	VOLUME=	"110",
	PAGES=	"112004",
	YEAR=	"2024",
}

@MISC{hepdata,
	HOWPUBLISHED=	"{HEPData} record for this analysis",
	DOI=	"10.17182/hepdata.170072",
	YEAR=	"2026",
}

@ARTICLE{CMS:2008xjf,
	AUTHOR=	"{CMS Collaboration}",
	TITLE=	"The {CMS} experiment at the {CERN} {LHC}",
	DOI=	"10.1088/1748-0221/3/08/S08004",
	JOURNAL=	"JINST",
	VOLUME=	"3",
	PAGES=	"S08004",
	YEAR=	"2008",
}

@ARTICLE{CMS:2023gfb,
	AUTHOR=	"{CMS Collaboration}",
	TITLE=	"Development of the {CMS} detector for the {CERN} {LHC} {Run} 3",
	EPRINT=	"2309.05466",
	ARCHIVEPREFIX=	"arXiv",
	PRIMARYCLASS=	"physics.ins-det",
	REPORTNUMBER=	"CMS-PRF-21-001, CERN-EP-2023-136",
	DOI=	"10.1088/1748-0221/19/05/P05064",
	JOURNAL=	"JINST",
	VOLUME=	"19",
	PAGES=	"P05064",
	YEAR=	"2024",
}

@ARTICLE{CMS:2020cmk,
	AUTHOR=	"{CMS Collaboration}",
	TITLE=	"Performance of the {CMS} Level-1 trigger in proton-proton collisions at $\sqrt{s} = 13$ {TeV}",
	JOURNAL=	"JINST",
	VOLUME=	"15",
	PAGES=	"P10017",
	YEAR=	"2020",
	DOI=	"10.1088/1748-0221/15/10/P10017",
	EPRINT=	"2006.10165",
	ARCHIVEPREFIX=	"arXiv",
	PRIMARYCLASS=	"hep-ex",
	REPORTNUMBER=	"CMS-TRG-17-001, CERN-EP-2020-065",
}

@ARTICLE{CMS:2016ngn,
	AUTHOR=	"{CMS Collaboration}",
	TITLE=	"The {CMS} trigger system",
	JOURNAL=	"JINST",
	VOLUME=	"12",
	PAGES=	"P01020",
	DOI=	"10.1088/1748-0221/12/01/P01020",
	YEAR=	"2017",
	EPRINT=	"1609.02366",
	ARCHIVEPREFIX=	"arXiv",
	PRIMARYCLASS=	"physics.ins-det",
	REPORTNUMBER=	"CMS-TRG-12-001, CERN-EP-2016-160",
	SLACCITATION=	"%%CITATION = ARXIV:1609.02366;%%",
}

@ARTICLE{CMS:2024aqx,
	AUTHOR=	"{CMS Collaboration}",
	TITLE=	"Performance of the {CMS} high-level trigger during {LHC} {Run 2}",
	EPRINT=	"2410.17038",
	ARCHIVEPREFIX=	"arXiv",
	PRIMARYCLASS=	"physics.ins-det",
	REPORTNUMBER=	"CMS-TRG-19-001, CERN-EP-2024-259",
	DOI=	"10.1088/1748-0221/19/11/P11021",
	JOURNAL=	"JINST",
	VOLUME=	"19",
	PAGES=	"P11021",
	YEAR=	"2024",
}

@ARTICLE{CMS:2020uim,
	AUTHOR=	"{CMS Collaboration}",
	TITLE=	"Electron and photon reconstruction and identification with the {CMS} experiment at the {CERN} {LHC}",
	EPRINT=	"2012.06888",
	JOURNAL=	"JINST",
	VOLUME=	"16",
	PAGES=	"P05014",
	YEAR=	"2021",
	ARCHIVEPREFIX=	"arXiv",
	PRIMARYCLASS=	"hep-ex",
	REPORTNUMBER=	"CMS-EGM-17-001, CERN-EP-2020-219",
	DOI=	"10.1088/1748-0221/16/05/P05014",
}

@ARTICLE{CMS:2018rym,
	AUTHOR=	"{CMS Collaboration}",
	TITLE=	"Performance of the {CMS} muon detector and muon reconstruction with proton-proton collisions at $\sqrt{s}=$ 13 {TeV}",
	EPRINT=	"1804.04528",
	ARCHIVEPREFIX=	"arXiv",
	PRIMARYCLASS=	"physics.ins-det",
	REPORTNUMBER=	"CMS-MUO-16-001, CERN-EP-2018-058",
	DOI=	"10.1088/1748-0221/13/06/P06015",
	JOURNAL=	"JINST",
	VOLUME=	"13",
	PAGES=	"P06015",
	YEAR=	"2018",
}

@ARTICLE{CMS:2014pgm,
	AUTHOR=	"{CMS Collaboration}",
	TITLE=	"Description and performance of track and primary-vertex reconstruction with the {CMS} tracker",
	EPRINT=	"1405.6569",
	ARCHIVEPREFIX=	"arXiv",
	PRIMARYCLASS=	"physics.ins-det",
	REPORTNUMBER=	"CMS-TRK-11-001, CERN-PH-EP-2014-070",
	DOI=	"10.1088/1748-0221/9/10/P10009",
	JOURNAL=	"JINST",
	VOLUME=	"9",
	PAGES=	"P10009",
	YEAR=	"2014",
}

@ARTICLE{CMS:2017yfk,
	AUTHOR=	"{CMS Collaboration}",
	TITLE=	"Particle-flow reconstruction and global event description with the {CMS} detector",
	JOURNAL=	"JINST",
	VOLUME=	"12",
	YEAR=	"2017",
	PAGES=	"P10003",
	DOI=	"10.1088/1748-0221/12/10/P10003",
	EPRINT=	"1706.04965",
	ARCHIVEPREFIX=	"arXiv",
	PRIMARYCLASS=	"physics.ins-det",
	REPORTNUMBER=	"CMS-PRF-14-001, CERN-EP-2017-110",
	SLACCITATION=	"%%CITATION = ARXIV:1706.04965;%%",
}

@TECHREPORT{CMS-TDR-15-02,
	AUTHOR=	"{CMS Collaboration}",
	TITLE=	"Technical proposal for the {Phase-II} upgrade of the {Compact Muon Solenoid}",
	URL=	"https://cds.cern.ch/record/2020886",
	TYPE=	"CMS Technical Proposal",
	NUMBER=	"CERN-LHCC-2015-010, CMS-TDR-15-02",
	YEAR=	"2015",
	NOTE=	"\url{https://cds.cern.ch/record/2020886}",
}

@ARTICLE{MiniIso,
	AUTHOR=	"Rehermann, Keith and Tweedie, Brock",
	TITLE=	"Efficient identification of boosted semileptonic top quarks at the {LHC}",
	EPRINT=	"1007.2221",
	ARCHIVEPREFIX=	"arXiv",
	PRIMARYCLASS=	"hep-ph",
	DOI=	"10.1007/JHEP03(2011)059",
	JOURNAL=	"JHEP",
	VOLUME=	"03",
	PAGES=	"059",
	YEAR=	"2011",
}

@ARTICLE{GSF,
	DOI=	"10.1088/0954-3899/31/9/N01",
	YEAR=	"2005",
	VOLUME=	"31",
	PAGES=	"N9",
	AUTHOR=	"Adam, W and Fr{\"u}hwirth, R and Strandlie, A and Todorov, T",
	TITLE=	"Reconstruction of electrons with the {Gaussian}-sum filter in the {CMS} tracker at the {LHC}",
	JOURNAL=	"J. Phys. G",
}

@ARTICLE{Cacciari:2008gp,
	AUTHOR=	"Cacciari, Matteo and Salam, Gavin P. and Soyez, Gregory",
	TITLE=	"The anti-\kt jet clustering algorithm",
	JOURNAL=	"JHEP",
	VOLUME=	"04",
	YEAR=	"2008",
	PAGES=	"063",
	DOI=	"10.1088/1126-6708/2008/04/063",
	EPRINT=	"0802.1189",
	ARCHIVEPREFIX=	"arXiv",
	PRIMARYCLASS=	"hep-ex",
}

@ARTICLE{Cacciari:2011ma,
	AUTHOR=	"Cacciari, Matteo and Salam, Gavin P. and Soyez, Gregory",
	TITLE=	"{FastJet} user manual",
	JOURNAL=	"Eur. Phys. J. C",
	VOLUME=	"72",
	PAGES=	"1896",
	DOI=	"10.1140/epjc/s10052-012-1896-2",
	YEAR=	"2012",
	EPRINT=	"1111.6097",
	ARCHIVEPREFIX=	"arXiv",
	PRIMARYCLASS=	"hep-ph",
	REPORTNUMBER=	"CERN-PH-TH-2011-297",
	SLACCITATION=	"%%CITATION = ARXIV:1111.6097;%%",
}

@ARTICLE{CMS:2016lmd,
	AUTHOR=	"{CMS Collaboration}",
	TITLE=	"Jet energy scale and resolution in the {CMS} experiment in pp collisions at 8 {TeV}",
	JOURNAL=	"JINST",
	VOLUME=	"12",
	YEAR=	"2017",
	PAGES=	"P02014",
	DOI=	"10.1088/1748-0221/12/02/P02014",
	EPRINT=	"1607.03663",
	ARCHIVEPREFIX=	"arXiv",
	PRIMARYCLASS=	"hep-ex",
	REPORTNUMBER=	"CMS-JME-13-004, CERN-PH-EP-2015-305",
	SLACCITATION=	"%%CITATION = ARXIV:1607.03663;%%",
}

@ARTICLE{CMS:2020ebo,
	AUTHOR=	"{CMS Collaboration}",
	TITLE=	"Pileup mitigation at {CMS} in 13 {TeV} data",
	EPRINT=	"2003.00503",
	ARCHIVEPREFIX=	"arXiv",
	PRIMARYCLASS=	"hep-ex",
	REPORTNUMBER=	"CMS-JME-18-001, CERN-EP-2020-017",
	DOI=	"10.1088/1748-0221/15/09/P09018",
	JOURNAL=	"JINST",
	VOLUME=	"15",
	PAGES=	"P09018",
	YEAR=	"2020",
}

@TECHREPORT{CMS-DP-2020-020,
	AUTHOR=	"{CMS Collaboration}",
	TITLE=	"Performance of the pile up jet identification in {CMS} for {Run 2}",
	URL=	"https://cds.cern.ch/record/2715906",
	TYPE=	"CMS Detector Performance Note",
	NUMBER=	"CMS-DP-2020-020",
	YEAR=	"2020",
	NOTE=	"\url{https://cds.cern.ch/record/2715906}",
}

@ARTICLE{Bols:2020bkb,
	AUTHOR=	"Bols, Emil and Kieseler, Jan and Verzetti, Mauro and Stoye, Markus and Stakia, Anna",
	TITLE=	"Jet flavour classification using \textsc{DeepJet}",
	EPRINT=	"2008.10519",
	ARCHIVEPREFIX=	"arXiv",
	PRIMARYCLASS=	"hep-ex",
	DOI=	"10.1088/1748-0221/15/12/P12012",
	JOURNAL=	"JINST",
	VOLUME=	"15",
	PAGES=	"P12012",
	YEAR=	"2020",
}

@TECHREPORT{CMS-DP-2018-058,
	AUTHOR=	"{CMS Collaboration}",
	TITLE=	"Performance of the \textsc{DeepJet} b tagging algorithm using 41.9/fb of data from proton-proton collisions at 13 {TeV} with {Phase 1} {CMS} detector",
	URL=	"https://cds.cern.ch/record/2646773",
	TYPE=	"CMS Detector Performance Note",
	NUMBER=	"CMS-DP-2018-058",
	YEAR=	"2018",
	NOTE=	"\url{https://cds.cern.ch/record/2646773}",
}

@ARTICLE{BTV-16-002,
	AUTHOR=	"{CMS Collaboration}",
	TITLE=	"Identification of heavy-flavour jets with the {CMS} detector in pp collisions at 13 {TeV}",
	JOURNAL=	"JINST",
	VOLUME=	"13",
	PAGES=	"P05011",
	DOI=	"10.1088/1748-0221/13/05/P05011",
	YEAR=	"2018",
	EPRINT=	"1712.07158",
	ARCHIVEPREFIX=	"arXiv",
	PRIMARYCLASS=	"physics.ins-det",
	REPORTNUMBER=	"CMS-BTV-16-002, CERN-EP-2017-326",
	SLACCITATION=	"%%CITATION = ARXIV:1712.07158;%%",
}

@ARTICLE{CMS:2019ctu,
	AUTHOR=	"{CMS Collaboration}",
	TITLE=	"Performance of missing transverse momentum reconstruction in proton-proton collisions at $\sqrt{s} = 13$\,{TeV} using the {CMS} detector",
	JOURNAL=	"JINST",
	VOLUME=	"14",
	YEAR=	"2019",
	PAGES=	"P07004",
	DOI=	"10.1088/1748-0221/14/07/P07004",
	EPRINT=	"1903.06078",
	ARCHIVEPREFIX=	"arXiv",
	PRIMARYCLASS=	"hep-ex",
	REPORTNUMBER=	"CMS-JME-17-001, CERN-EP-2018-335",
	SLACCITATION=	"%%CITATION = ARXIV:1903.06078;%%",
}

@ARTICLE{CMS-LUM-17-003,
	AUTHOR=	"{CMS Collaboration}",
	TITLE=	"Precision luminosity measurement in proton-proton collisions at $\sqrt{s} =$ 13 {TeV} in 2015 and 2016 at {CMS}",
	EPRINT=	"2104.01927",
	ARCHIVEPREFIX=	"arXiv",
	PRIMARYCLASS=	"hep-ex",
	REPORTNUMBER=	"CMS-LUM-17-003, CERN-EP-2021-033",
	YEAR=	"2021",
	DOI=	"10.1140/epjc/s10052-021-09538-2",
	JOURNAL=	"Eur. Phys. J. C",
	VOLUME=	"81",
	PAGES=	"800",
}

@TECHREPORT{CMS-PAS-LUM-17-004,
	TITLE=	"{CMS} luminosity measurement for the 2017 data-taking period at $\sqrt{s}$ = 13 {TeV}",
	AUTHOR=	"{CMS Collaboration}",
	URL=	"https://cds.cern.ch/record/2621960/",
	TYPE=	"CMS Physics Analysis Summary",
	NUMBER=	"CMS-PAS-LUM-17-004",
	YEAR=	"2018",
	NOTE=	"\url{https://cds.cern.ch/record/2621960/}",
}

@TECHREPORT{CMS-PAS-LUM-18-002,
	TITLE=	"{CMS} luminosity measurement for the 2018 data-taking period at $\sqrt{s}$ = 13 {TeV}",
	AUTHOR=	"{CMS Collaboration}",
	URL=	"https://cds.cern.ch/record/2676164/",
	TYPE=	"CMS Physics Analysis Summary",
	NUMBER=	"CMS-PAS-LUM-18-002",
	YEAR=	"2019",
	NOTE=	"\url{https://cds.cern.ch/record/2676164/}",
}

@ARTICLE{MG5_aMCatNLO,
	AUTHOR=	"Alwall, J. and Frederix, R. and Frixione, S. and Hirschi, V. and Maltoni, F. and Mattelaer, O. and Shao, H. -S. and Stelzer, T. and Torrielli, P. and Zaro, M.",
	TITLE=	"The automated computation of tree-level and next-to-leading order differential cross sections, and their matching to parton shower simulations",
	EPRINT=	"1405.0301",
	ARCHIVEPREFIX=	"arXiv",
	PRIMARYCLASS=	"hep-ph",
	REPORTNUMBER=	"CERN-PH-TH-2014-064, CP3-14-18, LPN14-066, MCNET-14-09, ZU-TH-14-14",
	DOI=	"10.1007/JHEP07(2014)079",
	JOURNAL=	"JHEP",
	VOLUME=	"07",
	PAGES=	"079",
	YEAR=	"2014",
}

@ARTICLE{UFO1,
	AUTHOR=	"Alva, Daniel and Han, Tao and Ruiz, Richard",
	TITLE=	"Heavy {Majorana} neutrinos from {W}$\gamma$ fusion at hadron colliders",
	EPRINT=	"1411.7305",
	ARCHIVEPREFIX=	"arXiv",
	PRIMARYCLASS=	"hep-ph",
	REPORTNUMBER=	"PITT-PACC-1407",
	DOI=	"10.1007/JHEP02(2015)072",
	JOURNAL=	"JHEP",
	VOLUME=	"02",
	PAGES=	"072",
	YEAR=	"2015",
}

@ARTICLE{UFO2,
	AUTHOR=	"Degrande, Celine and Mattelaer, Olivier and Ruiz, Richard and Turner, Jessica",
	TITLE=	"Fully-automated precision predictions for heavy neutrino production mechanisms at hadron colliders",
	EPRINT=	"1602.06957",
	ARCHIVEPREFIX=	"arXiv",
	PRIMARYCLASS=	"hep-ph",
	REPORTNUMBER=	"IPPP-16-13, MCNET-16-05",
	DOI=	"10.1103/PhysRevD.94.053002",
	JOURNAL=	"Phys. Rev. D",
	VOLUME=	"94",
	PAGES=	"053002",
	YEAR=	"2016",
}

@ARTICLE{FeynRules,
	AUTHOR=	"Alloul, Adam and Christensen, Neil D. and Degrande, C{\'e}line and Duhr, Claude and Fuks, Benjamin",
	TITLE=	"\textsc{FeynRules} 2.0 --- a complete toolbox for tree-level phenomenology",
	EPRINT=	"1310.1921",
	ARCHIVEPREFIX=	"arXiv",
	PRIMARYCLASS=	"hep-ph",
	REPORTNUMBER=	"CERN-PH-TH-2013-239, MCNET-13-14, IPPP-13-71, DCPT-13-142, PITT-PACC-1308",
	DOI=	"10.1016/j.cpc.2014.04.012",
	JOURNAL=	"Comput. Phys. Commun.",
	VOLUME=	"185",
	PAGES=	"2250",
	YEAR=	"2014",
}

@ARTICLE{tt1,
	AUTHOR=	"Czakon, Michal and Mitov, Alexander",
	TITLE=	"{\TOPpp}: a program for the calculation of the top-pair cross-section at hadron colliders",
	EPRINT=	"1112.5675",
	ARCHIVEPREFIX=	"arXiv",
	PRIMARYCLASS=	"hep-ph",
	REPORTNUMBER=	"CERN-PH-TH-2011-303, TTK-11-58",
	DOI=	"10.1016/j.cpc.2014.06.021",
	JOURNAL=	"Comput. Phys. Commun.",
	VOLUME=	"185",
	PAGES=	"2930",
	YEAR=	"2014",
}

@INPROCEEDINGS{MLM,
	AUTHOR=	"Hoeche, Stefan and Krauss, Frank and Lavesson, Nils and Lonnblad, Leif and Mangano, Michelangelo and Schalicke, Andreas and Schumann, Steffen",
	TITLE=	"Matching parton showers and matrix elements",
	BOOKTITLE=	"Proc. {HERA} and the {LHC}: A Workshop on the Implications of {HERA} for {LHC} Physics",
	EPRINT=	"hep-ph/0602031",
	ARCHIVEPREFIX=	"arXiv",
	DOI=	"10.5170/CERN-2005-014.288",
	PAGES=	"288",
	YEAR=	"2005",
}

@ARTICLE{Madspin,
	AUTHOR=	"Artoisenet, Pierre and Frederix, Rikkert and Mattelaer, Olivier and Rietkerk, Robbert",
	TITLE=	"Automatic spin-entangled decays of heavy resonances in {Monte Carlo} simulations",
	EPRINT=	"1212.3460",
	ARCHIVEPREFIX=	"arXiv",
	PRIMARYCLASS=	"hep-ph",
	REPORTNUMBER=	"NIKHEF-2012-021, CERN-PH-TH-2012-329",
	DOI=	"10.1007/JHEP03(2013)015",
	JOURNAL=	"JHEP",
	VOLUME=	"03",
	PAGES=	"015",
	YEAR=	"2013",
}

@ARTICLE{Powheg1,
	AUTHOR=	"Nason, Paolo",
	TITLE=	"A new method for combining {NLO QCD} with shower {Monte Carlo} algorithms",
	EPRINT=	"hep-ph/0409146",
	ARCHIVEPREFIX=	"arXiv",
	REPORTNUMBER=	"BICOCCA-FT-04-11",
	DOI=	"10.1088/1126-6708/2004/11/040",
	JOURNAL=	"JHEP",
	VOLUME=	"11",
	PAGES=	"040",
	YEAR=	"2004",
}

@ARTICLE{Powheg2,
	AUTHOR=	"Frixione, Stefano and Nason, Paolo and Oleari, Carlo",
	TITLE=	"Matching {NLO QCD} computations with parton shower simulations: the {POWHEG} method",
	EPRINT=	"0709.2092",
	ARCHIVEPREFIX=	"arXiv",
	PRIMARYCLASS=	"hep-ph",
	REPORTNUMBER=	"BICOCCA-FT-07-9, GEF-TH-21-2007",
	DOI=	"10.1088/1126-6708/2007/11/070",
	JOURNAL=	"JHEP",
	VOLUME=	"11",
	PAGES=	"070",
	YEAR=	"2007",
}

@ARTICLE{Powheg3,
	AUTHOR=	"Alioli, Simone and Nason, Paolo and Oleari, Carlo and Re, Emanuele",
	TITLE=	"A general framework for implementing {NLO} calculations in shower {Monte Carlo} programs: the {POWHEG BOX}",
	EPRINT=	"1002.2581",
	ARCHIVEPREFIX=	"arXiv",
	PRIMARYCLASS=	"hep-ph",
	REPORTNUMBER=	"DESY-10-018, SFB-CPP-10-22, IPPP-10-11, DCPT-10-22",
	DOI=	"10.1007/JHEP06(2010)043",
	JOURNAL=	"JHEP",
	VOLUME=	"06",
	PAGES=	"043",
	YEAR=	"2010",
}

@ARTICLE{PowhegVV,
	AUTHOR=	"Melia, Tom and Nason, Paolo and Rontsch, Raoul and Zanderighi, Giulia",
	TITLE=	"{W}$^+${W}$^-$, {WZ} and {ZZ} production in the {POWHEG BOX}",
	EPRINT=	"1107.5051",
	ARCHIVEPREFIX=	"arXiv",
	PRIMARYCLASS=	"hep-ph",
	DOI=	"10.1007/JHEP11(2011)078",
	JOURNAL=	"JHEP",
	VOLUME=	"11",
	PAGES=	"078",
	YEAR=	"2011",
}

@ARTICLE{PowhegttH,
	AUTHOR=	"Hartanto, Heribertus B. and Jager, Barbara and Reina, Laura and Wackeroth, Doreen",
	TITLE=	"Higgs boson production in association with top quarks in the {POWHEG BOX}",
	EPRINT=	"1501.04498",
	ARCHIVEPREFIX=	"arXiv",
	PRIMARYCLASS=	"hep-ph",
	DOI=	"10.1103/PhysRevD.91.094003",
	JOURNAL=	"Phys. Rev. D",
	VOLUME=	"91",
	PAGES=	"094003",
	YEAR=	"2015",
}

@ARTICLE{PowhegggH,
	AUTHOR=	"Bagnaschi, E. and Degrassi, G. and Slavich, P. and Vicini, A.",
	TITLE=	"Higgs production via gluon fusion in the {POWHEG} approach in the {SM} and in the {MSSM}",
	EPRINT=	"1111.2854",
	ARCHIVEPREFIX=	"arXiv",
	PRIMARYCLASS=	"hep-ph",
	REPORTNUMBER=	"RM3-TH-11-5, IFUM-990-FT",
	DOI=	"10.1007/JHEP02(2012)088",
	JOURNAL=	"JHEP",
	VOLUME=	"02",
	PAGES=	"088",
	YEAR=	"2012",
}

@ARTICLE{PowhegVBFH,
	AUTHOR=	"Nason, Paolo and Oleari, Carlo",
	TITLE=	"{NLO} {Higgs} boson production via vector-boson fusion matched with shower in {POWHEG}",
	EPRINT=	"0911.5299",
	ARCHIVEPREFIX=	"arXiv",
	PRIMARYCLASS=	"hep-ph",
	DOI=	"10.1007/JHEP02(2010)037",
	JOURNAL=	"JHEP",
	VOLUME=	"02",
	PAGES=	"037",
	YEAR=	"2010",
}

@ARTICLE{PowhegtW,
	AUTHOR=	"Re, Emanuele",
	TITLE=	"Single-top {W}t-channel production matched with parton showers using the {POWHEG} method",
	EPRINT=	"1009.2450",
	ARCHIVEPREFIX=	"arXiv",
	PRIMARYCLASS=	"hep-ph",
	REPORTNUMBER=	"IPPP-10-74, DCPT-10-148",
	DOI=	"10.1140/epjc/s10052-011-1547-z",
	JOURNAL=	"Eur. Phys. J. C",
	VOLUME=	"71",
	PAGES=	"1547",
	YEAR=	"2011",
}

@ARTICLE{Powhegtt,
	AUTHOR=	"Frixione, Stefano and Nason, Paolo and Ridolfi, Giovanni",
	TITLE=	"A positive-weight next-to-leading-order {Monte Carlo} for heavy flavour hadroproduction",
	EPRINT=	"0707.3088",
	ARCHIVEPREFIX=	"arXiv",
	PRIMARYCLASS=	"hep-ph",
	REPORTNUMBER=	"BICOCCA-FT-07-12, GEF-TH-19-2007",
	DOI=	"10.1088/1126-6708/2007/09/126",
	JOURNAL=	"JHEP",
	VOLUME=	"09",
	PAGES=	"126",
	YEAR=	"2007",
}

@ARTICLE{ZZNNLOCalc,
	AUTHOR=	"Cascioli, F. and Gehrmann, T. and Grazzini, M. and Kallweit, S. and Maierh{\"o}fer, P. and von Manteuffel, A. and Pozzorini, S. and Rathlev, D. and Tancredi, L. and Weihs, E.",
	TITLE=	"{ZZ} production at hadron colliders in {NNLO QCD}",
	EPRINT=	"1405.2219",
	ARCHIVEPREFIX=	"arXiv",
	PRIMARYCLASS=	"hep-ph",
	REPORTNUMBER=	"ZU-TH-19-14, MITP-14-033",
	DOI=	"10.1016/j.physletb.2014.06.056",
	JOURNAL=	"Phys. Lett. B",
	VOLUME=	"735",
	PAGES=	"311",
	YEAR=	"2014",
}

@ARTICLE{JHUGen,
	AUTHOR=	"Gao, Yanyan and Gritsan, Andrei V. and Guo, Zijin and Melnikov, Kirill and Schulze, Markus and Tran, Nhan V.",
	TITLE=	"Spin determination of single-produced resonances at hadron colliders",
	EPRINT=	"1001.3396",
	ARCHIVEPREFIX=	"arXiv",
	PRIMARYCLASS=	"hep-ph",
	REPORTNUMBER=	"FERMILAB-PUB-10-011-E",
	DOI=	"10.1103/PhysRevD.81.075022",
	JOURNAL=	"Phys. Rev. D",
	VOLUME=	"81",
	PAGES=	"075022",
	YEAR=	"2010",
}

@ARTICLE{Pythia8,
	AUTHOR=	"Sj{\"o}strand, Torbj{\"o}rn and Ask, Stefan and Christiansen, Jesper R. and Corke, Richard and Desai, Nishita and Ilten, Philip and Mrenna, Stephen and Prestel, Stefan and Rasmussen, Christine O. and Skands, Peter Z.",
	TITLE=	"An introduction to {PYTHIA} 8.2",
	EPRINT=	"1410.3012",
	ARCHIVEPREFIX=	"arXiv",
	PRIMARYCLASS=	"hep-ph",
	REPORTNUMBER=	"LU-TP-14-36, MCNET-14-22, CERN-PH-TH-2014-190, FERMILAB-PUB-14-316-CD, DESY-14-178, SLAC-PUB-16122",
	DOI=	"10.1016/j.cpc.2015.01.024",
	JOURNAL=	"Comput. Phys. Commun.",
	VOLUME=	"191",
	PAGES=	"159",
	YEAR=	"2015",
}

@ARTICLE{CPTunes,
	AUTHOR=	"{CMS Collaboration}",
	TITLE=	"Extraction and validation of a new set of {CMS} {PYTHIA8} tunes from underlying-event measurements",
	EPRINT=	"1903.12179",
	ARCHIVEPREFIX=	"arXiv",
	PRIMARYCLASS=	"hep-ex",
	REPORTNUMBER=	"CMS-GEN-17-001, CERN-EP-2019-007",
	DOI=	"10.1140/epjc/s10052-019-7499-4",
	JOURNAL=	"Eur. Phys. J. C",
	VOLUME=	"80",
	PAGES=	"4",
	YEAR=	"2020",
}

@ARTICLE{NNPDF31,
	AUTHOR=	"{NNPDF Collaboration}",
	COLLABORATION=	"NNPDF",
	TITLE=	"Parton distributions from high-precision collider data",
	EPRINT=	"1706.00428",
	ARCHIVEPREFIX=	"arXiv",
	PRIMARYCLASS=	"hep-ph",
	REPORTNUMBER=	"EDINBURGH-2017-08, NIKHEF-2017-006, OUTP-17-04P, TIF-UNIMI-2017-3, CAVENDISH-HEP-17-06, CERN-TH-2017-077, Edinburgh 2017/08, Nikhef/2017-006, OUTP-17-04P,TIF-UNIMI-2017-3",
	DOI=	"10.1140/epjc/s10052-017-5199-5",
	JOURNAL=	"Eur. Phys. J. C",
	VOLUME=	"77",
	PAGES=	"663",
	YEAR=	"2017",
}

@ARTICLE{GEANT4,
	AUTHOR=	"Agostinelli, S. and others",
	COLLABORATION=	"GEANT4",
	TITLE=	"{GEANT4}---a simulation toolkit",
	REPORTNUMBER=	"SLAC-PUB-9350, FERMILAB-PUB-03-339, CERN-IT-2002-003",
	DOI=	"10.1016/S0168-9002(03)01368-8",
	JOURNAL=	"Nucl. Instrum. Meth. A",
	VOLUME=	"506",
	PAGES=	"250",
	YEAR=	"2003",
}

@ARTICLE{CMSFRMethod,
	AUTHOR=	"{CMS Collaboration}",
	TITLE=	"Search for new physics with same-sign isolated dilepton events with jets and missing transverse energy at the {LHC}",
	EPRINT=	"1104.3168",
	ARCHIVEPREFIX=	"arXiv",
	PRIMARYCLASS=	"hep-ex",
	REPORTNUMBER=	"CERN-PH-EP-2011-033, CMS-SUS-10-004",
	DOI=	"10.1007/JHEP06(2011)077",
	JOURNAL=	"JHEP",
	VOLUME=	"06",
	PAGES=	"077",
	YEAR=	"2011",
}

@ARTICLE{CMS:2011bet,
	AUTHOR=	"{CMS Collaboration}",
	TITLE=	"Measurement of the lepton charge asymmetry in inclusive {W} production in pp collisions at $\sqrt{s} = 7$ {TeV}",
	EPRINT=	"1103.3470",
	ARCHIVEPREFIX=	"arXiv",
	PRIMARYCLASS=	"hep-ex",
	REPORTNUMBER=	"CERN-PH-EP-2011-024, CMS-EWK-10-006",
	DOI=	"10.1007/JHEP04(2011)050",
	JOURNAL=	"JHEP",
	VOLUME=	"04",
	PAGES=	"050",
	YEAR=	"2011",
}

@ARTICLE{Bethe:1934za,
	AUTHOR=	"Bethe, H. and Heitler, W.",
	TITLE=	"On the stopping of fast particles and on the creation of positive electrons",
	DOI=	"10.1098/rspa.1934.0140",
	JOURNAL=	"Proc. Roy. Soc. Lond. A",
	VOLUME=	"146",
	PAGES=	"83",
	YEAR=	"1934",
}

@TECHREPORT{tagandprobe,
	AUTHOR=	"{CMS Collaboration}",
	TITLE=	"Measurement of tracking efficiency",
	NUMBER=	"CMS-PAS-TRK-10-002",
	YEAR=	"2010",
	TYPE=	"CMS Physics Analysis Summary",
	URL=	"https://cds.cern.ch/record/1279139",
	NOTE=	"\url{https://cds.cern.ch/record/1279139}",
}

@ARTICLE{CMS:2018mlc,
	AUTHOR=	"{CMS Collaboration}",
	TITLE=	"Measurement of the inelastic proton-proton cross section at $ \sqrt{s}=13 $ {TeV}",
	EPRINT=	"1802.02613",
	ARCHIVEPREFIX=	"arXiv",
	PRIMARYCLASS=	"hep-ex",
	REPORTNUMBER=	"CMS-FSQ-15-005, CERN-EP-2018-004",
	DOI=	"10.1007/JHEP07(2018)161",
	JOURNAL=	"JHEP",
	VOLUME=	"07",
	PAGES=	"161",
	YEAR=	"2018",
}

@ARTICLE{L1Prefire,
	AUTHOR=	"{CMS Collaboration}",
	TITLE=	"Performance of the {CMS} electromagnetic calorimeter in pp collisions at $ \sqrt{s}=13 $ {TeV}",
	EPRINT=	"2403.15518",
	ARCHIVEPREFIX=	"arXiv",
	PRIMARYCLASS=	"physics.ins-det",
	REPORTNUMBER=	"CMS-EGM-18-002, CERN-EP-2024-014",
	DOI=	"10.1088/1748-0221/19/09/P09004",
	JOURNAL=	"JINST",
	VOLUME=	"19",
	PAGES=	"P09004",
	YEAR=	"2024",
}

@ARTICLE{WZNNLOCalc,
	AUTHOR=	"Grazzini, Massimiliano and Kallweit, Stefan and Rathlev, Dirk and Wiesemann, Marius",
	TITLE=	"W$^{\pm}${Z} production at hadron colliders in {NNLO QCD}",
	EPRINT=	"1604.08576",
	ARCHIVEPREFIX=	"arXiv",
	PRIMARYCLASS=	"hep-ph",
	REPORTNUMBER=	"ZU-TH-15-16, MITP-16-037, NSF-KITP-16-046, DESY-16-074",
	DOI=	"10.1016/j.physletb.2016.08.017",
	JOURNAL=	"Phys. Lett. B",
	VOLUME=	"761",
	PAGES=	"179",
	YEAR=	"2016",
}

@ARTICLE{YR4,
	AUTHOR=	"{LHC Higgs Cross Section Working Group}",
	TITLE=	"Handbook of {LHC Higgs} cross sections: 4. {Deciphering} the nature of the {Higgs} sector",
	EPRINT=	"1610.07922",
	ARCHIVEPREFIX=	"arXiv",
	PRIMARYCLASS=	"hep-ph",
	REPORTNUMBER=	"CERN-2017-002-M, CERN-2017-002",
	DOI=	"10.23731/CYRM-2017-002",
	JOURNAL=	"CERN Yellow Rep. Monogr.",
	VOLUME=	"2",
	PAGES=	"1",
	YEAR=	"2017",
}

@ARTICLE{tXNLOCalc,
	AUTHOR=	"Degrande, Celine and Maltoni, Fabio and Mimasu, Ken and Vryonidou, Eleni and Zhang, Cen",
	TITLE=	"Single-top associated production with a {Z} or {H} boson at the {LHC}: the {SMEFT} interpretation",
	EPRINT=	"1804.07773",
	ARCHIVEPREFIX=	"arXiv",
	PRIMARYCLASS=	"hep-ph",
	REPORTNUMBER=	"CERN-TH-2018-092",
	DOI=	"10.1007/JHEP10(2018)005",
	JOURNAL=	"JHEP",
	VOLUME=	"10",
	PAGES=	"005",
	YEAR=	"2018",
}

@ARTICLE{CMS:2024onh,
	AUTHOR=	"{CMS Collaboration}",
	TITLE=	"The {CMS} statistical analysis and combination tool: {\textsc{Combine}}",
	EPRINT=	"2404.06614",
	ARCHIVEPREFIX=	"arXiv",
	PRIMARYCLASS=	"physics.data-an",
	REPORTNUMBER=	"CMS-CAT-23-001, CERN-EP-2024-078",
	DOI=	"10.1007/s41781-024-00121-4",
	JOURNAL=	"Comput. Softw. Big Sci.",
	VOLUME=	"8",
	PAGES=	"19",
	YEAR=	"2024",
}

@INPROCEEDINGS{Verkerke:2003ir,
	AUTHOR=	"Verkerke, Wouter and Kirkby, David P.",
	TITLE=	"The {RooFit} toolkit for data modeling",
	EPRINT=	"physics/0306116",
	ARCHIVEPREFIX=	"arXiv",
	REPORTNUMBER=	"CHEP-2003-MOLT007",
	BOOKTITLE=	"Proc. 13th International Conference for Computing in High-Energy and Nuclear Physics ({CHEP} 2003)",
	YEAR=	"2003",
	URL=	"https://arxiv.org/abs/physics/0306116",
}

@ARTICLE{Moneta:2010pm,
	AUTHOR=	"Moneta, Lorenzo and Cranmer, Kyle and Schott, Gregory and Verkerke, Wouter",
	TITLE=	"The {RooStats} project",
	DOI=	"10.22323/1.093.0057",
	JOURNAL=	"PoS",
	YEAR=	"2011",
	VOLUME=	"ACAT2010",
	PAGES=	"057",
}

@TECHREPORT{CMS-NOTE-2011-005,
	AUTHOR=	"{ATLAS and CMS Collaborations, and LHC Higgs Combination Group}",
	TYPE=	"CMS Note",
	TITLE=	"Procedure for the {LHC} {Higgs} boson search combination in summer 2011",
	NUMBER=	"CMS-NOTE-2011-005, ATL-PHYS-PUB-2011-11",
	YEAR=	"2011",
	URL=	"https://cds.cern.ch/record/1379837",
	NOTE=	"\url{https://cds.cern.ch/record/1379837}",
}

@ARTICLE{BARLOW1993219,
	AUTHOR=	"Barlow, Roger J. and Beeston, Christine",
	TITLE=	"Fitting using finite {Monte Carlo} samples",
	REPORTNUMBER=	"MAN-HEP-93-1",
	DOI=	"10.1016/0010-4655(93)90005-W",
	JOURNAL=	"Comput. Phys. Commun.",
	VOLUME=	"77",
	PAGES=	"219",
	YEAR=	"1993",
}

@INPROCEEDINGS{pvalue,
	AUTHOR=	"Demortier, Luc",
	TITLE=	"{P} values and nuisance parameters",
	BOOKTITLE=	"Proc. {PHYSTAT LHC} {Workshop} on Statistical Issues for {LHC} Physics",
	ADDRESS=	"Geneva, Switzerland",
	YEAR=	"2008",
	PAGES=	"23",
	DOI=	"10.5170/CERN-2008-001",
	REPORTNUMBER=	"CERN-2008-001",
	URL=	"http://cds.cern.ch/record/1099967/files/p23.pdf",
	SLACCITATION=	"%%CITATION = CERN-2008-001;%%",
}

@ARTICLE{Junk_1999,
	AUTHOR=	"Junk, Thomas",
	TITLE=	"Confidence level computation for combining searches with small statistics",
	EPRINT=	"hep-ex/9902006",
	ARCHIVEPREFIX=	"arXiv",
	REPORTNUMBER=	"CARLETON-OPAL-PHYS-99-01, CERN-EP-99-041",
	DOI=	"10.1016/S0168-9002(99)00498-2",
	JOURNAL=	"Nucl. Instrum. Meth. A",
	VOLUME=	"434",
	PAGES=	"435",
	YEAR=	"1999",
}

@ARTICLE{Read_2002,
	AUTHOR=	"Read, Alexander L.",
	EDITOR=	"Whalley, M. R. and Lyons, L.",
	TITLE=	"Presentation of search results: the {CL$_\text{s}$} technique",
	DOI=	"10.1088/0954-3899/28/10/313",
	JOURNAL=	"J. Phys. G",
	VOLUME=	"28",
	PAGES=	"2693",
	YEAR=	"2002",
}

\cleardoublepage \appendix\section{The CMS Collaboration \label{app:collab}}\begin{sloppypar}\hyphenpenalty=5000\widowpenalty=500\clubpenalty=5000\cmsinstitute{Yerevan Physics Institute, Yerevan, Armenia}
{\tolerance=6000
A.~Gevorgyan\cmsorcid{0000-0003-2751-9489}, A.~Hayrapetyan, V.~Makarenko\cmsorcid{0000-0002-8406-8605}, A.~Tumasyan\cmsAuthorMark{1}\cmsorcid{0009-0000-0684-6742}
\par}
\cmsinstitute{Marietta Blau Institute for Particle Physics, Vienna, Austria}
{\tolerance=6000
W.~Adam\cmsorcid{0000-0001-9099-4341}, L.~Benato\cmsorcid{0000-0001-5135-7489}, T.~Bergauer\cmsorcid{0000-0002-5786-0293}, M.~Dragicevic\cmsorcid{0000-0003-1967-6783}, S.~Gundacker\cmsorcid{0000-0003-2087-3266}, A.K.~Guven\cmsorcid{0009-0004-5670-5138}, P.S.~Hussain\cmsorcid{0000-0002-4825-5278}, M.~Jeitler\cmsAuthorMark{2}\cmsorcid{0000-0002-5141-9560}, N.~Krammer\cmsorcid{0000-0002-0548-0985}, A.~Li\cmsorcid{0000-0002-4547-116X}, D.~Liko\cmsorcid{0000-0002-3380-473X}, M.~Matthewman, J.~Schieck\cmsAuthorMark{2}\cmsorcid{0000-0002-1058-8093}, R.~Sch\"{o}fbeck\cmsAuthorMark{2}\cmsorcid{0000-0002-2332-8784}, M.~Shooshtari\cmsorcid{0009-0004-8882-4887}, M.~Sonawane\cmsorcid{0000-0003-0510-7010}, N.~Van~Den~Bossche\cmsorcid{0000-0003-2973-4991}, W.~Waltenberger\cmsorcid{0000-0002-6215-7228}, C.E.~Wulz\cmsAuthorMark{2}\cmsorcid{0000-0001-9226-5812}
% ...
\par}
$^{1}$Also at Yerevan State University, Yerevan, Armenia\\
$^{2}$Also at Technische Universit\"{a}t Wien, Vienna, Austria\\
% ...
\end{sloppypar}
\end{document}